\pdfoutput=1
\documentclass[cernpreprint, atlasdraft=false, texlive=2023, UKenglish, texmf, orcidlogo]{atlasdoc}
 
\usepackage{atlaspackage}
\usepackage[backref=false]{atlasbiblatex}
 
\usepackage{atlasphysics}
 
\graphicspath{{logos/}{figures/}}
 
\usepackage{ANA-EXOT-2018-52-PAPER-defs}

\AtlasTitle{Search for single production of vector-like $T$ quarks decaying into
$Ht$ or $Zt$ in $pp$ collisions at $\sqrt{s}=13$~TeV with the ATLAS detector}
 
\AtlasAbstract{
 
This paper describes a search for the single production of an up-type vector-like quark ($T$) decaying
as $T \rightarrow Ht$ or $T \rightarrow Zt$.
The search utilises a dataset of $pp$ collisions at $\sqrt{s}=13$~TeV collected with the ATLAS detector
during the 2015--2018 data-taking period of the Large Hadron Collider, corresponding to an integrated luminosity
of 139~\ifb. Data are
analysed in final states containing a single lepton with multiple jets and $b$-jets. The presence of
boosted heavy resonances in the event is exploited to discriminate the signal from the Standard
Model background. No significant excess above the Standard Model expectation is observed, and 95\% CL upper
limits are set on the production cross section of $T$ quarks in different decay channels.
The results are interpreted in several benchmark scenarios to set limits on the mass and universal coupling
strength ($\kappa$) of the vector-like quark. For singlet $T$ quarks, $\kappa$ values above 0.53 are excluded
for all masses below 2.3~TeV. At a mass of 1.6~TeV, $\kappa$ values as low as 0.35 are excluded. For $T$ quarks
in the doublet scenario, where the production cross section is much lower, $\kappa$ values above 0.72 are excluded
for all masses below 1.7~TeV, and this exclusion is extended to $\kappa$ above 0.55 for low masses around 1.0~TeV.
}
\author{The ATLAS Collaboration}
 
\AtlasJournalRef{JHEP 08 (2023) 153}
 
\PreprintIdNumber{CERN-EP-2023-058}

\AtlasDOI{10.1007/JHEP08(2023)153}

\hypersetup{pdftitle={ATLAS document},pdfauthor={The ATLAS Collaboration}}
 
\begin{document}
 
\maketitle

\section{Introduction}
\label{sec:intro}
 
The discovery of a new particle consistent with the Standard Model (SM) Higgs boson by the
ATLAS~\cite{Aad:2012tfa} and CMS~\cite{Chatrchyan:2012ufa} experiments at the Large Hadron
Collider (LHC) represents a milestone in high-energy physics. A comprehensive programme of
measurements of the Higgs boson's properties to unravel its nature is underway at the LHC, so
far yielding results compatible with the SM predictions. However, a pressing question remains
as to why the electroweak mass scale (and the Higgs boson mass along with it) is so small
compared to the Planck scale, a situation known as the hierarchy problem. Naturalness
arguments~\cite{Susskind:1978ms} require that quadratic divergences that
arise from radiative corrections to the Higgs boson mass are cancelled out by some new mechanism
in order to avoid fine-tuning. To that effect, several explanations have been proposed in theories
beyond the SM (BSM theories).
 
One such solution involves the existence of a new strongly interacting sector, in which the
Higgs boson would be a pseudo Nambu--Goldstone boson (pNGB)~\cite{Hill:2002ap} of a spontaneously
broken global symmetry. The Composite
Higgs~\cite{Dimopoulos:1981xc,Kaplan:1983fs,Kaplan:1983sm} model is a particular realisation
of this scenario, which also addresses additional open 
questions in the SM, including the hierarchy in the mass spectrum of the SM particles.
A key prediction is the existence of new fermionic resonances
referred to as vector-like quarks (VLQs), which are also common in many other BSM scenarios
~\cite{Schmaltz:2005ky,Perelstein:2003wd,Gherghetta:2000qt,Martin:2009bg}.
Vector-like quarks are defined as colour-triplet spin-1/2 fermions whose left- and right-handed
chiral components have the same transformation properties under the weak-isospin SU(2) gauge
group~\cite{delAguila:1982fs,Aguilar-Saavedra:2013wba}. Assuming an unchanged scalar sector with
respect to the Standard Model, theoretical constraints on
renormalisability and gauge completeness restrict the SU(2) representation of the vector-like
quarks to seven possible multiplets: singlets ($T^{2/3}$) or ($B^{-1/3}$), doublets ($X^{5/3}~T^{2/3}$) or
($T^{2/3}~B^{-1/3}$), or triplets ($X^{5/3}~T^{2/3}~B^{-1/3}$) or ($T^{2/3}~B^{-1/3}~Y^{-4/3}$).\footnote{The
indices in superscript indicate the electric charges of the vector-like quarks. These indices
are omitted in the notation for the rest of the paper}
The vector-like quarks in these models are expected to couple preferentially
to third-generation quarks and can have flavour-changing neutral-current decays in addition
to the charged-current decays characteristic of chiral quarks~\cite{delAguila:1982fs,AguilarSaavedra:2009es}.
Thus, the up-type $T$ quark can decay into a $W$ boson and a $b$-quark, and also into a top quark and
a $Z$ or Higgs boson. The relative couplings to the massive bosons of the Standard
Model are determined by the gauge representation of the vector-like quarks.
The relative couplings to the $W$, $Z$ and Higgs bosons can be expressed in terms of
the $\xi_{W}$, $\xi_{Z}$ and $\xi_{H}$ parameters, respectively~\cite{Buchkremer:2013bha}. In
the asymptotic limit of large VLQ mass, these $\xi$ parameters correspond to the branching ratios
of the $T$ quark into their respective decay modes. The asymptotic limit holds to a very good approximation
for VLQ masses above 1~\TeV.
For a $T$ singlet, $\xi_{W}=0.5$ and $\xi_{Z}=\xi_{H}=0.25$. For $T$ quarks that are in an ($X~T$) doublet,
or in a ($T~B$) doublet with mixing only to up-type SM singlets~\cite{Buchkremer:2013bha,AguilarSaavedra:2009es},
$\xi_{W}=0$, and $\xi_{Z}=\xi_{H}=0.5$.
 
At the LHC, vector-like quarks with masses below $\sim$1$~\tev$ would be produced mostly in
pairs via the strong interaction. For higher masses, however, single production, mediated by the
electroweak interaction, may dominate depending on the coupling strength of the interaction between
the vector-like quark and the SM quarks~\cite{Aguilar-Saavedra:2013wba}. The single-production channel for vector-like quarks
provides a unique opportunity to probe the universal coupling strength ($\kappa$) in addition to
the relative couplings and branching ratios that can be probed in pair production searches.
The universal coupling strength controls both the production cross section and the resonance
width of the VLQ. For a VLQ with mass $m_{T}$, the resonance width $\Gamma_{T}$
scales as $\Gamma_{T}\propto\kappa^{2}m_{T}^{3}$~\cite{Buchkremer:2013bha}. Thus, the relative width ($\Gamma_{T}/m_{T}$)
of the VLQ resonance scales quadratically
with both $\kappa$ and $m_{T}$, and is independent of the multiplet representation.
 
\begin{figure*}[htb]
\centering
\subfloat[]{\includegraphics[width=0.5\textwidth, trim={3cm 5cm 3cm 3cm}]{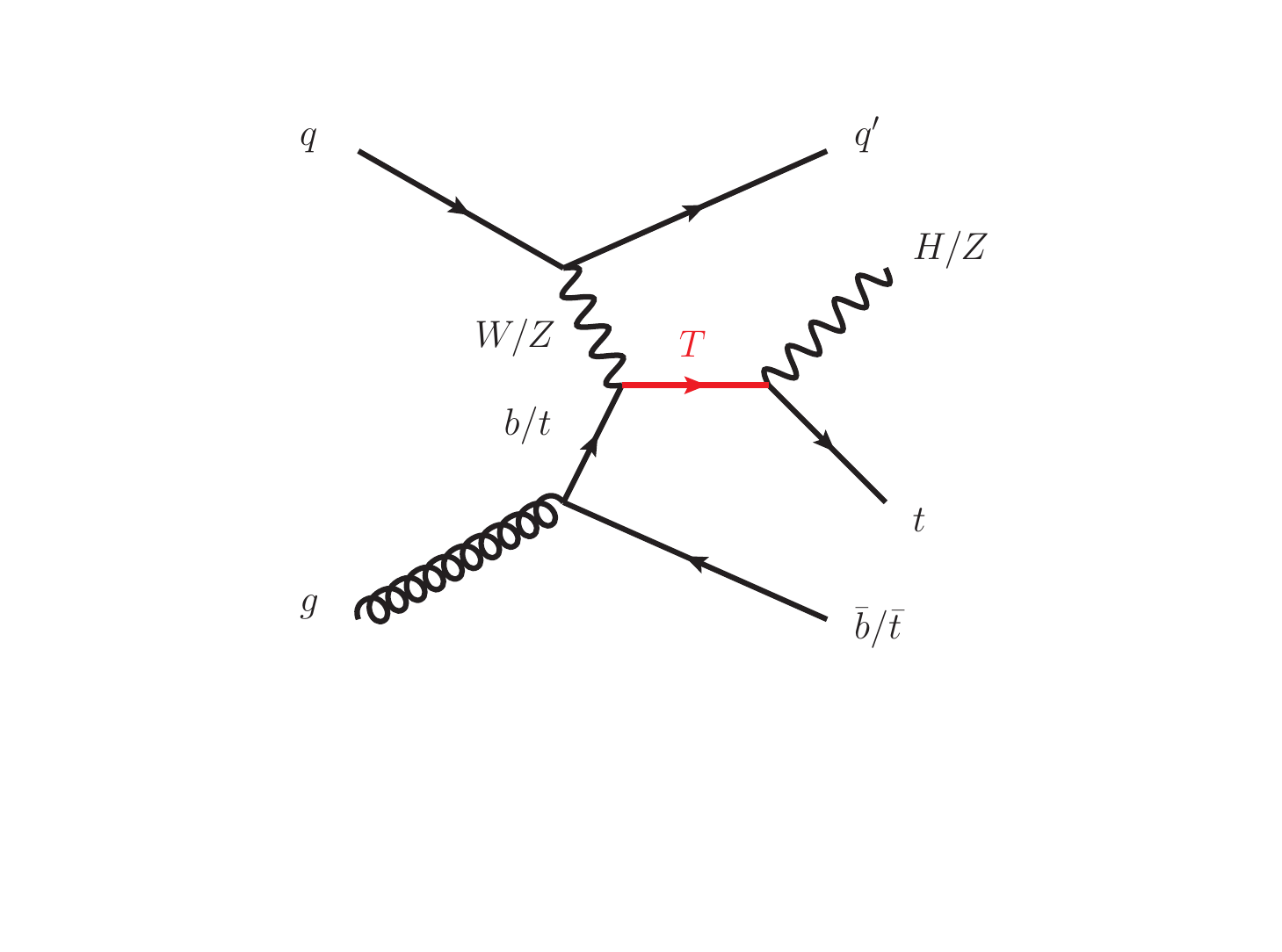}\label{subfig:singleT_resonant}}
\subfloat[]{\includegraphics[width=0.5\textwidth, trim={2cm 4cm 2cm 2cm}]{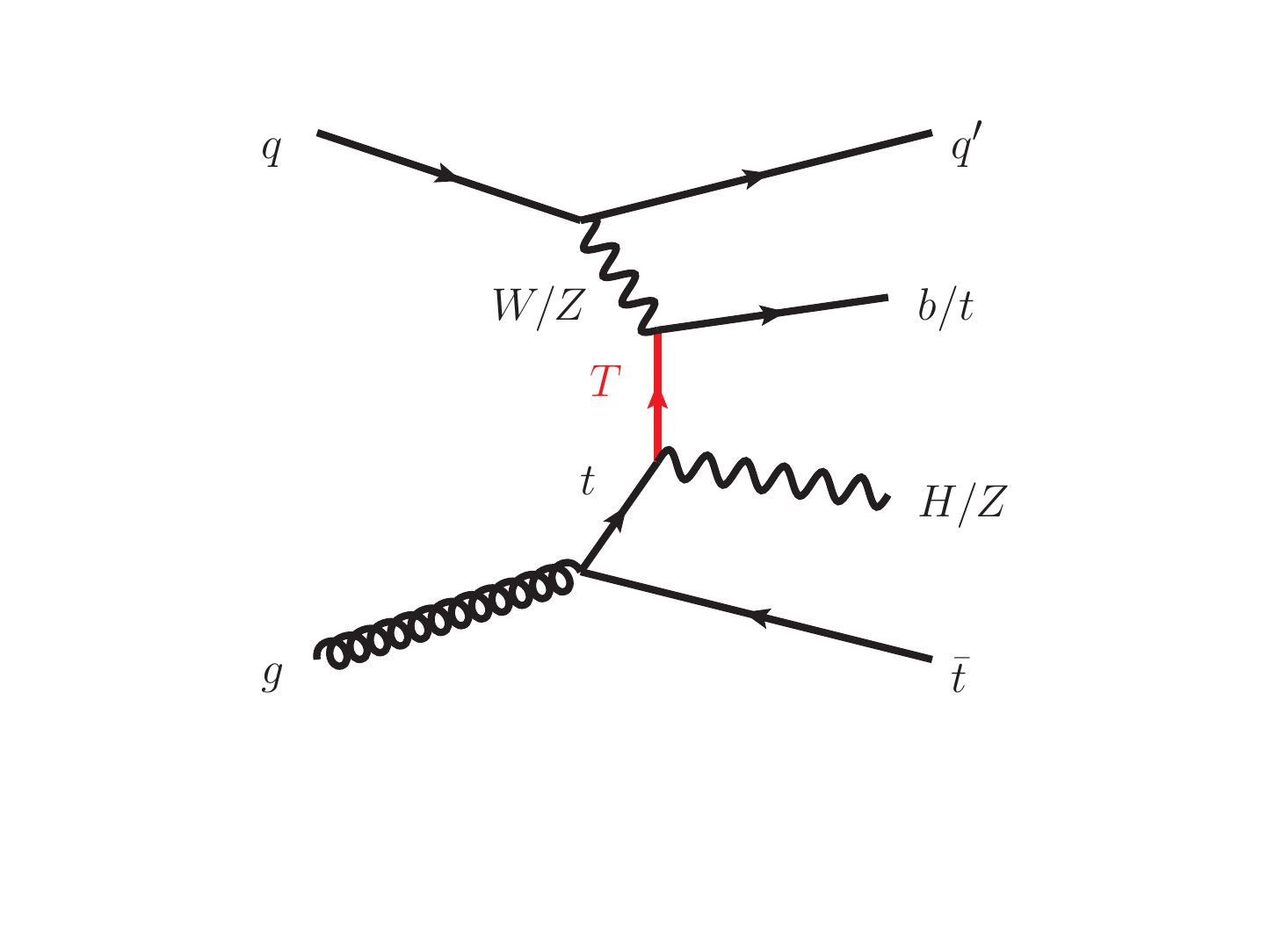}\label{subfig:singleT_nonresonant}}
\caption{Example of leading-order Feynman diagrams of single vector-like $T$ production in association with
a $b$-quark or $t$-quark and subsequent decay into either $Ht$ or $Zt$ in the (a) resonant and (b) non-resonant modes.}
\label{fig:singleT}
\end{figure*}

The dominant channel for resonant production of a single $T$ quark is $t$-channel production
mediated by a gauge boson (Figure~\ref{subfig:singleT_resonant}). The final state is characterised by
the presence of multiple ($b$-tagged) jets from the decay of the produced heavy quarks and bosons,
along with the recoiling initial-state quark, which typically manifests as a forward jet.
In the four-flavour scheme, and assuming
that the $T$ quark couples only to SM quarks of the third generation, this process requires an initial-state
gluon to split into a \bbbar\ or \ttbar\ pair. Given the difference in masses between the
top and bottom quarks, $b$-associated (or $W$-mediated) $T$-quark production is kinematically
favoured over $t$-associated (or $Z$-mediated) production. However, in certain gauge representations,
such as for a ($T~B$) doublet with mixing only to up-type SM singlets~\cite{Buchkremer:2013bha,AguilarSaavedra:2009es},
or for an ($X~T$) doublet, the coupling to $W$ bosons vanishes, and the $t$-associated mode
is the only allowed production channel. Therefore, both production modes are theoretically
interesting, even though the parameter space in the limit of $\xi_{W} \rightarrow 0$ is difficult
to probe due to small production cross sections. Non-resonant $T$-quark production
(Figure~\ref{subfig:singleT_nonresonant}) is subdominant compared to resonant production in the $b$-associated mode.
By contrast, in the $t$-associated mode, both the resonant and non-resonant $T$-quark production processes
have comparable cross sections, as both require the splitting of an initial-state gluon into a $\ttbar$ pair.
The relative contribution of non-resonant production grows with resonance width, and is therefore larger at higher coupling strengths.
 
Physical realisations of Composite Higgs models require the presence of additional scalar~\cite{Chala:2017xgc}
and vector bosons~\cite{Criado:2019mvu,Chala:2014mma,Araque:2015cna} for UV-completeness, thus opening
up production and decay channels for the VLQs in addition to the ones discussed
above. However, processes involving additional BSM particles are not considered in this paper, and results
are interpreted in the context of a minimal extension of the Standard Model including only one VLQ multiplet~\cite{Buchkremer:2013bha}.
 
The ATLAS and CMS collaborations have a broad programme of searches that has largely focused on
pair production of vector-like quarks, targeting different decay modes and final states separately.
A combination of all ATLAS pair production analyses using the data collected by
the ATLAS detector in 2015 and 2016 delivered the most stringent limits to date on pair-produced
vector-like quarks~\cite{Aaboud:2018pii}, with masses observed to be excluded below 1.31~\TeV\ for $T$
and 1.03~\TeV\ for $B$ for any combination of decay modes. The $T$ ($B$) singlet configuration was excluded
for masses up to 1.31 (1.21)~\TeV, and $T$ ($B$) quarks in the ($T~B$) doublet configuration were
excluded up to masses of 1.37 (1.35)~\TeV. Searches by both ATLAS~\cite{ATLAS:2022ozf,Aaboud:2018zpr,Aaboud:2018ifs,Aaboud:2018saj}
and CMS~\cite{Sirunyan:2017tfc,Khachatryan:2016vph,Sirunyan:2017ynj,CMS:2018dcw,CMS-B2G-16-005,CMS-B2G-18-003} targeting
single vector-like quark production have set limits
on the allowed vector-like quark parameter space in terms of model-dependent parameters such as
coupling strengths and mixing angles. These searches have mostly focused on the $b$-associated single
production modes for VLQs. A search by ATLAS for single VLQ production in the $T \rightarrow Wb$ channel has
excluded $\sigma \times \BR(T\to Wb)$ above $\sim$100~fb for $b$-associated $T$ production in the mass
range of 1.0--1.9~\TeV~\cite{Aaboud:2018ifs}. An ATLAS search~\cite{Aaboud:2018saj} in the
$T \rightarrow Z (\ell\ell)t$ channel has excluded $\sigma \times \BR(T\to Zt)$ for $b$-associated $T$
production above $\sim$90~fb ($\sim$40~fb) at a mass of 1.0~\TeV\ (2.0~\TeV) in the singlet hypothesis.
A CMS search for $t$-associated $T$ production, also in $T \rightarrow Z (\ell\ell)t$ final
states~\cite{Sirunyan:2017ynj}, has excluded $\sigma \times \BR(T\to Zt)$ above $\sim$100~fb ($\sim$40~fb)
at a mass of 0.8~\TeV\ (1.7~\TeV) under the doublet hypothesis. Conversely, the CMS search for $T$ production
in the $T \rightarrow Z (\nu\nu)t$ channel~\cite{CMS:2022yxp} excluded $\sigma \times \BR(T\to Zt)$ down
to $\sim$200~fb ($\sim$20~fb) at a mass of 0.6~\TeV\ (1.8~\TeV) under the singlet hypothesis, for a decay width
of 30\%. A recently published search by ATLAS in the all-hadronic final state~\cite{ATLAS:2022ozf} excludes
$\sigma \times \BR(T\to Ht)$ for $b$-associated $T$ production above $\sim$30--100~fb in the 1.0--2.3~\TeV\
mass range for a wide range of coupling strengths.
 
These exclusion limits were all derived at 95\% CL, assuming that the VLQs couple exclusively to SM
quarks in the third generation, and production modes involving other BSM particles were not considered.
 
This paper presents a search for the single production of the up-type vector-like quark $T$, with
subsequent decays into $Ht$ with $H \to \bbbar$, or into $Zt$ with $Z \to q\bar{q}$
(Figure~\ref{fig:singleT}). Both the $b$- and $t$-associated production modes are considered in this search.
The search uses 139~fb$^{-1}$ of $pp$ collision data at
$\sqrt{s}=13~\tev$ recorded
by the ATLAS Collaboration during Run 2 of the Large Hadron Collider (LHC). Data are analysed in
the lepton+jets final state, characterised by an isolated electron or muon with high transverse
momentum and multiple jets and $b$-jets. The presence of heavy hadronic boosted objects in the events
is used as an important distinguishing characteristic of the signal. In the absence of a significant excess above the SM expectation, the results are used to set upper limits
on the single production of $T$ quarks for several scenarios of the mass, the universal coupling strength
$\kappa$, and the relative couplings to $W$, $Z$, and Higgs bosons.


\newcommand{\AtlasCoordFootnote}{
ATLAS uses a right-handed coordinate system with its origin at the nominal interaction point (IP)
in the centre of the detector and the \(z\)-axis along the beam pipe.
The \(x\)-axis points from the IP to the centre of the LHC ring,
and the \(y\)-axis points upwards.
Cylindrical coordinates \((r,\phi)\) are used in the transverse plane,
\(\phi\) being the azimuthal angle around the \(z\)-axis.
The pseudorapidity is defined in terms of the polar angle \(\theta\) as \(\eta = -\ln \tan(\theta/2)\).
Angular distance is measured in units of \(\Delta R \equiv \sqrt{(\Delta\eta)^{2} + (\Delta\phi)^{2}}\).}

\section{ATLAS detector}
\label{sec:detector}
 
The ATLAS detector~\cite{PERF-2007-01} at the LHC covers nearly the entire solid angle around the collision point.\footnote{\AtlasCoordFootnote}
It consists of an inner tracking detector surrounded by a thin superconducting solenoid, electromagnetic and hadron calorimeters,
and a muon spectrometer incorporating three large superconducting air-core toroidal magnets.
 
The inner-detector system is immersed in a \SI{2}{\tesla} axial magnetic field
and provides charged-particle tracking in the range \(|\eta| < 2.5\).
The high-granularity silicon pixel detector covers the vertex region and typically provides four measurements per track,
the first hit normally being in the insertable B-layer installed before Run~2~\cite{ATLAS-TDR-19,PIX-2018-001}.
It is followed by the silicon microstrip tracker, which usually provides eight measurements per track.
These silicon detectors are complemented by the transition radiation tracker (TRT),
which enables radially extended track reconstruction up to \(|\eta| = 2.0\).
The TRT also provides electron identification information
based on the fraction of hits (typically 30 in total) above a higher energy-deposit threshold corresponding to transition radiation.
 
The calorimeter system covers the pseudorapidity range \(|\eta| < 4.9\).
Within the region \(|\eta|< 3.2\), electromagnetic calorimetry is provided by barrel and
endcap high-granularity lead/liquid-argon (LAr) calorimeters,
with an additional thin LAr presampler covering \(|\eta| < 1.8\)
to correct for energy loss in material upstream of the calorimeters.
Hadron calorimetry is provided by the steel/scintillator-tile calorimeter,
segmented into three barrel structures within \(|\eta| < 1.7\), and two copper/LAr hadron endcap calorimeters.
The solid angle coverage is completed with forward copper/LAr and tungsten/LAr calorimeter modules
optimised for electromagnetic and hadronic energy measurements respectively.
 
The muon spectrometer (MS) comprises separate trigger and
high-precision tracking chambers measuring the deflection of muons in a magnetic field generated by the superconducting air-core toroidal magnets.
The field integral of the toroids ranges between \num{2.0} and \SI{6.0}{\tesla\metre}
across most of the detector.
Three layers of precision chambers, each consisting of layers of monitored drift tubes, cover the region \(|\eta| < 2.7\),
complemented by cathode-strip chambers in the forward region, where the background is highest.
The muon trigger system covers the range \(|\eta| < 2.4\) with resistive-plate chambers in the barrel, and thin-gap chambers in the endcap regions.
 
Interesting events are selected by the first-level trigger system implemented in custom hardware,
followed by selections made by algorithms implemented in software in the high-level trigger~\cite{TRIG-2016-01}.
The first-level trigger accepts events from the \SI{40}{\MHz} bunch crossings at a rate below \SI{100}{\kHz},
which the high-level trigger further reduces in order to record events to disk at about \SI{1}{\kHz}.
 
An extensive software suite~\cite{ATL-SOFT-PUB-2021-001} is used in data simulation, in the reconstruction
and analysis of real and simulated data, in detector operations, and in the trigger and data acquisition
systems of the experiment.


\section{Object reconstruction}
\label{sec:obj_reco}
 
Interaction vertices from proton--proton collisions are reconstructed from at least two tracks
with transverse momentum ($\pt$) larger than $500~\mev$ that are consistent with originating from
the beam collision region in the $x$--$y$ plane. If more than one primary vertex candidate is found, the
candidate whose associated tracks form the largest sum of squared $\pt$~\cite{ATL-PHYS-PUB-2015-026}
is selected as the hard-scatter primary vertex.
 
Electron candidates~\cite{PERF-2017-01} are reconstructed from energy
clusters in the electromagnetic calorimeter associated with reconstructed tracks in the inner
detector. They are required to have $\pt>30~\gev$ and $|\eta_{\textrm{cluster}}| < 2.47$.
Candidates in the transition region between the electromagnetic barrel and endcap calorimeters
($1.37 < |\eta_{\textrm{cluster}}| < 1.52$) are excluded. They are required to satisfy the \enquote{tight}
likelihood-based identification criteria~\cite{PERF-2017-01} based on calorimeter, tracking
and combined variables that provide separation between electrons and jets.
Muon candidates~\cite{ATLAS:2020auj} are reconstructed by matching tracks in the MS
to those found in the inner detector. The resulting muon candidates are
re-fitted using the complete track information from both detector systems. Muon candidates are
required to satisfy the \enquote{medium} identification criteria~\cite{ATLAS:2020auj}, and to have
$\pt>30~\gev$ and $|\eta|<2.5$. Electron (muon) candidates are matched to the primary vertex by
requiring that the significance of their transverse impact parameter, $d_0$, satisfies
$|d_0/\sigma(d_0)|<5(3)$, where $\sigma(d_0)$ is the measured uncertainty in $d_0$,
and by requiring that their longitudinal impact parameter, $z_0$, satisfies $|z_0 \sin\theta|<0.5$~mm.
To further reduce the background from non-prompt leptons, photon conversions and hadrons, the lepton candidates
are also required to be isolated in the tracking system and the calorimeter.
 
Candidate jets are reconstructed with the anti-$k_t$ algorithm~\cite{Cacciari:2008gp,Fastjet} with a
radius parameter $R=0.4$ (referred to as \enquote{small-$R$ jets}). The input constituents for jet reconstruction
are built by combining topological clusters of energy in the calorimeter~\cite{Aad:2016upy} with measured tracks in the
inner detector using the particle-flow algorithm~\cite{PERF-2015-09}. The reconstructed jets are then calibrated to the
particle level by the application of a jet energy scale derived from simulation and {\textit{in situ}} corrections
based on $\sqrt{s}=13~\tev$ data~\cite{Aaboud:2017jcu}. Calibrated jets are required to have $\pt > 25~\gev$ and $|\eta| < 4.5$.
Quality criteria are imposed to reject events that contain any jets arising from non-collision sources
or detector noise~\cite{ATLAS-CONF-2015-029}. Jets with $|\eta| < 2.5$ are labelled \enquote{central} jets for the purposes
of event selection and categorisation, while jets with $2.5 < |\eta| < 4.5$ are called \enquote{forward} jets. Requirements
based on the jet-vertex tagger (JVT)~\cite{PERF-2014-03} algorithm are imposed on central jets with $\pt<60~\gev$
and $|\eta| < 2.4$ to suppress contamination from jets that originate from pile-up interactions. Pile-up contamination
for jets with $|\eta| > 2.4$ is reduced by requirements on the closely related forward JVT (fJVT) algorithm for jets with
$\pt<120~\gev$~\cite{PERF-2016-06}.
 
In order to identify $b$-hadrons in the event, variable-$R$~\cite{Krohn:2009zg} jets built from reconstructed tracks in the
inner detector are used. Track-jets containing $b$-hadrons are identified ($b$-tagged) via the multivariate
\enquote{DL1} algorithm~\cite{FTAG-2019-07}, which uses information about the kinematic and topological properties
of displaced tracks associated with the jet, and of secondary and tertiary decay vertices reconstructed from these
tracks. For each jet, a value of the multivariate $b$-tagging discriminant is
calculated. In this analysis, a jet is considered $b$-tagged if this value is above the threshold corresponding to
an average 77\% efficiency to tag a $b$-quark jet, with a light-jet\footnote{Light-jet refers to a jet originating
from the hadronisation of a light quark ($u$, $d$, $s$) or a gluon.} rejection factor of $\sim$112 and a charm-jet
rejection factor of $\sim$5, as determined for jets with $\pt >20~\gev$ and $|\eta|<2.5$ in simulated $\ttbar$ events.
 
Overlaps between candidate objects are removed sequentially.  Firstly, electron candidates that lie within $\Delta R = 0.01$ of a muon
candidate are removed to suppress contributions from muon bremsstrahlung.  Overlaps between electron and jet candidates are resolved
next, and finally, overlaps between remaining jet candidates and muon candidates are removed.
Clusters associated with identified electrons are not excluded during jet reconstruction.
In order to avoid double-counting of electrons as jets, the closest jet whose axis is within ${\Delta}R = 0.2$ of an
electron is discarded. If the electron is within ${\Delta}R = 0.4$ of the axis of any jet after this initial removal,
the jet is retained and the electron is removed.
The overlap removal procedure applied to the remaining jet candidates and muon candidates is designed to remove those muons
that are likely to have arisen in the decay chain of hadrons and to retain the overlapping jet instead.
Jets and muons may also appear in close proximity when the jet results from high-$\pt$ muon bremsstrahlung,
and in such cases the jet should be removed and the muon retained. Such jets are characterised by having very
few matching inner-detector tracks. Selected muons that satisfy $\Delta R(\mu,{\textrm{jet}}) < 0.04+10~\gev/\pt^\mu$ are rejected
if the jet has at least three tracks originating from the primary vertex; otherwise the jet is removed and the muon is kept.
 
The candidate small-$R$ jets surviving the overlap removal procedure discussed above are used as inputs for further
jet reclustering~\cite{Nachman:2014kla} using the variable-$R$ jet reconstruction algorithm with $\rho=550~\gev$.
The parameter $\rho$ controls the evolution of the effective size of the reclustered jet as $R = \rho/\pt$.
Since the input constituents of these reclustered jets are already calibrated, no further calibration of them is
required. Uncertainties in the energy and mass scales and resolutions of the constituent small-$R$ jets are propagated
to the kinematics of the reclustered (RC) jets. In order to suppress contributions from pile-up and
soft radiation, the RC jets are trimmed~\cite{Krohn:2009th} by removing all small-$R$ (sub)jets within a RC jet
that have $\pt$ below 5\% of the $\pt$ of the reclustered jet.
Due to the pile-up suppression and $\pt>25~\gev$ requirements imposed on the small-$R$ jets, the average fraction of small-$R$ jets removed
by the trimming requirement is less than 1\%.
The resulting RC jets are required to have $|\eta|<$~2.0 and are used to identify high-$\pt$ hadronically decaying
top quark, Higgs boson or $W$/$Z$ boson candidates by placing requirements on their transverse momentum, mass, and number of constituents.
Hadronically decaying top-quark candidates are reconstructed as RC jets with $\pt > 400~\gev$ and mass larger than 140~\gev.
Top-quark candidates with $\pt<700~\gev$ are required to contain at least two constituent subjets. Higgs boson candidates
are reconstructed as RC jets with $\pt > 350~\gev$, a mass between 105 and
140~\gev, and a $\pt$-dependent requirement on the number of subjets (exactly two for $\pt < 600~\gev$,
and one or two for $\pt >600~\gev$). RC jets are tagged as arising from a $W$ or $Z$ boson if they have $\pt > 350~\gev$,
and a mass between 70 and 105~\gev. Furthermore, candidate $W$/$Z$ ($V$ boson) RC jets are required to have exactly two subjets
if they have $\pt < 450~\gev$, while candidates with one or two subjets are allowed at higher $\pt$.
In the following, these are referred to as \enquote{$t$-tagged}, \enquote{$H$-tagged} and \enquote{$V$-tagged} jets, respectively, while
the term \enquote{jet} without further qualifiers is used to refer to central small-$R$ jets.
 
The missing transverse momentum $\mpt$ (with magnitude $\met$) is defined as the negative vector sum of the
$\vec{\pt}$ of all selected and calibrated objects in the event, including a term to account for energy from soft particles
in the event which are not associated with any of the selected objects.
This soft term is calculated from inner-detector tracks matched to the selected primary vertex to make it more resilient to
contamination from pile-up interactions~\cite{Aaboud:2018tkc}.
 
To reconstruct boosted leptonic top-quark candidates in the event, the lepton and $\mpt$ in the event are first
used to construct a leptonic $W$-boson candidate. The azimuthal direction and \pt\ of
the neutrino daughter of the $W$ boson are then chosen to be the same as the azimuthal
direction and magnitude of the missing transverse momentum, and the neutrino $p_{z}$
is determined by imposing a $W$ mass constraint on the lepton--neutrino system. This amounts to solving a
quadratic equation. If two solutions exist to the quadratic equation, the one with the lowest neutrino $p_{z}$ is chosen. If no solution exists, the \MET value is shifted
until the equation has a unique solution. The leptonic $W$-boson candidate thus constructed is combined with a candidate $b$-jet
in a two-step process. First, the $b$-tagged track-jet that is closest to the leptonic $W$-boson candidate in
$\eta$--$\phi$ space is identified. Then, the closest small-$R$ jet to this $b$-tagged jet with ${\Delta}R<0.4$
is found and combined with the leptonic $W$-boson candidate. The $b$-jet is required to be within $\Delta R=1.5$ of the
leptonic $W$-boson candidate. Furthermore, in order to avoid double-counting, any $b$-jet within $\Delta R=1.0$ of a $t$-tagged,
$H$-tagged or $V$-tagged jet is not considered for leptonic top-quark
reconstruction. The reconstructed leptonic top-quark candidate is required to have $\pt >300~\gev$. The average
reconstruction efficiency with these requirements in signal events is around 50\%.


\section{Data sample and event preselection}
\label{sec:data_presel}
 
As described in Section~\ref{sec:intro}, this search is based on a dataset of $pp$
collisions at $\sqrt{s}=13~\tev$ with 25~ns
bunch spacing collected by the ATLAS experiment during the 2015--2018 data-taking period,
corresponding to an integrated luminosity of $139~\ifb$. Only events recorded with a
single-electron trigger, a single-muon trigger, or a $\met$ trigger under stable beam
conditions and for which all detector subsystems were operational are considered~\cite{ATLAS:2016wtr}.
 
Single-lepton triggers~\cite{TRIG-2018-01,TRIG-2018-05} with low $\pt$ threshold and lepton isolation requirements are
combined in a logical OR with higher-threshold triggers without isolation requirements
to give maximum efficiency. For muons, triggers with a $\pt$ threshold of 20 (26)~\gev\
in 2015 (2016--2018) and isolation requirements, are combined with triggers with a 50~\gev\
threshold with no isolation requirement. A trigger with a 60~\gev\ threshold was added
for the 2017--2018 data-taking period. The lowest $\pt$ threshold for electron triggers
was 24~(26)~\gev\ in 2015 (2016--2018), while the highest $\pt$ threshold
varied from 120~\gev\ to 300~\gev\ during the data-taking period. The {\met}
trigger~\cite{TRIG-2019-01} used an $\met$ threshold of 70~\gev\ in the
HLT in 2015 and a run-period-dependent {\met} threshold varying between
90~\gev\ and 110~\gev\ in other years.
 
Events satisfying the trigger selection are required to have at least one primary vertex
candidate, and exactly one selected electron or muon. For events that only pass a single-lepton
trigger, the selected lepton is required to match the lepton reconstructed by the trigger
within $\Delta R < 0.15$. Events that do not satisfy the single-lepton trigger acceptance or
matching conditions are selected only if they pass the {\met} trigger requirements and
the offline reconstructed {\met} is above 200~\gev. The combination of single-lepton and
{\met} triggers maximises the efficiency of the trigger selection.
 
In addition to the above, events are required to
have at least three small-$R$ jets and at least one $b$-tagged variable-radius jet.
All selected small-$R$
jets are considered for the purpose of event selection and categorisation, including those
used to build RC jets.
 
The background from multijet production is suppressed by placing requirements on $\met$
as well as on the transverse mass of the lepton and $\met$ system ($\mtw$):\footnote{$\mtw =
\sqrt{2 p^\ell_{\mathrm T} \met (1-\cos\Delta\phi)}$, where $p^\ell_{\mathrm T}$
is the transverse momentum (energy) of the muon (electron) and $\Delta\phi$ is the azimuthal
angle separation between the lepton and the direction of the missing transverse momentum.}
$\met >20~\gev$ and $\met +\mtw>60~\gev$.
 
In order to select events that are kinematically close to those expected from signal, an
additional selection is placed on the \enquote{effective mass} ({\meff}) observable,
which is defined as the scalar sum of the \pt\ of all central small-$R$ jets and leptons in the event and the \met\ .
All events considered in the analysis are required to have $\meff > 600~\gev$.
 
The above requirements are referred to as the \enquote{preselection} and are summarised in Table~\ref{tab:preselection}.
 
\begin{table*}[t!]
\begin{center}
\caption{\small{Summary of preselection requirements.}}
\label{tab:preselection}
\begin{tabular}{c}
\toprule\toprule
\multicolumn{1}{c}{Preselection requirements } \\
\midrule
Single-lepton or $\met$ trigger \\
=1 isolated $e$ OR $\mu$ \\
$\geq$3 jets  \\
$\geq$1 $b$-tagged jets \\
$\met >20~\gev$ \\
$\met +\mtw>60~\gev$ \\
$\meff >600~\gev$ \\
\bottomrule\bottomrule
\end{tabular}
\end{center}
\end{table*}


\section{Signal and background modelling}
\label{sec:sig_bkgd_modelling}
 
Signal and background processes were modelled using Monte Carlo (MC) simulations.
In the simulation, the masses of the top quark and Higgs boson were set to 172.5~\GeV\ and
125~\GeV, respectively. All simulated samples, except those produced with the
{\Sherpa}~\cite{Gleisberg:2008ta} event generator, utilised the \EVTGEN program~\cite{Lange:2001uf}
to model the decays of heavy-flavour hadrons. While
\EVTGEN[1.2.0] was used for \ttW\ and \ttZ\ samples, \EVTGEN[1.6.0] was used for all the other samples.
In samples where the parton showering and hadronisation were modelled with \Pythia[8]~\cite{Sjostrand:2007gs}, a set of tuned parameters called the
A14 tune~\cite{ATL-PHYS-PUB-2014-021} was used for underlying-event modelling, and
the \NNPDF[2.3lo] parton distribution function (PDF) set~\cite{Ball:2012cx} was used for the showering and hadronisation processes. The
H7UE parameter tune~\cite{Bellm:2015jjp} was used instead for samples that
employ \Herwig[7] for hadronisation and showering, with the \MMHT[lo] PDF set~\cite{Harland-Lang:2014zoa}.
 
To model the effects of pile-up, events from minimum-bias interactions were generated
using the \Pythia[8.186] event generator and overlaid
on the simulated hard-scatter events according to the luminosity profile of the
recorded data. The generated events were processed through a simulation~\cite{Aad:2010ah}
of the ATLAS detector geometry and response using \GEANT~\cite{Agostinelli:2002hh}.
A faster simulation, where the full \GEANT simulation of the calorimeter
response is replaced by a detailed parameterisation of the shower shapes~\cite{FastCaloSim},
was adopted for some of the samples used to estimate systematic uncertainties. In these
cases, the systematic uncertainties were estimated by comparing these alternative samples with
versions of the nominal samples that were also processed through the same fast detector simulation.
Simulated events are processed through the same reconstruction software as the data, and
corrections are applied so that the object identification efficiencies, energy
scales and energy resolutions match those determined from data control samples.

\subsection{Signal modelling}
\label{sec:sig_modelling}
 
Single production of $T$ quarks was simulated with samples produced at leading order in the four-flavour
scheme with \MGNLO[2.3.3]~\cite{Alwall:2014hca}, using \NNPDF[3.0lo]~\cite{Ball:2014uwa} PDF sets. The generator
was interfaced to \Pythia[8.212]~\cite{Sjostrand:2014zea}
for the modelling of parton showering and hadronisation. The matrix elements were calculated according
to the phenomenological model given in Ref.~\cite{Buchkremer:2013bha} and all tree-level processes are included. The VLQs
are assumed to couple exclusively to SM quarks of the third generation. Separate samples are generated for \WTHt, \WTZt,
\ZTHt, and \ZTZt\ processes in the 1.0--2.3~\tev\ mass range at fixed values of mass and $\kappa$.
Matrix-element-based event weights~\cite{Mattelaer:2016gcx} calculated during the generation are used to reweight the events in each sample to other
values of mass and $\kappa$, to fill out a grid in the mass--$\kappa$ plane.
Samples at specific values of the relative couplings $\xi_{W}$, $\xi_{Z}$ and $\xi_{H}$ are
obtained by reweighting
the samples for the individual production and decay modes according to their corresponding branching
fractions and combining them. All signal samples are normalised to cross sections calculated at next-to-leading order
(NLO)~\cite{Cacciapaglia:2018qep} in quantum chromodynamics (QCD). Since the NLO cross sections were computed in
a narrow-width
approximation, a correction factor is applied to them to account for finite-width effects~\cite{Roy:2020fqf} and
an additional correction is applied to account for non-resonant $T$-quark production~\cite{Roy:2022wjj}.
A change in the dynamic scale in \MADGRAPH at the threshold $\Gamma_{T}/m_{T}=0.1$ leads to a discontinuity in the
computed cross section~\cite{Roy:2022wjj}. As a result, two different parameterisations of the cross section are
available: $\sigma_{\mathrm{low}}(\Gamma_{T}/m_{T})$ for the $\Gamma_{T}/m_{T}<0.1$ regime, and $\sigma_{\mathrm{high}}(\Gamma_{T}/m_{T})$
for $\Gamma_{T}/m_{T}>0.1$.
An averaging procedure is used in order to obtain a smooth cross section $\sigma(\Gamma_{T}/m_{T})$ across the mass and
coupling grid:\\
 
\[
\sigma(\Gamma_{T}/m_{T})=
\begin{cases}
\sigma_{\mathrm{low}}(\Gamma_{T}/m_{T}) + \frac{1}{2}(\sigma_{\mathrm{high}}(0.1) - \sigma_{\mathrm{low}}(0.1)),& \text{if } \Gamma_{T}/m_{T} < 0.1\\
\sigma_{\mathrm{high}}(\Gamma_{T}/m_{T}) - \frac{1}{2}(\sigma_{\mathrm{high}}(0.1) - \sigma_{\mathrm{low}}(0.1)),& \text{if }\Gamma_{T}/m_{T} \geq 0.1
\end{cases}
\]
 
An additional uncertainty of $\frac{1}{2}(\sigma_{\mathrm{high}}(0.1) - \sigma_{\mathrm{low}}(0.1))$ is
assigned to the cross section at every point to account for this choice.


\subsection{Background modelling}
\label{sec:bkgd_modelling}
 
After preselection, the main source of background is the production of \ttbar.
The production of a top quark in association with a $W$ boson ($Wt$) makes
significant contributions in regimes of high transverse momentum. The remaining background
contributions originate mostly from \Wjets\ processes.

 
The production of \ttbar\ and single-top-quark events was modelled using the
\POWHEGBOX[v2]~\cite{Frixione:2007nw,Nason:2004rx,Frixione:2007vw,Alioli:2010xd}
generator at NLO with the \NNPDF[3.0nlo]~\cite{Ball:2014uwa} PDF set.
The \hdamp\ parameter\footnote{The
\hdamp\ parameter is a resummation damping factor and one of the
parameters that controls the matching of \POWHEGBOX matrix elements to
the parton shower and thus effectively regulates the
high-\pt\ radiation against which the \ttbar\ system recoils.} for \ttbar\ samples
was set
to 1.5\,\mtop~\cite{ATL-PHYS-PUB-2016-020}.
The events were interfaced
to \Pythia[8.230]~\cite{Sjostrand:2014zea} to model the parton shower,
hadronisation, and underlying event.
 
Samples to model $Wt$ production were
generated using the diagram removal scheme~\cite{Frixione:2008yi}, which is designed to
remove interference and overlap with \ttbar\ production. The related uncertainty is
estimated by comparison with an alternative sample
generated using the diagram subtraction scheme~\cite{Frixione:2008yi,ATL-PHYS-PUB-2016-020}
and the same generator set-up as the nominal sample.
Separate samples were generated to model $s$-channel and $t$-channel single top-quark
production.
 
The uncertainty associated with using the chosen parton shower and hadronisation model is evaluated
by comparing the sample from the nominal generator set-up with a sample also produced with
the \POWHEGBOX[v2]~\cite{Frixione:2007nw,Nason:2004rx,Frixione:2007vw,Alioli:2010xd}
generator using the \NNPDF[3.0nlo]~\cite{Ball:2014uwa} PDF set, but with the
events interfaced with \Herwig[7.04]~\cite{Bahr:2008pv,Bellm:2015jjp}
to model the parton shower and hadronisation.
 
To assess the uncertainty in the matching of NLO matrix elements and
parton showers for \ttbar\ samples, the \POWHEGBOX sample is compared with a sample of events
generated with \MGNLO[2.6.0] interfaced with
\Pythia[8.230]~\cite{Sjostrand:2014zea}. The samples used to estimate this
uncertainty for single top-quark production were generated with \MGNLO[2.6.2], also interfaced with
\Pythia[8.230]~\cite{Sjostrand:2014zea}. For both samples, the \NNPDF[3.0nlo] set of PDFs~\cite{Ball:2014uwa}
was used in the matrix-element calculations.
 
The $\ttbar$ samples were generated inclusively, but events are categorised depending
on the flavour content of additional particle jets not originating from the decay of
the $\ttbar$ system (see Ref.~\cite{Aad:2015gra} for details).
Events labelled as either \ttbin\ or \ttcin\ are generically referred to in the rest of the paper
as $\ttbar$+HF events, where HF stands for \enquote{heavy flavour}. The remaining events are
labelled as $\ttbar$+light-jets events, including those with no additional jets.
 
The inclusive cross section for $\ttbar$ production was corrected to the theory prediction
at next-to-next-to-leading order (NNLO)
in QCD including the resummation of next-to-next-to-leading logarithmic (NNLL) soft-gluon
terms calculated using \TOPpp[2.0]~\cite{Beneke:2011mq,Cacciari:2011hy,Baernreuther:2012ws,
Czakon:2012zr,Czakon:2012pz,Czakon:2013goa,Czakon:2011xx}. For proton--proton collisions at
a centre-of-mass energy of $\rts = \SI{13}{\TeV}$, this cross section corresponds to
$\sigma(\ttbar)_{\textrm{NNLO+NNLL}} = \mathrm{832\pm51~pb}$ using a top-quark mass
of $\mtop = 172.5\,\GeV$. The cross-section uncertainties due to the PDF and
\alphas were calculated using the \PDFforLHC[15] prescription~\cite{Butterworth:2015oua}
with the \MSTW[nnlo]~\cite{Martin:2009iq,Martin:2009bu}, \CT[10nnlo]~\cite{Lai:2010vv,Gao:2013xoa}
and \NNPDF[2.3lo]~\cite{Ball:2012cx} PDF sets in the five-flavour scheme, and were added
in quadrature to the effect of the factorisation and renormalisation scale uncertainties.


 
The production of $W/Z$($V$)+jets was simulated with the
\Sherpa[2.2.1]~\cite{Bothmann:2019yzt}
generator using NLO matrix elements for up to two partons, and
leading-order (LO) matrix elements
for up to four partons, calculated with the Comix~\cite{Gleisberg:2008fv}
and \OPENLOOPS[1]~\cite{Buccioni:2019sur,Cascioli:2011va,Denner:2016kdg} libraries. They
were matched with the \Sherpa parton shower~\cite{Schumann:2007mg} using the MEPS@NLO
prescription~\cite{Hoeche:2011fd,Hoeche:2012yf,Catani:2001cc,Hoeche:2009rj}
and the set of tuned parameters developed by the \Sherpa authors.
The \NNPDF[3.0nnlo] set of PDFs~\cite{Ball:2014uwa} was used and the samples
are normalised to a NNLO
prediction~\cite{Anastasiou:2003ds}.

Samples of diboson final states ($VV$) were simulated with the
\Sherpa[2.2.1] or 2.2.2~\cite{Bothmann:2019yzt} generator depending on the process,
including off-shell effects and Higgs boson contributions, where appropriate. Only
processes with at least one lepton in the final state are considered in this search.
Fully leptonic final states and semileptonic final states, where one boson
decays leptonically and the other hadronically, were generated using
matrix elements at NLO accuracy in QCD for up to one additional parton
and at LO accuracy for up to three additional parton
emissions. Samples for the loop-induced processes $gg \to VV$ were
generated using LO-accurate matrix elements for up to one
additional parton emission for both the cases of fully leptonic and
semileptonic final states. The matrix-element calculations were matched
and merged with the \Sherpa parton shower based on Catani--Seymour
dipole factorisation~\cite{Gleisberg:2008fv,Schumann:2007mg} using the MEPS@NLO
prescription~\cite{Hoeche:2011fd,Hoeche:2012yf,Catani:2001cc,Hoeche:2009rj}.
The virtual QCD corrections were provided by the
\OPENLOOPS[1] library~\cite{Buccioni:2019sur,Cascioli:2011va,Denner:2016kdg}. The
\NNPDF[3.0nnlo] set of PDFs was used~\cite{Ball:2014uwa}, along with the
dedicated set of tuned parton-shower parameters developed by the
\Sherpa authors.
 
The production of \ttW\ and \ttZ\ events was modelled using the
\MGNLO[2.3.3]~\cite{Alwall:2014hca} generator at NLO with the
\NNPDF[3.0nlo]~\cite{Ball:2014uwa} PDF set.
The events were interfaced to \Pythia[8.210]~\cite{Sjostrand:2014zea}.
The production of \ttH\ events was modelled using the
\POWHEGBOX[v2]~\cite{Frixione:2007nw,Nason:2004rx,Frixione:2007vw,Alioli:2010xd,Hartanto:2015uka}
generator at NLO with the \NNPDF[3.0nlo]~\cite{Ball:2014uwa} PDF set.
The production of $tH$ events was modelled using the \MGNLO[2.3.3]~\cite{Alwall:2014hca}
generator at NLO with the \NNPDF[3.0nlo]~\cite{Ball:2014uwa} PDF set.
Events in both of these samples were interfaced with \Pythia[8.230]~\cite{Sjostrand:2014zea} for
showering and hadronisation.
The production of four-top-quark events in the SM was simulated at LO using \MGNLO[2.2.2]
and the \NNPDF[2.3lo] PDF set, interfaced to \Pythia[8.186] in combination with the A14 underlying-event tune,
and normalised to the NLO theoretical cross section.
 
Multijet events were generated using \Pythia[8.230]~\cite{Sjostrand:2014zea} with leading-order
matrix elements for dijet production and interfaced to a \pt-ordered parton shower. Events from this
simulation are normalised to data in a multijet-enriched region obtained by inverting the requirements on
the \met\ and \mtw\ observables in the preselection, and also requiring the lepton $\pt$ to be below 100~\gev.




\section{Analysis strategy}
\label{sec:strategy}
 
The search described in this paper targets the production of a single $T$ quark that
decays to a leptonically decaying top quark and a hadronically decaying Higgs or $Z$ boson.
Both $b$-associated and $t$-associated single-$T$ production are considered
in the search. The four resulting production and decay modes are:
\begin{enumerate}
\item \WTZt: $b$-associated $T$ production with $T \rightarrow Zt$ decay
\item \WTHt: $b$-associated $T$ production with $T \rightarrow Ht$ decay
\item \ZTZt: $t$-associated $T$ production with $T \rightarrow Zt$ decay
\item \ZTHt: $t$-associated $T$ production with $T \rightarrow Ht$ decay
\end{enumerate}
 
To test for the presence of signal, a likelihood fit is performed on the distribution of the
\meff\ variable (defined in Section~\ref{sec:data_presel}) across a set of 24 \enquote{fit regions}
constructed from events in the preselection sample. These regions are summarised in Table~\ref{tab:fullSR}.
The fit regions are designed to be pure in one or
more of the four targeted signal modes, or in specific background
processes. The combined use of these regions in the
fit allows the search to retain sensitivity to all of the processes that can occur
simultaneously in a benchmark model, and the signal-depleted regions serve to improve the description
of the expected background. In the following, the strategy and motivation behind
the event categorisation model is described.
 
\begin{table*}[t!]
\begingroup
\renewcommand{\arraystretch}{1.5} 
\begin{center}
\caption{\small{Definition of the 24 regions (referred to as \enquote{fit regions}) that enter the likelihood fit.
Events are categorised in terms of jet (j), $b$-tagged jet (b), forward jet (fj), $V$-tagged jet (V),
Higgs-tagged jet (H), hadronic top-tagged jet (t$_h$), and leptonic top-quark candidate (t$_l$) multiplicities.}}
\label{tab:fullSR}
\begin{tabular}{c|c|c|c}
\toprule\toprule
\multicolumn{4}{c}{Fit regions} \\
\midrule
Jet mult. & $b$-tag mult. & Region & Targeted signal / bkg\\
\midrule
\multirow{11}{*}{3--5} & \multirow{2}{*}{1} & LJ, 1b, $\geq$1fj, 0(t$_{h}$+t$_{l}$), 0H, $\geq$1V         & \multirow{4}{*}{\WTZt}\\
& & LJ, 1b, $\geq$1fj, 0t$_{h}$, $\geq$1t$_{l}$, 0H, $\geq$1V         & \\ \cdashline{2-3}
& \multirow{2}{*}{2} & LJ, 2b, $\geq$1fj, 0(t$_{h}$+t$_{l}$), 0H, $\geq$1V         & \\
& & LJ, 2b, $\geq$1fj, 0t$_{h}$, $\geq$1t$_{l}$, 0H, $\geq$1V         & \\ \cdashline{2-4}
& \multirow{3}{*}{3} & LJ, 3b, $\geq$1fj, 0(t$_{h}$+t$_{l}$), $\geq$1H, 0V   & \multirow{6}{*}{\WTHt}\\
& & LJ, 3b, $\geq$1fj, 0t$_{h}$, $\geq$1t$_{l}$, $\geq$1H, 0V   & \\
& & LJ, 3b, $\geq$1fj, $\geq$1t$_{h}$, 0t$_{l}$, $\geq$1H, 0V   & \\ \cdashline{2-3}
& \multirow{4}{*}{$\geq$4} & LJ, $\geq$4b, $\geq$1fj, 0(t$_{h}$+t$_{l}$), $\geq$1H, 0V   & \\
& & LJ, $\geq$4b, $\geq$1fj, 0t$_{h}$, $\geq$1t$_{l}$, $\geq$1H, 0V   & \\
& & LJ, $\geq$4b, $\geq$1fj, $\geq$1t$_{h}$, 0t$_{l}$, $\geq$1H, 0V   & \\ \cdashline{3-4}
& & LJ, $\geq$4b, 0fj, $\geq$1t$_{l}$, 0H, 0(V+t$_{h}$)   & \ttbin, \ttcin\\
\midrule
\multirow{13}{*}{$\geq$6} & \multirow{3}{*}{1} & HJ, 1b, $\geq$1fj, 0t$_{h}$, 1t$_{l}$, 0H, $\geq$1V     & \multirow{6}{*}{\ZTZt}\\
& & HJ, 1b, $\geq$1fj, 1t$_{h}$, 0t$_{l}$, 0H, $\geq$1V      & \\
& & HJ, 1b, $\geq$1fj, $\geq$2(t$_{h}$+t$_{l}$), 0H, $\geq$1V     & \\ \cdashline{2-3}
& \multirow{3}{*}{2} & HJ, 2b, $\geq$1fj, 0t$_{h}$, 1t$_{l}$, 0H, $\geq$1V     & \\
& & HJ, 2b, $\geq$1fj, 1t$_{h}$, 0t$_{l}$, 0H, $\geq$1V     & \\
& & HJ, 2b, $\geq$1fj, $\geq$2(t$_{h}$+t$_{l}$), 0H, $\geq$1V     & \\ \cdashline{2-4}
& \multirow{3}{*}{3} & HJ, 3b, $\geq$1fj, 1t$_{l}$, $\geq$1H, 0(V+t$_{h}$) & \multirow{6}{*}{\ZTHt}\\
& & HJ, 3b, $\geq$1fj, 0t$_{l}$, $\geq$1H, 1(V+t$_{h}$) & \\
& & HJ, 3b, $\geq$1fj, $\geq$1H, $\geq$2(V+t$_{l}$+t$_{h}$) & \\ \cdashline{2-3}
& \multirow{4}{*}{$\geq$4} & HJ, $\geq$4b, $\geq$1fj, 1t$_{l}$, $\geq$1H, 0(V+t$_{h}$) & \\
& & HJ, $\geq$4b, $\geq$1fj, 0t$_{l}$, $\geq$1H, 1(V+t$_{h}$) & \\
& & HJ, $\geq$4b, $\geq$1fj, $\geq$1H, $\geq$2(V+t$_{l}$+t$_{h}$) & \\ \cdashline{3-4}
& & HJ, $\geq$4b, 0fj, $\geq$1t$_{l}$, 0H, 0(V+t$_{h}$)   & \ttbin, \ttcin\\
\bottomrule\bottomrule
\end{tabular}
\end{center}
\endgroup
\end{table*}
 
\begin{figure*}[ht!]
\centering
\subfloat[]{\includegraphics[width=0.45\textwidth]{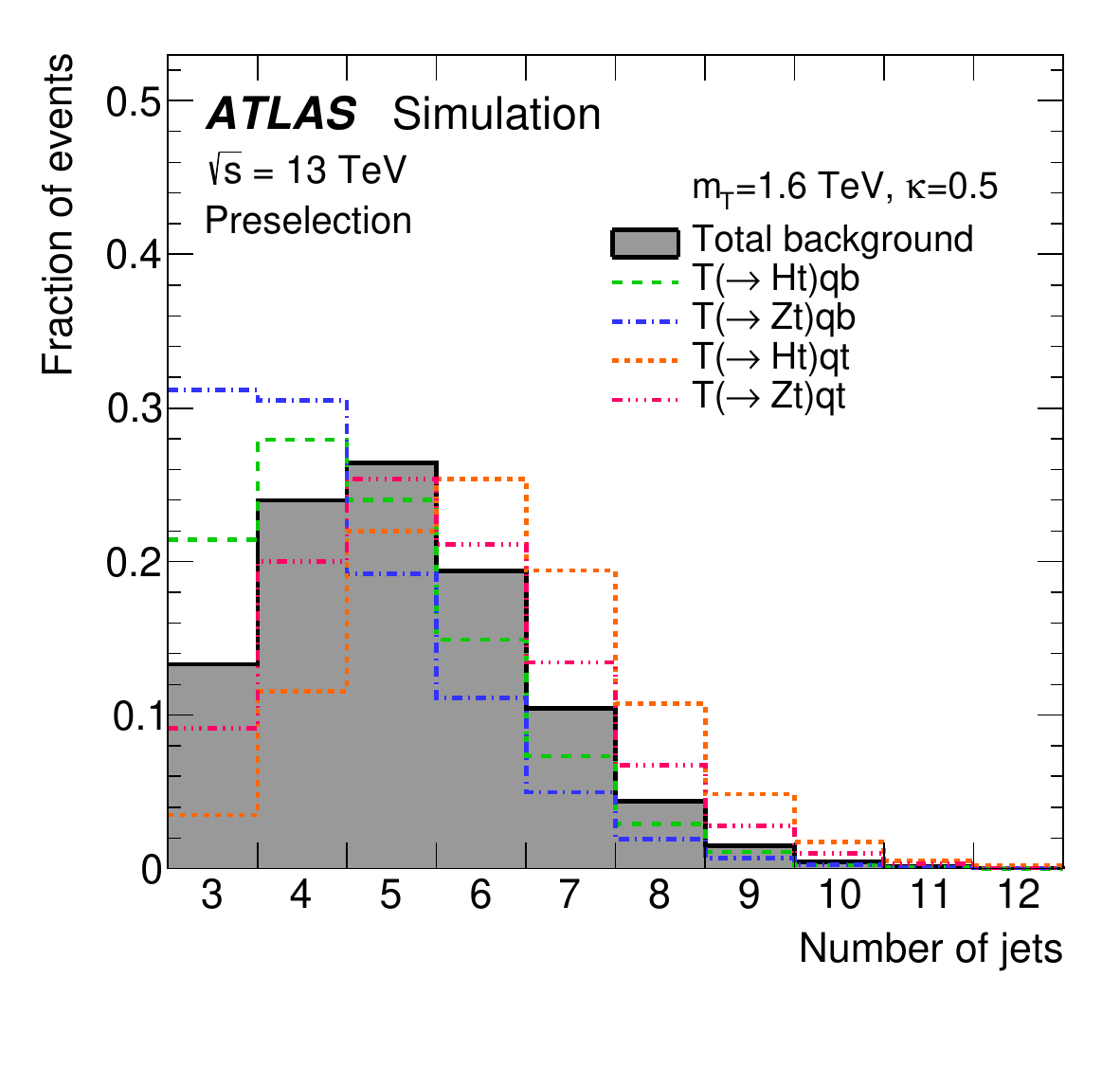}\label{subfig:presel_njet}}
\subfloat[]{\includegraphics[width=0.45\textwidth]{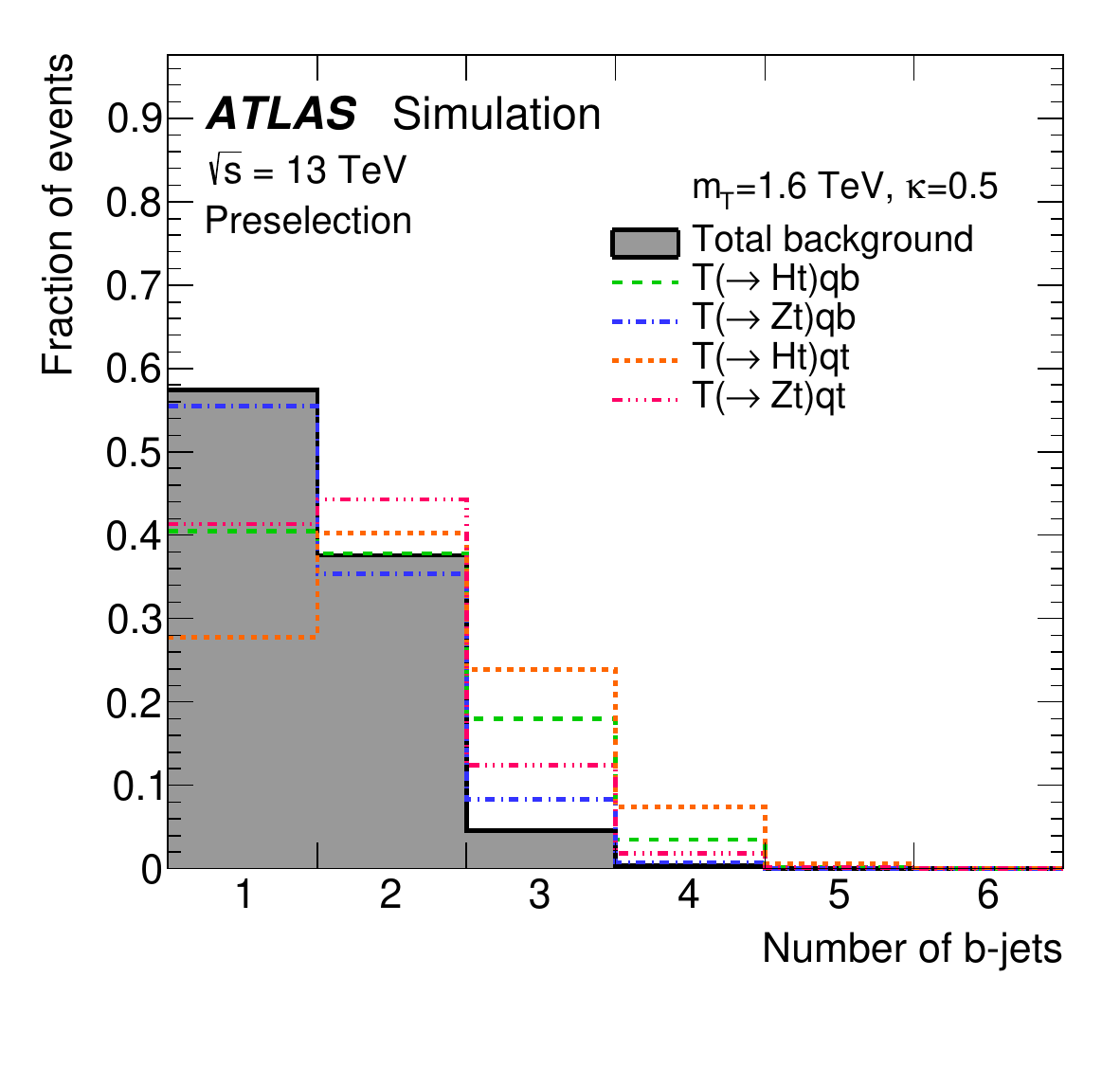}\label{subfig:presel_nbjet}}
\caption{Comparison of the shape of the distribution of (a) the central jet multiplicity, and (b) the $b$-tagged jet multiplicity in the preselection region, between the total background (filled area) and the signal scenarios considered in this search for $T$ quarks with a mass of 1.6~\TeV\ and $\kappa = 0.5$. The last bin in each distribution contains the overflow.}
\label{fig:presel_njet_nbjet_shape}
\end{figure*}
 
\begin{figure*}[ht!]
\captionsetup[subfloat]{farskip=2pt,captionskip=1pt}
\centering
\subfloat[]{\includegraphics[width=0.4\textwidth]{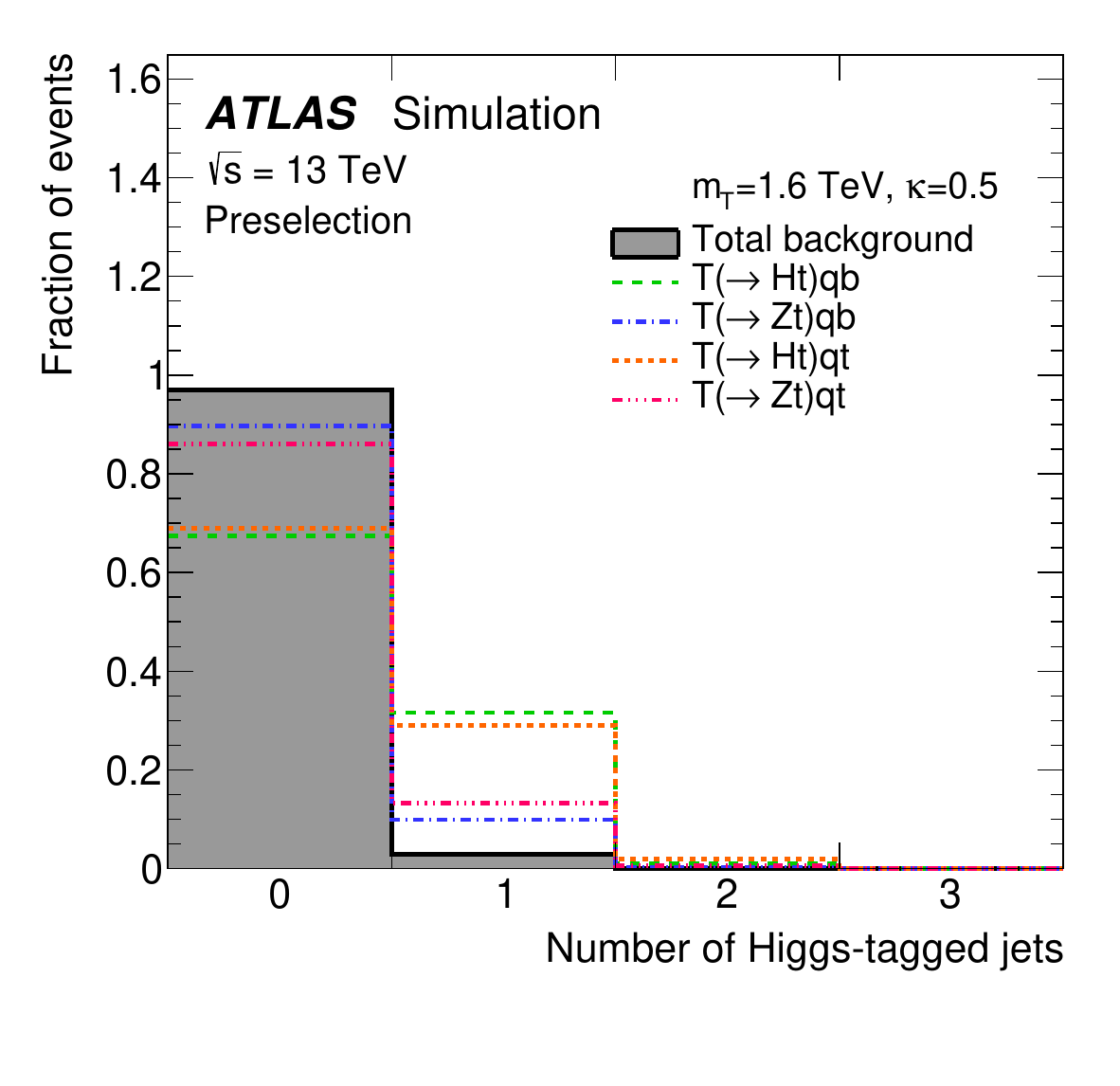}\label{subfig:presel_nhiggs}}
\subfloat[]{\includegraphics[width=0.4\textwidth]{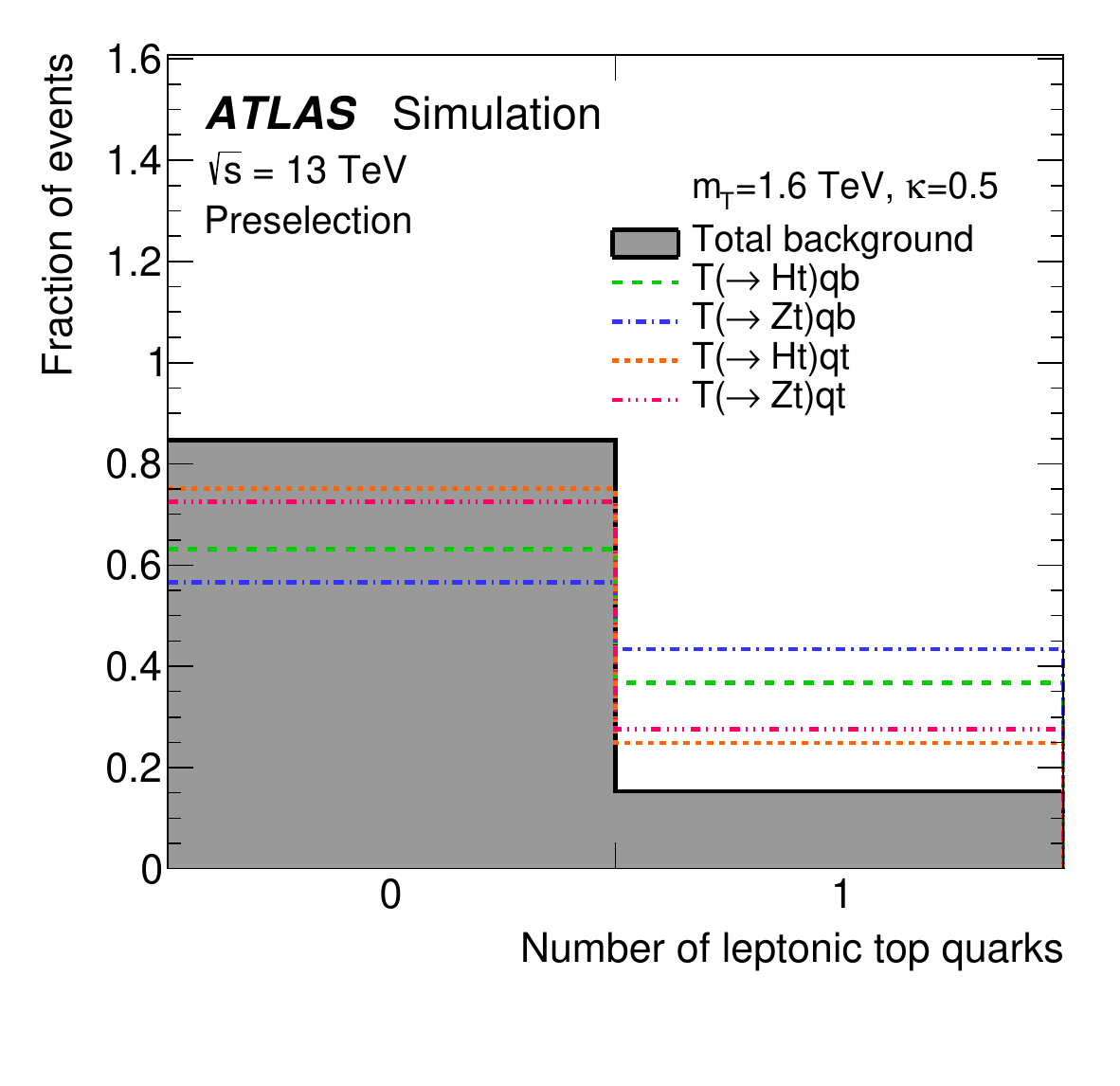}}\\
\subfloat[]{\includegraphics[width=0.4\textwidth]{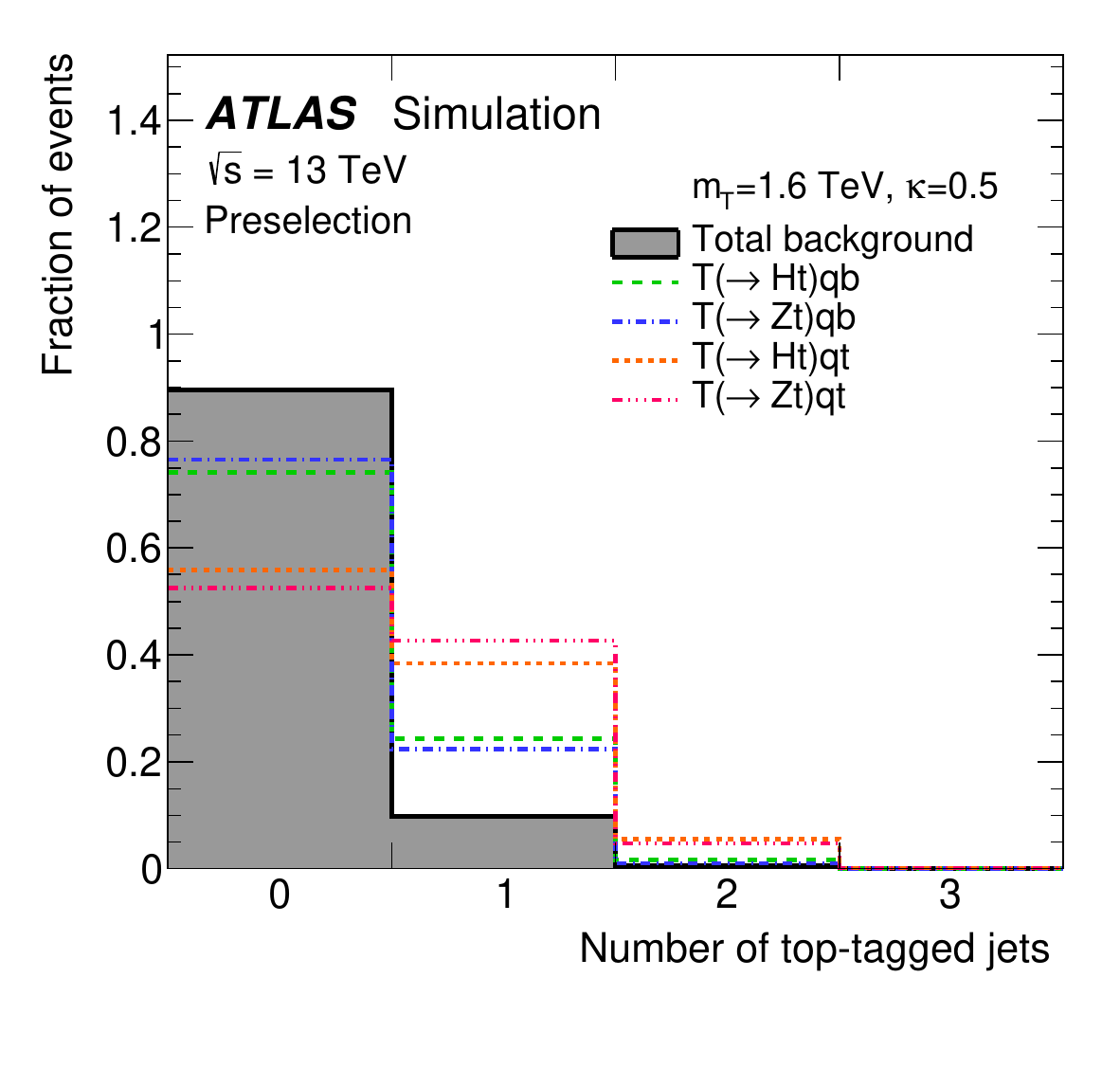}}
\subfloat[]{\includegraphics[width=0.4\textwidth]{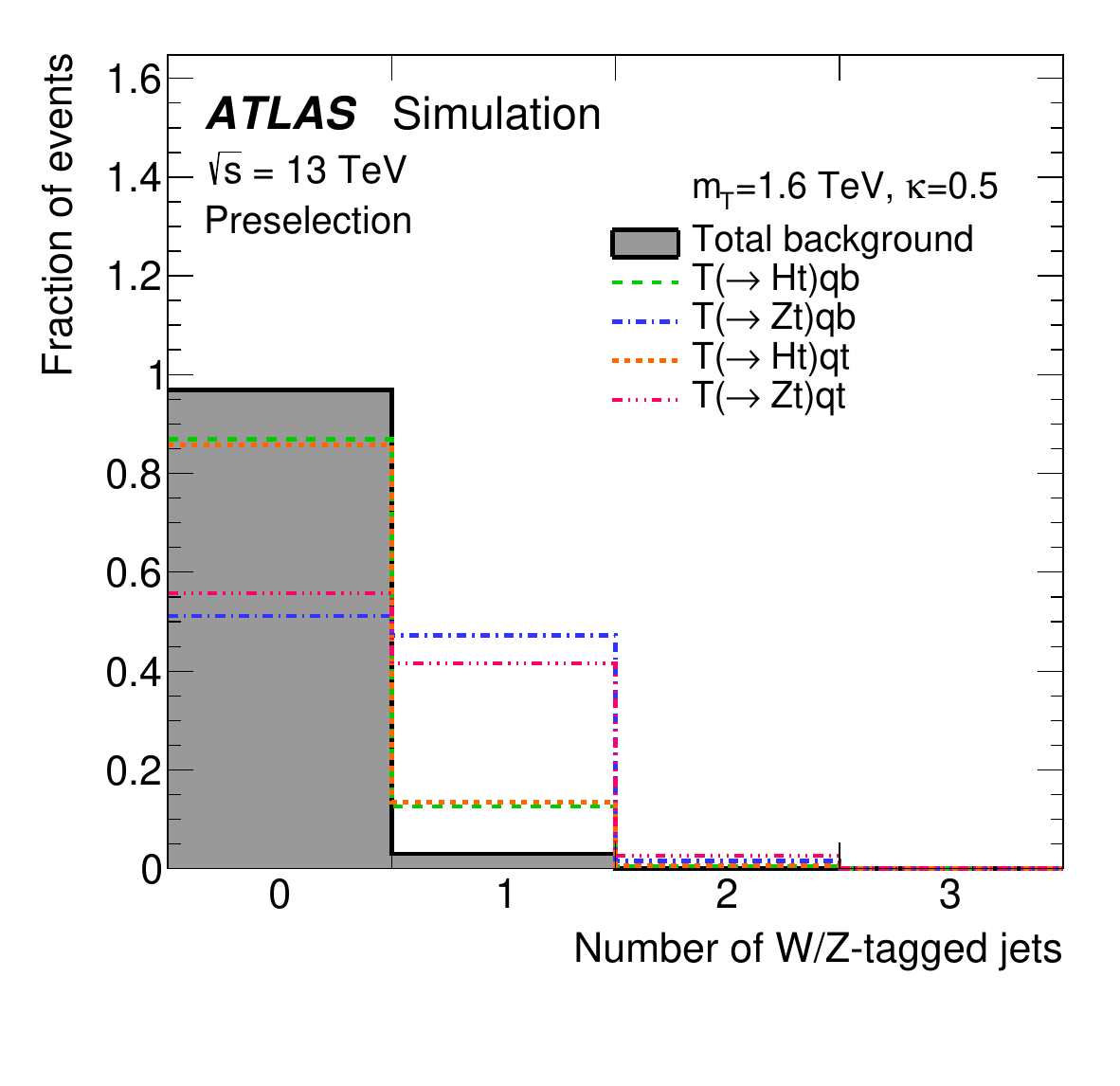}\label{subfig:presel_nV}}\\
\subfloat[]{\includegraphics[width=0.4\textwidth]{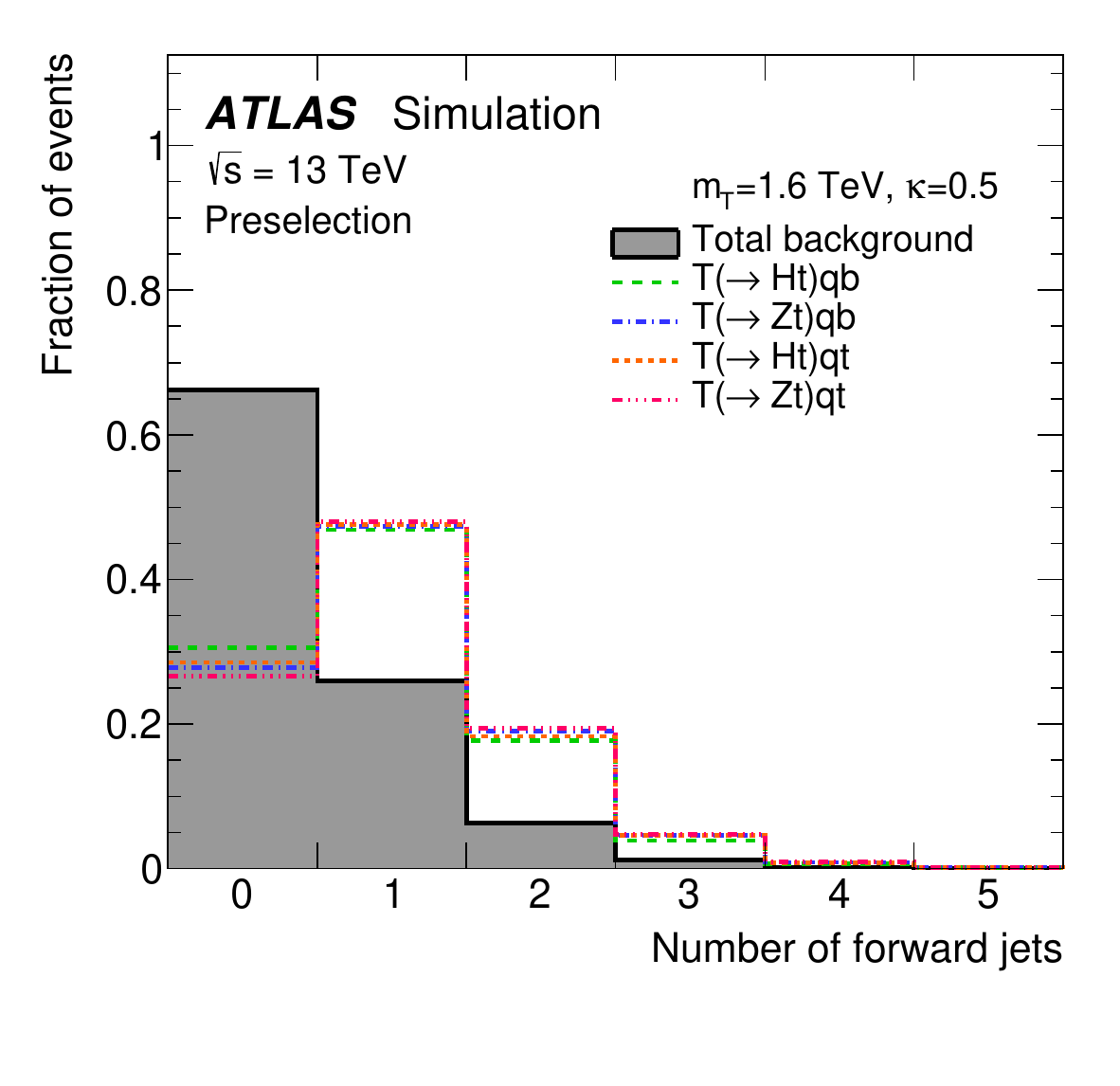}\label{subfig:presel_nfwdjets}}
\subfloat[]{\includegraphics[width=0.4\textwidth]{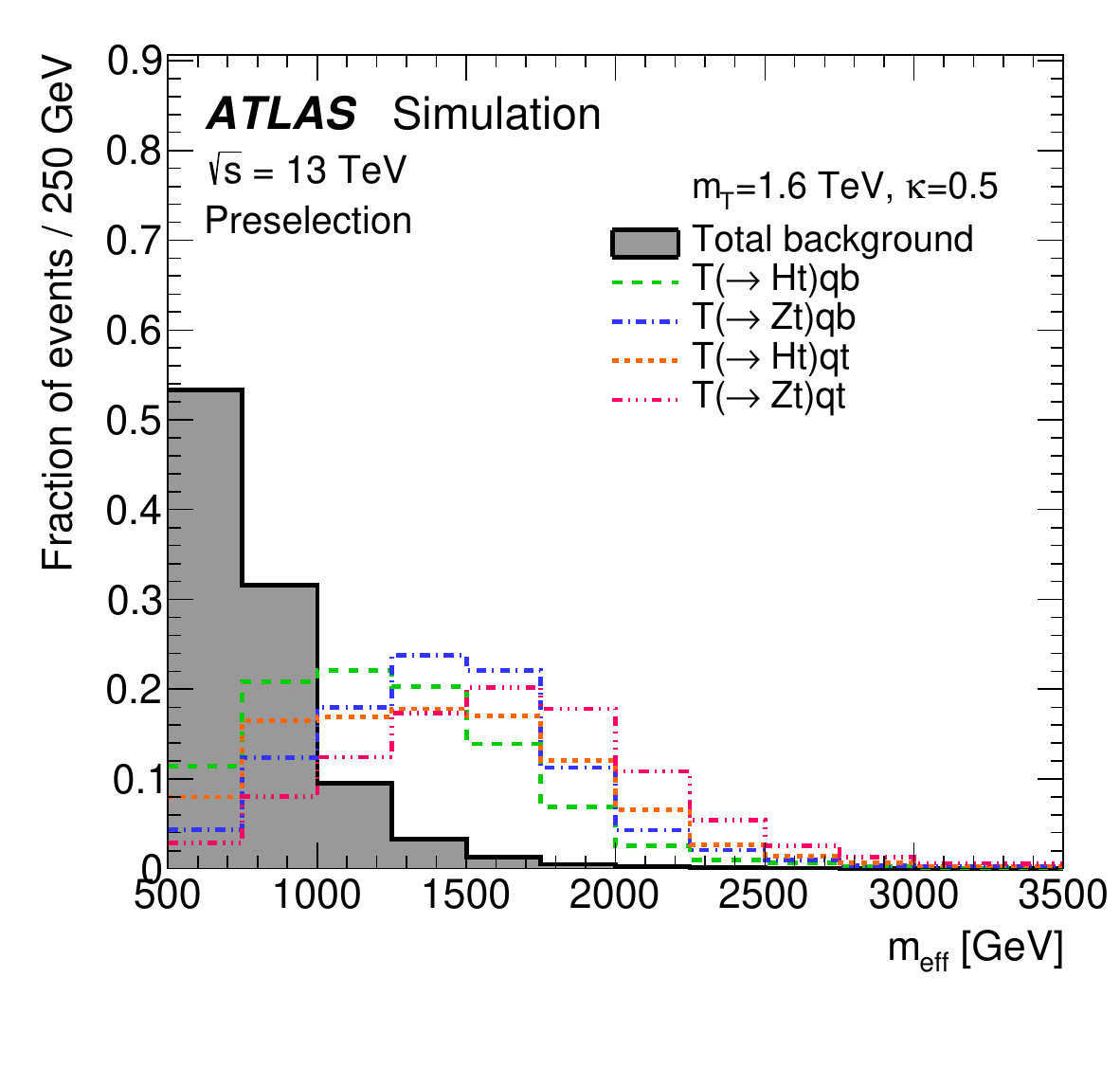}\label{subfig:presel_meff}}
\caption{Comparison of the shape of the distribution of (a) Higgs-tagged jet multiplicity,
(b) leptonic top-quark candidate multiplicity, (c) hadronic top-tagged jet multiplicity, (d) $W$/$Z$-tagged jet multiplicity,
(e) the forward jet multiplicity, and (f) the effective mass \meff,
between the total background (filled area) and the signal scenarios considered in this search for a $T$ with a mass of 1.6~TeV
and $\kappa = 0.5$, in the preselection region. The last bin in each distribution contains the overflow.}
\label{fig:presel_nboosted_meff_shape}
\end{figure*}
 
Different $T$-quark production modes can be distinguished by the number of jets in the final state, as illustrated
in Figure~\ref{subfig:presel_njet}. For $b$-associated
signal production with a leptonically decaying top quark and hadronically decaying Higgs or $Z$ boson,
four central jets are expected in the final state at leading order in the four-flavour scheme.
However, some of these jets can merge due to the collimation of the decay daughters from boosted
Higgs or $Z$ bosons. Furthermore, the initial $b$-quark from gluon splitting can
sometimes be produced at high pseudorapidity and therefore not be reconstructed in the central region.
Conversely, since exactly one lepton is required to be present in the final state, one of
the top quarks in events with $t$-associated production can be expected to decay hadronically,
producing extra jets. 
Based on these arguments, the basic category of regions targeting $b$-associated production modes
(henceforth labelled as LJ, standing for \enquote{Low Jet multiplicity}) are required to have 3--5 jets,
while regions targeting $t$-associated production modes (henceforth labelled as HJ, standing
for \enquote{High Jet multiplicity}) are required to have $\geq$6 jets.
 
The different $T$-quark decay modes are characterised by differences in the multiplicity of
$b$-tagged jets (Figure~\ref{subfig:presel_nbjet}).
Events with $T \rightarrow Zt$ decays typically contain fewer $b$-jets than events with $T \rightarrow Ht$
decays, due to the dominant Higgs boson decay to bottom-quark pairs. Accordingly, regions targeting
$T \rightarrow Zt$ decays are required to have exactly 1 or 2 $b$-tagged jets,
while events with exactly 3 or $\geq$4 $b$-tagged jets are used to target $T \rightarrow Ht$ decays.
 
Since the top quark, $Z$ boson or Higgs boson daughters of the heavy $T$ quark are often produced
at high-$\pt$ in boosted states, signal events are characterised by high multiplicities
of tagged boosted objects.
The distribution of the multiplicities of $H$-tagged ($H$) jets, leptonic
top-quark ($t_{l}$) candidates, $t$-tagged ($t_{h}$) jets, and $V$-tagged ($V$) jets
in the preselection region are shown in Figures~\ref{subfig:presel_nhiggs}--\ref{subfig:presel_nV}.
The specific requirements on the multiplicities of these boosted objects that are applied in each fit
region are tailored to the signal or background process targeted by that particular region.
For example, regions in the $\geq$3b categories are designed to be sensitive to $T \rightarrow Ht$ signals,
and therefore at least one $H$-tagged jet is required in these regions. Conversely, at least one $V$-tagged
jet is required in all regions of the 1--2b category, since these regions target $T \rightarrow Zt$
decays. Signal events with $t$-associated $T$-quark production contain an additional top quark that can
decay hadronically. Therefore, the presence of a hadronic top-tagged jet is required in some
signal regions. Events can also have $V$-tagged jets arising from semi-boosted hadronic top quarks, in cases
where the decay products of the top quark are not collimated enough to produce a $t$-tagged jet.
Finally, an LJ region and HJ region are used in the fit to constrain the normalisation of \ttbin\ and \ttcin\
backgrounds, requiring $\geq$4b and no boosted hadronic objects.
 
As discussed in Section~\ref{sec:intro}, the initial quark recoiling from the gauge boson in
single-$T$ production often emerges at high pseudorapidity, resulting in the presence of jets in
the forward region (Figure~\ref{subfig:presel_nfwdjets}). Thus, at least one forward jet
(fj) is required in the signal-enriched fit regions to increase the signal purity. On the other hand,
a 0fj requirement is applied in the regions used to control the \ttbin\ and \ttcin\ background normalisations,
in order to deplete the signal fraction in those regions.
 
Background predictions in the fit regions are validated in a set of 20 validation regions, designed
to be statistically independent of the fit regions and yet kinematically
similar to them (Table~\ref{tab:fullVR}). These regions are created by adopting a combination of inverted
tagged boosted-object multiplicity requirements, or specifically vetoing forward jets. The maximum allowed
signal contamination in the overall yield in each validation region is required to be $<$10\%, assuming
a signal cross section of 100~fb. This value was chosen as an upper bound on the allowed cross section
in the considered mass range, corresponding to the observed exclusion limits from previous searches.

\begin{table*}[t!]
\begingroup
\renewcommand{\arraystretch}{1.5} 
\begin{center}
\caption{\small{Event categorisation into validation regions in terms of jet (j), $b$-tagged jet (b), forward jet (fj),
$V$-tagged jet (V), Higgs-tagged jet (H), hadronic top-tagged jet (t$_h$), and leptonic top-quark candidate (t$_l$) multiplicities.
The selection follows the same principle as the fit region categorisation defined in Table \ref{tab:fullSR}, but maintains
orthogonality by inverting either the forward jet cut or the cuts on tagged boosted objects. }}
\label{tab:fullVR}
\begin{tabular}{c|c|c}
\toprule\toprule
\multicolumn{3}{c}{Validation regions} \\
\midrule
Jet mult. & $b$-tag mult. & Region \\
\midrule
\multirow{10}{*}{3--5} & \multirow{4}{*}{1} & LJ, 1b, 0fj, 0t$_{h}$, 0t$_{l}$, 0H, $\geq$1V         \\
& & LJ, 1b, 0fj, 0t$_{h}$, $\geq$1t$_{l}$, 0H, $\geq$1V         \\
& & LJ, 1b, $\geq$1fj, $\geq$1(t$_{h}$+t$_{l}$), 0H, 0V         \\
& & LJ, 1b, $\geq$1fj, $\geq$1t$_{h}$, 0t$_{l}$, 0H, $\geq$1V         \\ \cdashline{2-3}
& \multirow{4}{*}{2} & LJ, 2b, 0fj, 0t$_{h}$, 0t$_{l}$, 0H, $\geq$1V         \\
& & LJ, 2b, 0fj, 0t$_{h}$, $\geq$1t$_{l}$, 0H, $\geq$1V         \\
& & LJ, 2b, $\geq$1fj, $\geq$1(t$_{h}$+t$_{l}$), 0H, 0V         \\
& & LJ, 2b, $\geq$1fj, $\geq$1t$_{h}$, 0t$_{l}$, 0H, $\geq$1V         \\ \cdashline{2-3}
& \multirow{2}{*}{$\geq$3} & LJ, $\geq$3b, 0fj, 0(t$_{h}$+t$_{l}$), $\geq$1H, 0V         \\
& & LJ, $\geq$3b, $\geq$1fj, 0H, $\geq$1(V+t$_{l}$+t$_{h}$)                 \\
\midrule
\multirow{10}{*}{$\geq$6} & \multirow{4}{*}{1} & HJ, 1b, 0fj, 1(t$_{h}$+t$_{l}$), 0H, $\geq$1V   \\
& & HJ, 1b, 0fj, $\geq$2(t$_{h}$+t$_{l}$), 0H, $\geq$1V   \\
& & HJ, 1b, $\geq$1fj, 0t$_{h}$, 0t$_{l}$, $\geq$1H, $\geq$1V   \\
& & HJ, 1b, $\geq$1fj, $\geq$2(t$_{h}$+t$_{l}$), $\geq$1H, 0V   \\ \cdashline{2-3}
& \multirow{4}{*}{2} & HJ, 2b, 0fj, 1(t$_{h}$+t$_{l}$), 0H, $\geq$1V   \\
& & HJ, 2b, 0fj, $\geq$2(t$_{h}$+t$_{l}$), 0H, $\geq$1V   \\
& & HJ, 2b, $\geq$1fj, 0t$_{h}$, 0t$_{l}$, $\geq$1H, $\geq$1V   \\
& & HJ, 2b, $\geq$1fj, $\geq$2(t$_{h}$+t$_{l}$), $\geq$1H, 0V   \\ \cdashline{2-3}
& \multirow{2}{*}{$\geq$3} & HJ, $\geq$3b, 0fj, $\geq$1H, $\geq$1(V+t$_{l}$+t$_{h}$)             \\
& & HJ, $\geq$3b, $\geq$1fj, 0H, $\geq$1(V+t$_{l}$+t$_{h}$)                 \\
\bottomrule\bottomrule
\end{tabular}
\end{center}
\endgroup
\end{table*}
 
The choice of \meff\ as the final fitted observable is driven by the strong signal discrimination
power of this observable, as can be seen in Figure~\ref{subfig:presel_meff}.
Due to the presence of highly energetic jets and leptons from the decay of the $T$ quark, the average \meff\
in signal events is much larger than in background events. The shape of the \meff\ distribution depends on
the assumed mass $m_{T}$ of the $T$ quark, as well as on the $T$-quark production and decay modes. In particular,
an extra factor of $m_{T}^{2}/s$ leads to an enhancement of the partonic cross section at low centre-of-mass
energies for the \WTHt and \ZTHt processes~\cite{Deandrea:2021vje}, and consequently the \meff\ distribution shifts to lower
values.\footnote{Again, $m_{T}$ refers to the mass of the $T$ quark, and $s$ is the usual
Mandelstam variable.} Nevertheless, the discrimination power of this observable is relatively
independent of the signal model parameters, and the search is therefore sensitive across a wide
range of parameter space.
 
\clearpage


\section{Kinematic reweighting of background}
\label{sec:bkgd_rw}
 
Recent measurements of differential cross sections have demonstrated that the current
simulations of $\ttbar$ processes overestimate the upper tail of the top-quark $\pt$
spectrum~\cite{TOPQ-2018-15,TOPQ-2018-17}. Conversely, the cross section for this process
is underestimated at high jet multiplicities~\cite{TOPQ-2018-15}. There are similar
issues of accuracy in the modelling of $W+$jets processes~\cite{ATL-PHYS-PUB-2017-006} in the
regime of high jet multiplicities and/or high-$H_\mathrm{T}$ (where $H_\mathrm{T}$ is defined as the scalar
sum of the $\pt$ of central jets in the event). This leads to discrepancies in the shapes of the
\meff\ and \njets\ spectra between the data and expected background in the kinematic
regimes relevant for this search. Data-driven reweighting factors are therefore used to
correct the observed discrepancies in these processes. These corrections are derived using the
following iterative procedure.
 
For a given group of background processes, a \enquote{reweighting source region} (RSR)
is first identified,
which is enriched in events from that process group, and depleted in signal events.
Then, the reweighting factor $R_{a}(x)$ for any observable $x$, for the process group $a$, can
be calculated as
\begin{equation*}
R_{a}(x) = \frac{\text{Data}(x) - \text{MC}^{\text{non-}a}(x)}{\text{MC}^{a}(x)}\,.
\end{equation*}
 
For each group of processes, reweighting factors are first derived from the jet multiplicity distribution.
After correcting the \njets\ distribution of the process with
this initial reweighting, a second set of reweighting factors is derived as a function of a reduced \meff\
variable, which is defined as $\meffred = \meff - (\njets-3)\times 50~\gev$.
The constant value of 50~\gev\ approximately corresponds to the average \pt\ of each additional
jet expected in $\ttbar$ events. Thus, this definition makes the shape of the reweighting functions
in the \meffred\ variable more consistent across $\njets$. The residual dependence on \njets\ is
addressed by deriving the reweighting in exclusive \njets\ bins. The \meffred\ reweighting factors
are parameterised by sigmoid functions to mitigate statistical fluctuations.
 
The subdominant $W(\ell\nu)+$jets background process is corrected first. However, since it is
difficult to isolate $W(\ell\nu)+$jets events in the preselection sample with $\geq$1$b$ selection,
the correction factors
for this process are extrapolated from a sample enriched in $Z(\ell\ell)+$jets instead. This
sample is obtained by selecting events containing exactly two leptons with the same flavour
and opposite electric charges, and requiring the invariant mass of the lepton pair ($m_{\ell\ell}$)
to be consistent with the $Z$ boson mass ($m_{Z}$). To match the kinematic regime of the
preselection more closely, events in this sample must also contain at least three small-$R$ jets,
at least one of which is $b$-tagged. Contamination from $\ttbar$ processes is suppressed by
requiring $\met<100~\gev$. The corrections derived from this sample are applied to both the $W+$jets
and $Z+$jets events in the preselection sample, before correction factors for the \ttbar\ background
are derived.
 
Since \ttbar\ events share the same final state as $tW$ production, these two
processes are grouped together for the purposes of deriving the reweighting factors. A sample enriched
in $\ttbar+$light-jets and $tW$ production, selected
by requiring exactly one lepton, at least three small-$R$ jets and exactly two
$b$-jets, is used to derive combined correction factors for both processes. The same correction
factors are also applied to the \ttbin\ and \ttcin\ processes in the analysis.
 
The selection requirements for the RSRs are summarised in
Table~\ref{tab:bkgd_rw_RSRs}.

\begin{table*}[ht!]
\begin{center}
\caption{\small{Reweighting source regions from which the reweighting functions for \ttbar\ and $tW$ production and $W/Z+$jets production are derived.}}
\label{tab:bkgd_rw_RSRs}
\begin{tabular}{ccc|c|c}
\toprule\toprule
\multicolumn{5}{c}{Reweighting source regions } \\
\midrule
Lepton multiplicity & Jet multiplicity & $b$-tag multiplicity & Other requirements & Targeted background \\
\midrule
1 & $\geq$3 & 2 & -- & \ttbar\ + $tW$ \\
\hdashline
\multirow{2}{*}{2} & \multirow{2}{*}{$\geq$3} & \multirow{2}{*}{1} & $|m_{\ell\ell}-m_{Z}| \leq 10~\gev$,  & \multirow{2}{*}{\Zjets} \\
&  &  & $\met<~100~\gev$ &  \\
\bottomrule\bottomrule
\end{tabular}
\end{center}
\end{table*}
 
The $\meff$ distribution in the preselection region, before and after the application of the full reweighting,
is shown in Figure~\ref{fig:presel_unrew_dataMC}. Since the preselection region is kinematically very close to
the RSRs, almost perfect agreement can be seen between the data and the expected background
after reweighting. As explained later in Section~\ref{sec:syst_unc}, dedicated
reweighting factors are derived
for the alternative simulations that are used to estimate the modelling uncertainties for $\ttbar$, $Wt$ and $W/Z+$jets
backgrounds. As a result, the modelling uncertainties of these background processes vanish in the RSRs,
and are small in regions that are kinematically close to them. This is reflected in the small uncertainty bands seen
in Figure~\ref{fig:presel_unrew_dataMC}. The reweighted modelling uncertainties in the fit and validation
regions are substantially larger, since they are kinematically further away from the RSRs.
 
\begin{figure*}[ht!]
\centering
\includegraphics[width=0.45\textwidth]{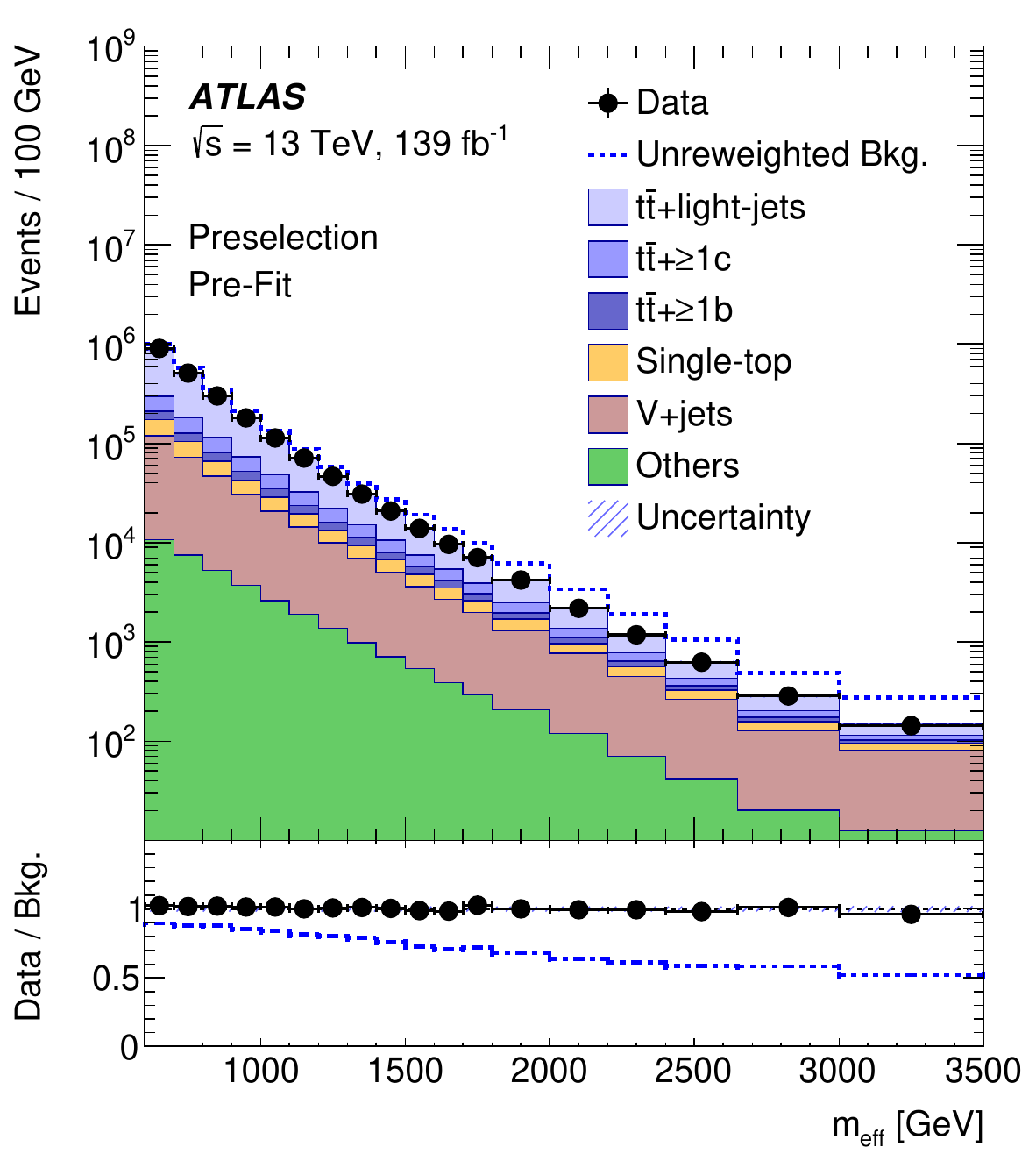}
\caption{Comparison of the \meff\ distribution between the data and unreweighted (blue dashed line) and
reweighted (stacked histograms) background predictions in the preselection region before the likelihood fit. The
\enquote{others} background includes the $\ttbar\ V/H$, \textit{VH}, $tZ$, \fourtop, diboson, and
multijet backgrounds. The last bin in the distribution contains the overflow. The bottom panel displays the ratios of data to the total background
predictions.}
\label{fig:presel_unrew_dataMC}
\end{figure*}


\section{Systematic uncertainties}
\label{sec:syst_unc}
 
The impact of several sources of systematic uncertainty on the normalisation of signal
and background and/or the shape of the $\meff$ distributions is considered. Each
systematic uncertainty is considered to be correlated across processes, channels, and bins of
the signal discriminant,
unless explicitly stated otherwise. Uncertainties from different sources are considered
to be uncorrelated with each other.
 
The leading sources of uncertainty in this search arise from the modelling
of the \ttbar\ and single-top $Wt$ backgrounds, the flavour-tagging efficiencies ($b$, $c$ and light),
and the jet mass resolution. The relative contributions of these uncertainties vary depending on the
analysis region, and therefore their relative impact on the search sensitivity depends on the
signal process being considered. The following sections describe each of the systematic uncertainties
considered in this analysis. An overview of the combined impact of the different groups of systematic uncertainties is furthermore given in Table~\ref{tab:GroupedImpact}.
 
\subsection{Experimental uncertainties}
\label{sec:syst_exp}
 
Uncertainties associated with leptons arise from the efficiencies of the trigger selection, reconstruction,
identification, and isolation criteria, as well as the lepton momentum scale and resolution. These are measured in data using
$Z\to \ell^+\ell^-$, $W\to \ell\nu$ and $J/\psi\to \ell^+\ell^-$ events~\cite{EGAM-2018-01,ATLAS:2020auj}. 
The combined effect of all these uncertainties results in an overall normalisation uncertainty in
signal and background of approximately 1\%.
 
Uncertainties associated with jets arise from the jet energy scale (JES) and resolution (JER),
the jet mass scale (JMS) and resolution (JMR), and the efficiency of the JVT requirements imposed to reject
jets from pile-up. The JES and JER uncertainties are estimated by combining information from collision data,
test-beam data and simulation~\cite{Aaboud:2017jcu,JETM-2018-05}. The JES (JER) uncertainties are split into 30 (8) uncorrelated
components, corresponding to different physical sources. The uncertainty in the JMS is estimated by
comparing each nominal sample with two corresponding alternative event samples, in which the mass of each jet
is either raised or lowered by 10\%, respectively. The uncertainty in the JMR is estimated by comparing each nominal sample with
an alternative event sample in which the mass of each jet is smeared by a Gaussian function whose width is shifted by 20\% relative to
the nominal JMR.
As previously mentioned, the above uncertainties associated with jets are propagated to the RC jets from which the hadronic tagged boosted objects are constructed.
 
The flavour-tagging efficiencies for $b$-, $c$-, and light-jets in simulation are corrected to match efficiencies
measured in data control samples. Uncertainties in these efficiencies are also estimated in these auxiliary
measurements, following the methods described in Refs.~\cite{FTAG-2018-01,FTAG-2020-08,ATLAS:2023lwk}.
A set of nine independent uncertainty sources are considered for $b$-jets, while five
sources are considered for $c$-jets, and six components are considered for light-flavour jets. These components
are taken to be uncorrelated among $b$-jets, $c$-jets, and light-jets. An additional
extrapolation uncertainty component is considered for high-\pt jets that are outside the kinematic reach of
the calibration data sample; it is taken to be correlated amongst the jet flavours. Finally, an uncertainty
related to the application of the $c$-jet scale factors to $\tau$-jets is considered, but has a negligible impact in this analysis.
 
\subsection{Background modelling uncertainties}
\label{sec:syst_modelling}
 
An uncertainty of $+5.5$/$-6.1\%$ is assigned to the inclusive $\ttbar$ production
cross section~\cite{Czakon:2011xx}, including contributions from varying the factorisation and renormalisation
scales, and from uncertainties in the PDF, \alphas, and the top-quark mass, all added in quadrature.
Normalisation uncertainties of 50\% each are assigned to the normalisation of \ttbin\ and \ttcin\ processes. These
uncertainties are motivated by the observed level of agreement between data and simulation
in dedicated measurements of the cross section of the \ttbin\ process~\cite{TOPQ-2017-12}. For single-top processes,
a $\pm$5\% uncertainty in the total cross section, estimated as a weighted average of the theoretical
uncertainties in $t$-, $Wt$- and $s$-channel production~\cite{Kidonakis:2011wy,Kidonakis:2010ux,Kidonakis:2010tc}, is included.
 
A number of sources of systematic uncertainty affecting the modelling of $\ttbar$+jets and single-top production are
considered. Uncertainties due to the choice of the NLO generator are estimated by comparing the nominal \POWPY[8] samples
with alternative samples generated by \MGNLO and showered by \Pythia[8]. The nominal samples are compared with
\POWHER[7] samples to estimate the uncertainties in the modelling of the parton showering and hadronisation
processes. The alternative samples used to evaluate modelling uncertainties are described in detail in
Section~\ref{sec:bkgd_modelling}. The uncertainty due to initial-state radiation (ISR) is estimated by simultaneously varying the \hdamp\
parameter and the renormalisation and factorisation scales, and choosing the Var3c up/down variants of the A14 tune as described in
Ref.~\cite{ATL-PHYS-PUB-2017-007}. The impact of final-state radiation (FSR) is evaluated by doubling or halving the renormalisation
scale for emissions from the parton shower. The \NNPDF[3.0lo] replicas are used to evaluate
the PDF uncertainties for the nominal PDF. These uncertainties are all considered to be uncorrelated among
\ttlight, \ttcin, \ttbin\ and single-top samples, but correlated across the subcategories of the single-top background
($s$-channel, $t$-channel or $tW$). Furthermore, these uncertainties are treated as uncorrelated amongst LJ and HJ analysis
regions and regions with 0, 1 or $\geq$2 tagged boosted objects. The impact of this correlation scheme on the search
sensitivity was studied and compared with a scheme where the modelling uncertainties are correlated across all fit regions.
The expected cross-section limits are around 10\% higher with the current correlation scheme for low VLQ masses, whereas
the impact is negligible for high mass signals.
 
An additional systematic uncertainty in $Wt$-channel production, concerning the separation
between $t\bar{t}$ and $Wt$ at NLO~\cite{Frixione:2008yi}, is assessed by comparing
the nominal sample, which uses the so-called \enquote{diagram removal} scheme, with an alternative sample
using the \enquote{diagram subtraction} scheme.
 
The uncertainties in the modelling of the $W$/$Z$+jets backgrounds are estimated by varying the values of the
internal renormalisation and factorisation scale parameters in \Sherpa. An additional $\pm$30\% normalisation
uncertainty for the $W$/$Z$+jets background is considered for events in analysis regions with different $b$-tag
multiplicity (1, 2, 3, $\geq$4), uncorrelated among $b$-tag multiplicities. This uncertainty is based
on variations of the factorisation and renormalisation scales and \Sherpa matching parameters~\cite{ATLAS:2021qou}.
Since a dedicated reweighting of the jet multiplicity spectrum is applied to $V$+jets events, these
uncertainties are considered to be correlated between LJ and HJ regions. Each of these uncertainties is also
considered to be correlated between $W$+jets and $Z$+jets processes.
 
The kinematic reweighting of the main background processes (described in Section~\ref{sec:bkgd_rw}) is also derived
for each of the modelling uncertainties described above, so that each alternative background model matches the data
(and the nominal background prediction) in the reweighting source regions. Thus, the role of the modelling uncertainties
after this reweighting is to account for extrapolations in kinematics and background composition between the
reweighting source region and the fit and validation regions.
 
Uncertainties in the reweighting procedure
itself arise mainly from the statistical uncertainties in the reweighting source regions and the choice of the functional form
for parameterisation.
These uncertainties are evaluated by shifting the fitted function to $\pm$2$\sigma$ from its nominal
value using the uncertainties of the fitted parameters and taking their internal correlations into account. These shifts
represent possible variations in both the scale and the shape of the reweighting function, although the functional form
of the parameterisation is not altered.
 
Since the diboson, $\ttbar W/Z$ and $\ttbar H$ processes contribute a small fraction of the events in the fit regions, their shape uncertainties
have a negligible impact on the results, and only cross-section uncertainties are considered.
The assigned  uncertainties, due to the PDF and scale uncertainties in the NLO calculation, are $\pm$15\%
for $\ttbar W/Z$ (treated as uncorrelated between LJ and HJ regions), and $+9$/$-12\%$ for $\ttbar H$.
Uncertainties in the diboson background include a 5\% uncertainty in the inclusive cross section calculated at
NLO~\cite{Campbell:1999ah}. An additional uncorrelated 24\% uncertainty in the production cross section is considered
for each additional jet in the event, based on a comparison among different algorithms for merging LO matrix elements
and parton showers~\cite{Alwall:2007fs}. The uncertainty estimate
is based on the average jet multiplicity in each fit region, which is approximately three in the LJ regions
(i.e.\ one additional jet from radiation) and six in the HJ regions (i.e.\ four additional jets from radiation). A $\pm$30\%
normalisation uncertainty is considered for the production of additional heavy-flavour jets. Since the leading two
$b$-tagged jets in diboson events are expected to arise from $W \to cs$ or $Z \to bb$ decays, this uncertainty is only
applied to fit regions with $\geq$3 $b$-jets. All of these
uncertainties are added in quadrature in each region, and the resulting uncertainty is treated as uncorrelated
between LJ and HJ regions, as well as between low-$b$ (1--2$b$) and high-$b$ ($\geq$3$b$) regions. The total
normalisation uncertainty for diboson processes in LJ regions is 24\% and 38\% in low- and high-$b$ regions,
respectively; in HJ regions it is 48\% and 56\% in low- and high-$b$ regions, respectively.


\section{Statistical analysis}
\label{sec:stat_analysis}
 
For each benchmark scenario considered in this search, the $\meff$ distributions
across all search regions are jointly analysed to test for the presence of
the predicted signal. A binned likelihood function ${\cal L}(\mu,\theta)$ is
constructed as a product of Poisson probability terms over all $\meff$ bins considered in
the search.
 
The likelihood function depends on the signal-strength parameter $\mu$, which
multiplies the predicted production cross section for signal, and $\theta$, a set
of nuisance parameters that encode the effect of systematic uncertainties in the
signal and background expectations. Therefore, the expected total number of events
in a given bin depends on $\mu$ and $\theta$.
 
Nuisance parameters corresponding to all systematic uncertainties are implemented in the
likelihood function with Gaussian constraints. In each bin of the $\meff$ distributions,
uncertainties due to the limited size of the simulated samples are also taken
into account by dedicated parameters in the fit that are independent across bins.
These parameters are implemented with Poisson constraints.
 
For a given value of $\mu$, variations in the nuisance parameters $\theta$ allow
the expectations for signal and background to change according to the corresponding
systematic uncertainties. The fitted values of $\theta$ correspond to deviations
from the nominal expectations that globally provide the best fit to the data. This
procedure reduces the impact of systematic uncertainties on the search sensitivity
and improves the background prediction by taking advantage of the highly populated
background-dominated regions included in the likelihood fit.
 
The improvement in the background prediction is verified by performing fits under the background-only
hypothesis and checking how well the data agrees with the post-fit background
in validation regions that are disjoint from the regions used in the fit.
 
The test statistic $q_\mu$, as implemented in a framework based on
RooStats~\cite{Verkerke:2003ir,Moneta:2010pm} and
HistFitter~\cite{Baak:2014wma}, is defined as the profile likelihood ratio:
$q_\mu = -2\ln({\cal L}(\mu,\hat{\hat{\theta}}_\mu)/{\cal L}(\hat{\mu},\hat{\theta}))$.
Here, $\hat{\mu}$ and $\hat{\theta}$ are the values of the parameters $\mu$ and $\theta$
that simultaneously maximise the likelihood function ${\cal L}(\mu,\theta)$ (subject
to the constraint $0\leq \hat{\mu} \leq \mu$), whereas $\hat{\hat{\theta}}_\mu$ are the
values of the nuisance parameters that maximise the likelihood function for a given value
of $\mu$. The statistic used for the discovery test, to test the compatibility of the
observed data with the background-only hypothesis, is obtained by setting $\mu=0$ in
the profile likelihood ratio and leaving $\hat{\mu}$ unconstrained:
$q_0 = -2\ln({\cal L}(0,\hat{\hat{\theta}}_0)/{\cal L}(\hat{\mu},\hat{\theta}))$.
The $p$-value of the discovery test is given by the integral of the probability
distribution of $q_{0}$ above the observed $q_{0}$ value when assuming the
background-only hypothesis, and it is computed using the asymptotic approximation detailed in
Refs.~\cite{Cowan:2010js,ErratumCowan:2010js}. For each signal scenario considered,
the upper limit on the signal production cross section is computed using $q_\mu$ in
the CL$_{\textrm{s}}$ method~\cite{Junk:1999kv,Read:2002hq}, also in the asymptotic
approximation. For a given signal scenario, values of the production cross section
(parameterised by $\mu$) yielding CL$_{\textrm{s}} < 0.05$ are excluded at $\geq 95\%$ CL.
The exclusion limits obtained using the asymptotic approximation are then compared with
limits calculated using pseudoexperiments for 2~\tev\ benchmark signal points close to
the obtained constraints. The central values of the limits computed using the two methods
agree within 5\%, while the uncertainty bands agree within 10\%--15\%.


\section{Results}
\label{sec:results}

\subsection{Likelihood fits to data}
\label{subsec:results_fit}
 
A likelihood fit, as described in Section~\ref{sec:stat_analysis}, is performed under the background-only hypothesis. 
A comparison between the overall observed and expected yields in each fit region is shown, before and after the fit to data, in Figure~\ref{fig:SR_prefit_postfit_summary}.
As can be seen, the combined impact of the systematic uncertainties has been constrained
as a result of the fit, using information from the large number of events in signal-depleted regions with different background
contributions. Consequently, an improved background prediction is obtained with reduced uncertainty across regions, including
those with significant fractions of expected signal events. The data and pre- and post-fit background
yields in four of the most sensitive search regions are given in Table~\ref{tab:prefit_yields_4SR} and
Table~\ref{tab:postfit_yields_4SR}, respectively.
 
The pre- and post-fit \meff\ distributions in these four regions are furthermore shown in Figure~\ref{fig:prepost_LJ} and Figure~\ref{fig:prepost_HJ}.
The post-fit agreement between the data and prediction in the fit regions is good overall.
A comparison of the observed and expected yields in all validation regions,
pre- and post-fit, is shown in Figure~\ref{fig:VR_prefit_postfit_summary}.
The pre- and post-fit \meff\ distributions in the validation regions closest to the four selected fit regions are
shown in Figure~\ref{fig:prepost_LJ_VR} and Figure~\ref{fig:prepost_HJ_VR}.
The expected background in these regions
agrees with the data within uncertainties, both before and after the fit. The general post-fit improvement in the estimated background
in the validation regions, which are not included in the fit, gives confidence in the background estimation procedure.
 
The impact of the different sources of uncertainty on the signal-strength parameter $\mu$ is reported in
Table~\ref{tab:GroupedImpact} for a $T$ singlet and $T$ doublet signal near the expected sensitivity of the search.
As can be seen, the results are dominated
by systematic uncertainties. The largest overall impact is related to the modelling of background processes, in
particular $t\bar{t}$+$\geq$1$b$ production. The systematic uncertainty associated with jets also has a significant impact.
The relative size of impacts from different sources of uncertainty differs between $T$ singlet and $T$ doublet
signals due to their kinematic differences.
Uncertainties related to $t\bar{t}$+$\geq$1$b$ modelling and $b$-tagging contribute significantly more in the $T$
doublet scenario, whereas single-top
modelling has a larger impact in the $T$ singlet scenario.
 
The data-driven kinematic reweighting procedure described in Section~\ref{sec:bkgd_rw} provides better
agreement between the data and the nominal pre-fit background, and, as described in Section~\ref{sec:syst_unc},
modelling uncertainties affecting the dominant background processes are also reweighted. This procedure
improves the stability of the likelihood fit. However, the fit is robust against these
changes to the pre-fit background model, and the post-fit values of the modelling uncertainties are
not significantly impacted by the reweighting procedure.
 
\newpage
\begin{figure*}[h!]
\centering
\includegraphics[width=0.8\textwidth]{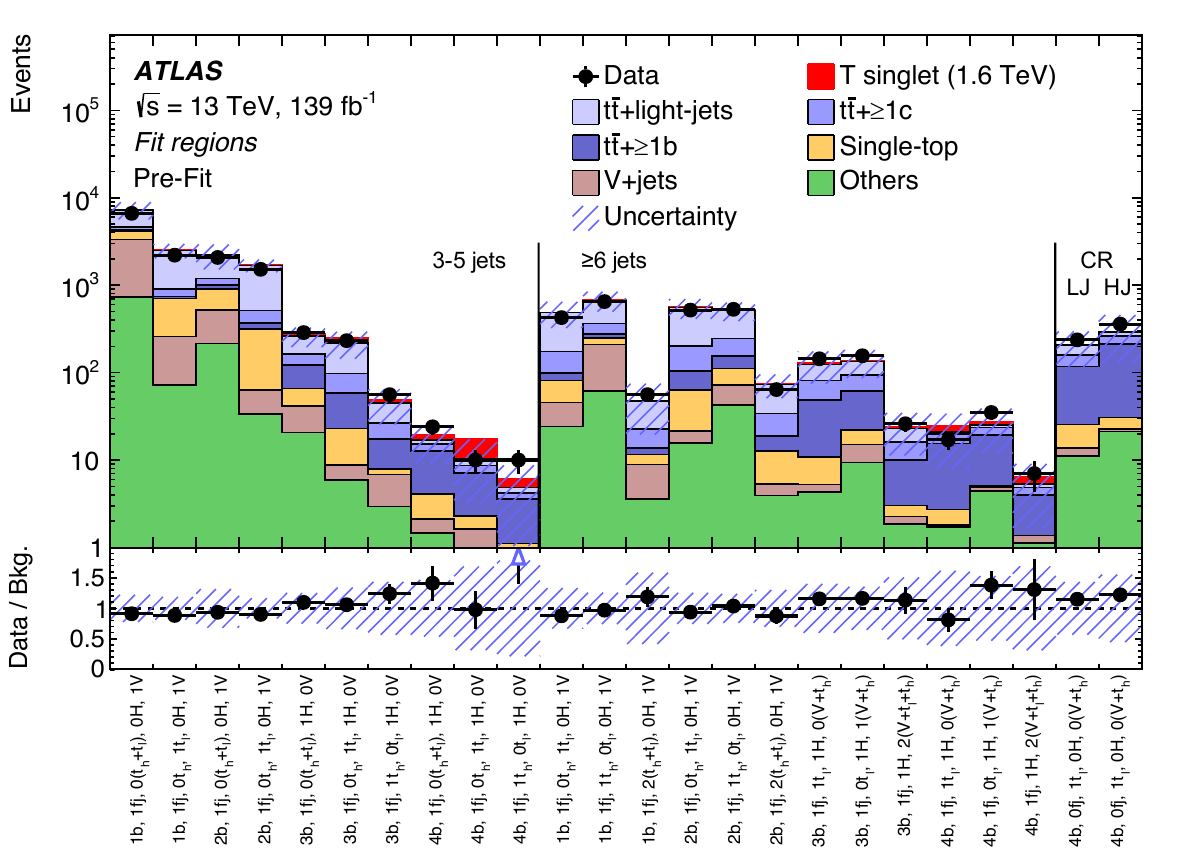}\\
\includegraphics[width=0.8\textwidth]{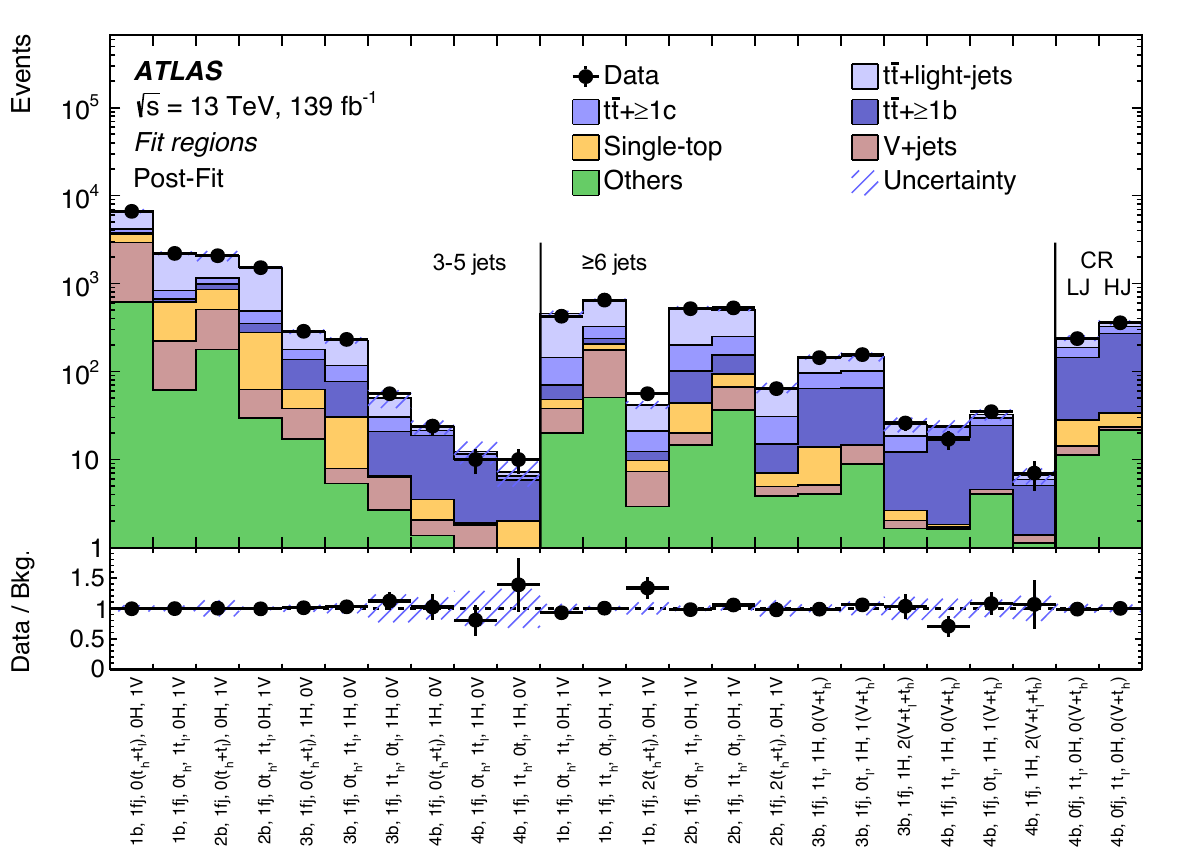}
\caption{\small{Comparison between the data and background prediction for the yields in each of the fit regions
considered (top) pre-fit and (bottom) post-fit, performed under the background-only hypothesis. The two rightmost
regions shown in the plot are the 3--5j (LJ) and $\geq$6j (HJ) control regions, respectively. The \enquote{others} background includes the $\ttbar\ V/H$, \textit{VH}, $tZ$, \fourtop, diboson, and multijet backgrounds. The expected $T$ singlet signal (solid red) for $m_{T}=1.6~\tev$ and $\kappa=0.5$ is included in the pre-fit figure. The bottom panels display the ratios of data to the total background prediction. The hashed area represents the total uncertainty in the background. }}
\label{fig:SR_prefit_postfit_summary}
\end{figure*}

\begin{table}
\small
\centering
\caption{Predicted and observed yields in four of the most sensitive search regions (depending on the signal scenario)
considered. The background prediction is shown before the fit to data.
The \enquote{rare backgrounds} category includes the \textit{VH}, $tZ$ and \fourtop backgrounds.
Also shown are the signal predictions for different
benchmark scenarios considered. The individual systematic uncertainties for the different background processes can be
correlated, and do not necessarily add in quadrature to equal the systematic uncertainty in the total background yield.
The quoted uncertainties  are the sum in quadrature of statistical and systematic uncertainties in the yields. }
\label{tab:prefit_yields_4SR}
\begin{tabularx}{1.\textwidth}{ l  Y  Y  Y  Y }
\toprule\toprule
&  LJ, 2b, $\geq$1fj, 0t$_{h}$, $\geq$1t$_{l}$, 0H, $\geq$1V  &  LJ, $\geq$4b, $\geq$1fj, 0t$_{h}$, $\geq$1t$_{l}$, $\geq$1H, 0V  &  HJ, 2b, $\geq$1fj, $\geq$2(t$_{h}$+t$_{l}$), 0H, $\geq$1V  &  HJ, $\geq$4b, $\geq$1fj, $\geq$1H, $\geq$2(V+t$_{l}$+t$_{h}$)  \\
\midrule\midrule{\,}  $T$ singlet ($m_{T}=1.6~\tev$, $\kappa=0.5$) & 31.8 $\pm$ 4.9 & 7.2 $\pm$ 3.5 & 1.3 $\pm$ 0.4 & 1.0 $\pm$ 0.5 \\
{\,}  $T$ doublet ($m_{T}=1.6~\tev$, $\kappa=0.5$) & 21.8 $\pm$ 2.4 & 8.5 $\pm$ 5.6 & 7.3 $\pm$ 2.1 & 7.1 $\pm$ 4.5 \\
\midrule\midrule
\ttlight 			& 1170 $\pm$ 210 			& 1.6 $\pm$ 2 	& 39.1 $\pm$ 9.5 	& 0.49 $\pm$ 0.29 \\
\ttcin 			& 143 $\pm$ 80 			& 1.5 $\pm$ 1.3 			& 15.3 $\pm$ 9.9 	& 0.86 $\pm$ 0.58 \\
\ttbin 			& 57 $\pm$ 32 	& 4.8 $\pm$ 3.9 			& 6.1 $\pm$ 4.3 			& 2.6 $\pm$ 2 \\
Single-top 		& 250 $\pm$ 50 			& 0.66 $\pm$ 0.87 			& 7.3 $\pm$ 7.5 			& $<$0.001 \\
$\ttbar W/Z$ 		& 13.2 $\pm$ 3.1 			& 0.33 $\pm$ 0.19 			& 2.5 $\pm$ 1.1 			& 0.22 $\pm$ 0.82 \\
\ttH  				& 1.5 $\pm$ 0.2 	& 0.51 $\pm$ 0.15 			& 0.34 $\pm$ 0.14 			& 0.42 $\pm$ 0.12 \\
\Wjets 			& 25.7 $\pm$ 9.4 			& 0.70 $\pm$ 1.3 	& 1.2 $\pm$ 1.1 			& 0.24 $\pm$ 0.15 \\
\Zjets 			& 4.4 $\pm$ 1.7 	& $<$0.001 				& 0.25 $\pm$ 0.10 			& 0.007 $\pm$ 0.007 \\
Dibosons 			& 3.8 $\pm$ 1.4 	& 0.02 $\pm$ 0.03 			& 0.21 $\pm$ 0.15 			& $<$0.001 \\
Multijet 			& 12.9 $\pm$ 7.3 			& 0.025 $\pm$ 0.017 		& 0.61 $\pm$ 0.46 			& 0.16 $\pm$ 0.14 \\
Rare backgrounds 	& 2.0 $\pm$ 0.3 	& 0.03 $\pm$ 0.04 			& 0.25 $\pm$ 0.14 			& 0.33 $\pm$ 0.06 \\
\midrule
Total background 	& 1690 $\pm$ 280 			& 10.2 $\pm$ 4.8 	& 73 $\pm$ 20 				& 5.4 $\pm$ 2.5 \\
\midrule
Data &  1519  &  10  &  64  &  7  \\
\bottomrule\bottomrule \\
\end{tabularx}
\end{table}


\begin{table}
\small
\centering
\caption{Predicted and observed yields in four of the most sensitive search regions (depending on the signal scenario)
considered. The background prediction is shown after the fit to data under the background-only hypothesis.
The ``rare backgrounds'' category includes the \textit{VH}, $tZ$ and \fourtop backgrounds. The individual
systematic uncertainties for the different background processes can be correlated, and do not necessarily add in quadrature
to equal the systematic uncertainty in the total background yield. The quoted uncertainties are computed after taking into
account correlations among nuisance parameters and among processes. The statistical uncertainty is added in quadrature to the
systematic uncertainties.}
\label{tab:postfit_yields_4SR}
\begin{tabularx}{1.\textwidth}{ l  Y  Y  Y  Y }
\toprule\toprule
&  LJ, 2b, $\geq$1fj, 0t$_{h}$, $\geq$1t$_{l}$, 0H, $\geq$1V  &  LJ, $\geq$4b, $\geq$1fj, 0t$_{h}$, $\geq$1t$_{l}$, $\geq$1H, 0V  &  HJ, 2b, $\geq$1fj, $\geq$2(t$_{h}$+t$_{l}$), 0H, $\geq$1V  &  HJ, $\geq$4b, $\geq$1fj, $\geq$1H, $\geq$2(V+t$_{l}$+t$_{h}$)  \\
\midrule\midrule
\ttlight 			& 1033 $\pm$ 72 	& 0.6 $\pm$ 0.8 			& 33.6 $\pm$ 4.5 	& 0.57 $\pm$ 0.24 \\
\ttcin 			& 144 $\pm$ 54 	& 1.5 $\pm$ 1.0 			& 15.6 $\pm$ 5.5 	& 0.82 $\pm$ 0.32 \\
\ttbin 			& 75 $\pm$ 22 				& 8 $\pm$ 3 				& 8.2 $\pm$ 2.3 			& 3.8 $\pm$ 1.1 \\
Single-top 		& 223 $\pm$ 55 	& 0.09 $\pm$ 0.55 			& 2.3 $\pm$ 4.5 			& $<$0.001 \\
$\ttbar W/Z$ 		& 12.1 $\pm$ 2.3 	& 0.36 $\pm$ 0.18 			& 2.3 $\pm$ 0.8 			& 0.62 $\pm$ 0.76 \\
\ttH 				& 1.46 $\pm$ 0.21 			& 0.51 $\pm$ 0.11 			& 0.29 $\pm$ 0.08 			& 0.40 $\pm$ 0.09 \\
\Wjets 			& 26.6 $\pm$ 7.1 	& 0.6 $\pm$ 1.0 			& 0.8 $\pm$ 0.5 			& 0.22 $\pm$ 0.13 \\
\Zjets 			& 4.5 $\pm$ 1.2 			& $<$0.001 				& 0.27 $\pm$ 0.08 			& 0.005 $\pm$ 0.006 \\
Dibosons 			& 3.4 $\pm$ 1.2 			& 0.017 $\pm$ 0.029 		& 0.17 $\pm$ 0.13 			& $<$0.001 \\
Multijet 			& 9.5 $\pm$ 5.7 			& 0.018 $\pm$ 0.015 		& 0.45 $\pm$ 0.41 			& 0.12 $\pm$ 0.12 \\
Rare backgrounds 	& 2.0 $\pm$ 0.2 			& 0.03 $\pm$ 0.03 			& 0.22 $\pm$ 0.08 			& 0.31 $\pm$ 0.05 \\
\midrule
Total background 	& 1534 $\pm$ 56 	& 12.1 $\pm$ 3.5 	& 64 $\pm$ 8 		& 6.8 $\pm$ 1.5 \\
\midrule
Data &  1519  &  10  &  64  &  7 \\
\bottomrule\bottomrule \\
\end{tabularx}
\end{table}


\begin{figure*}[t!]
\subfloat[]{\includegraphics[width=0.45\textwidth]{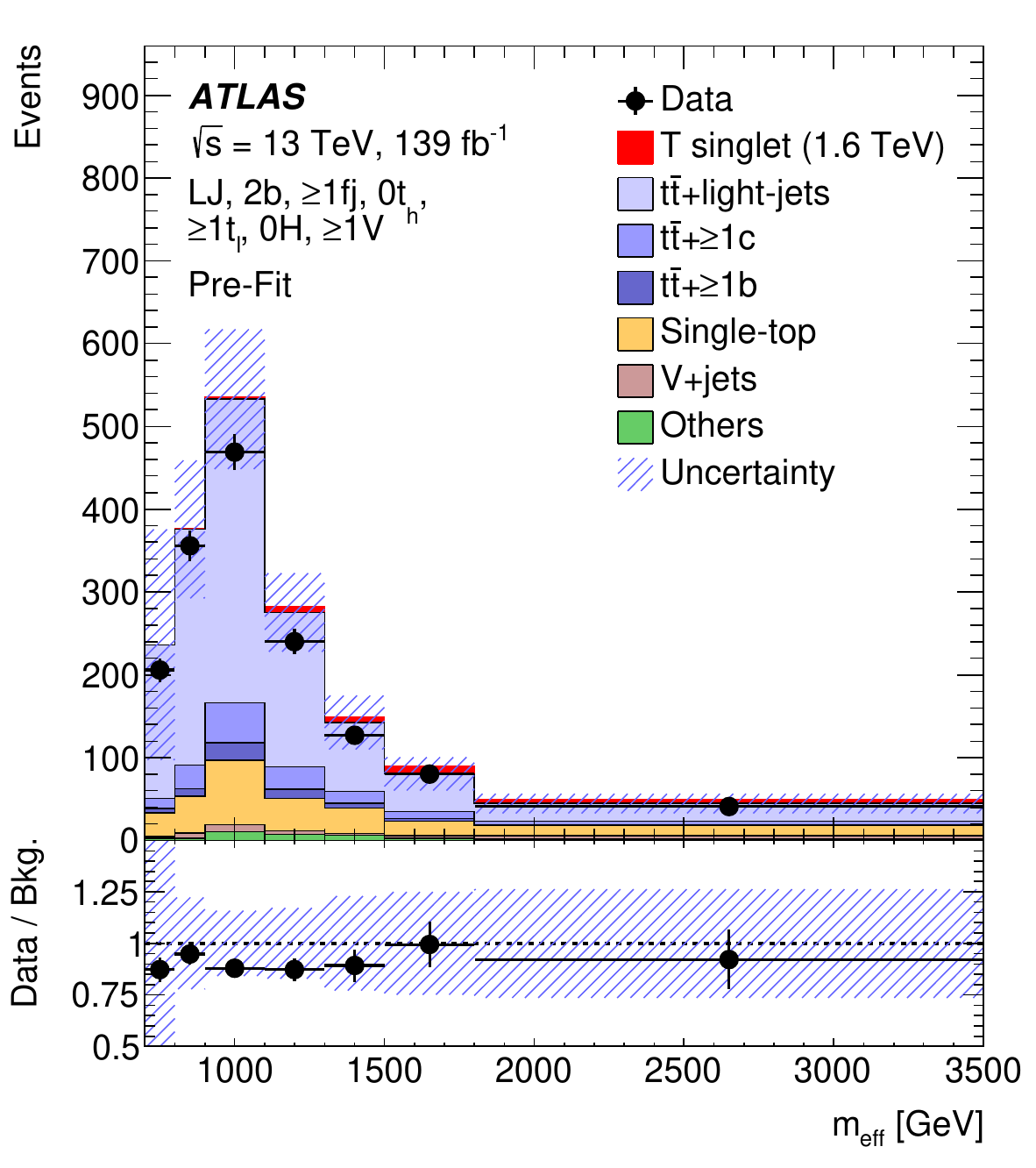}}
\subfloat[]{\includegraphics[width=0.45\textwidth]{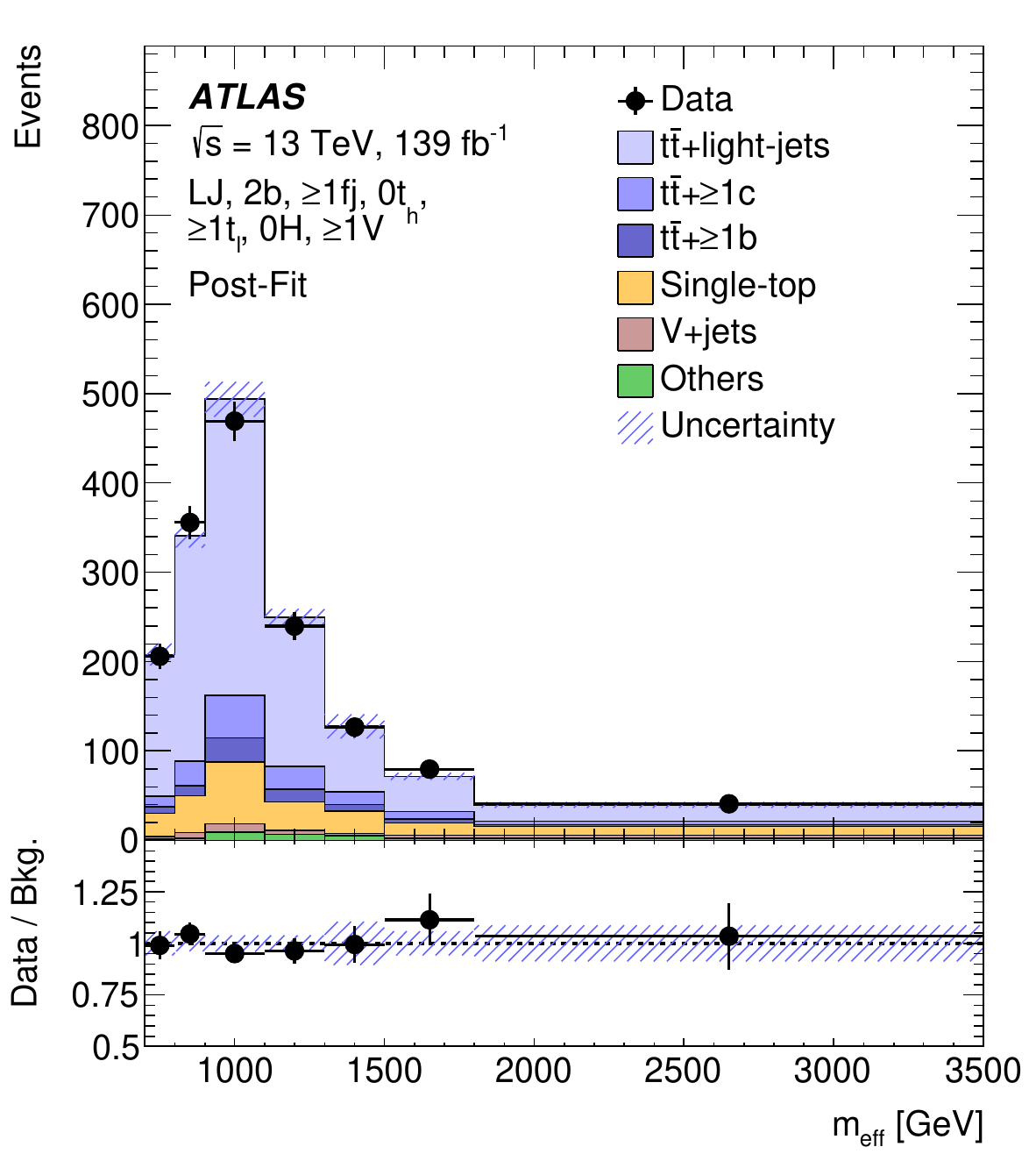}}\\
\subfloat[]{\includegraphics[width=0.45\textwidth]{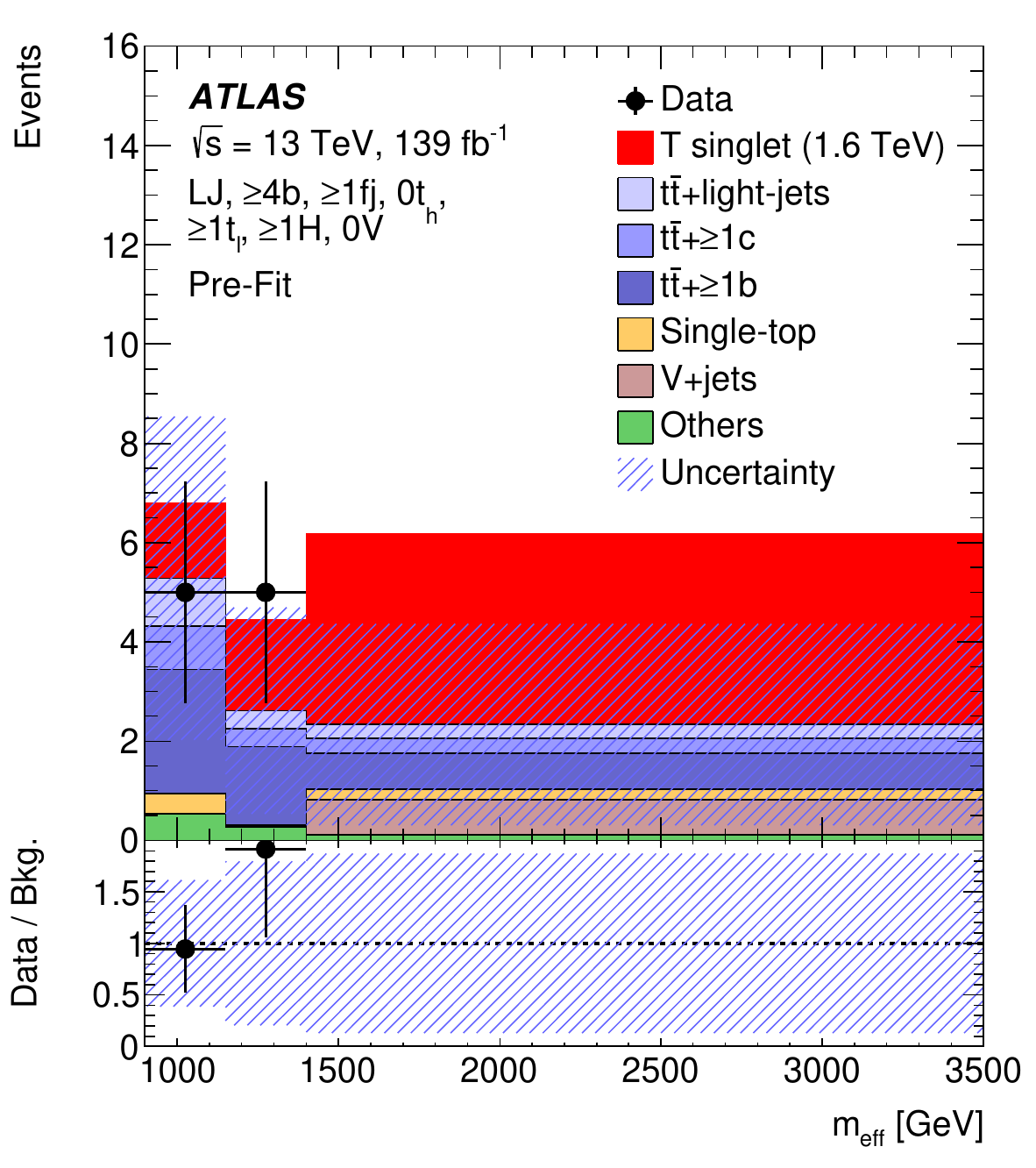}}
\subfloat[]{\includegraphics[width=0.45\textwidth]{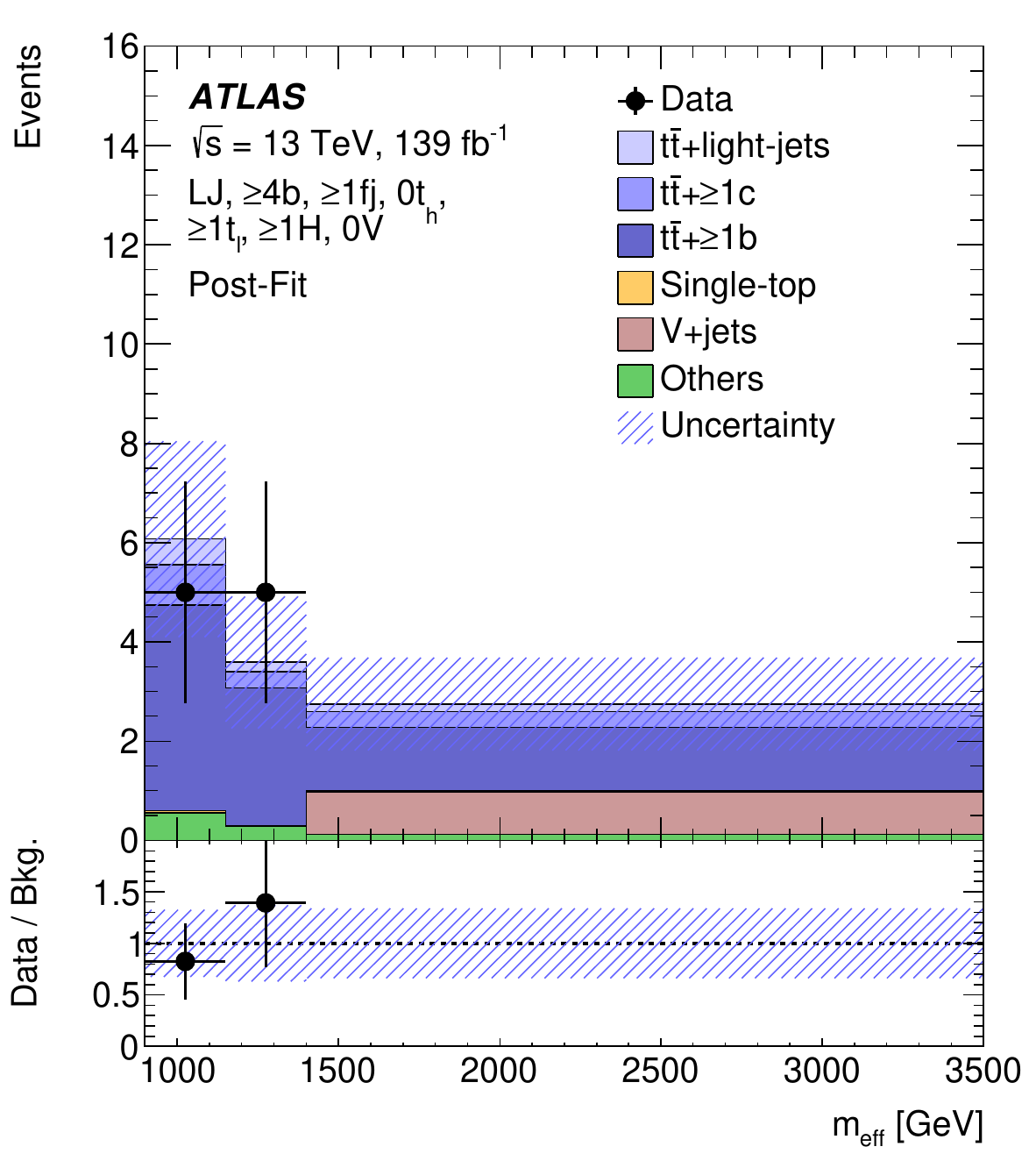}}\\
\caption{\small{Comparison between the data and prediction for the $\meff$ distribution under the background-only hypothesis, in the (LJ, 2b, $\geq$1fj, 0t$_h$, $\geq$1t$_l$, 0H, $\geq$1V) region (a) pre-fit and (b) post-fit, and the (LJ, $\geq$4b, $\geq$1fj, 0t$_h$, $\geq$1t$_l$, $\geq$1H, 0V) region (c) pre-fit and (d) post-fit. The expected $T$ singlet signal (solid red) for $m_{T}=1.6~\tev$ and $\kappa=0.5$ is included in the pre-fit figures. The \enquote{others} background includes the $\ttbar\ V/H$, \textit{VH}, $tZ$, \fourtop, diboson, and multijet backgrounds. The bottom panels display the ratios of data to the total background prediction. The hashed area represents the total uncertainty in the background. The last bin in each distribution contains the overflow.}}
\label{fig:prepost_LJ}
\end{figure*}
\begin{figure*}[t!]
\subfloat[]{\includegraphics[width=0.45\textwidth]{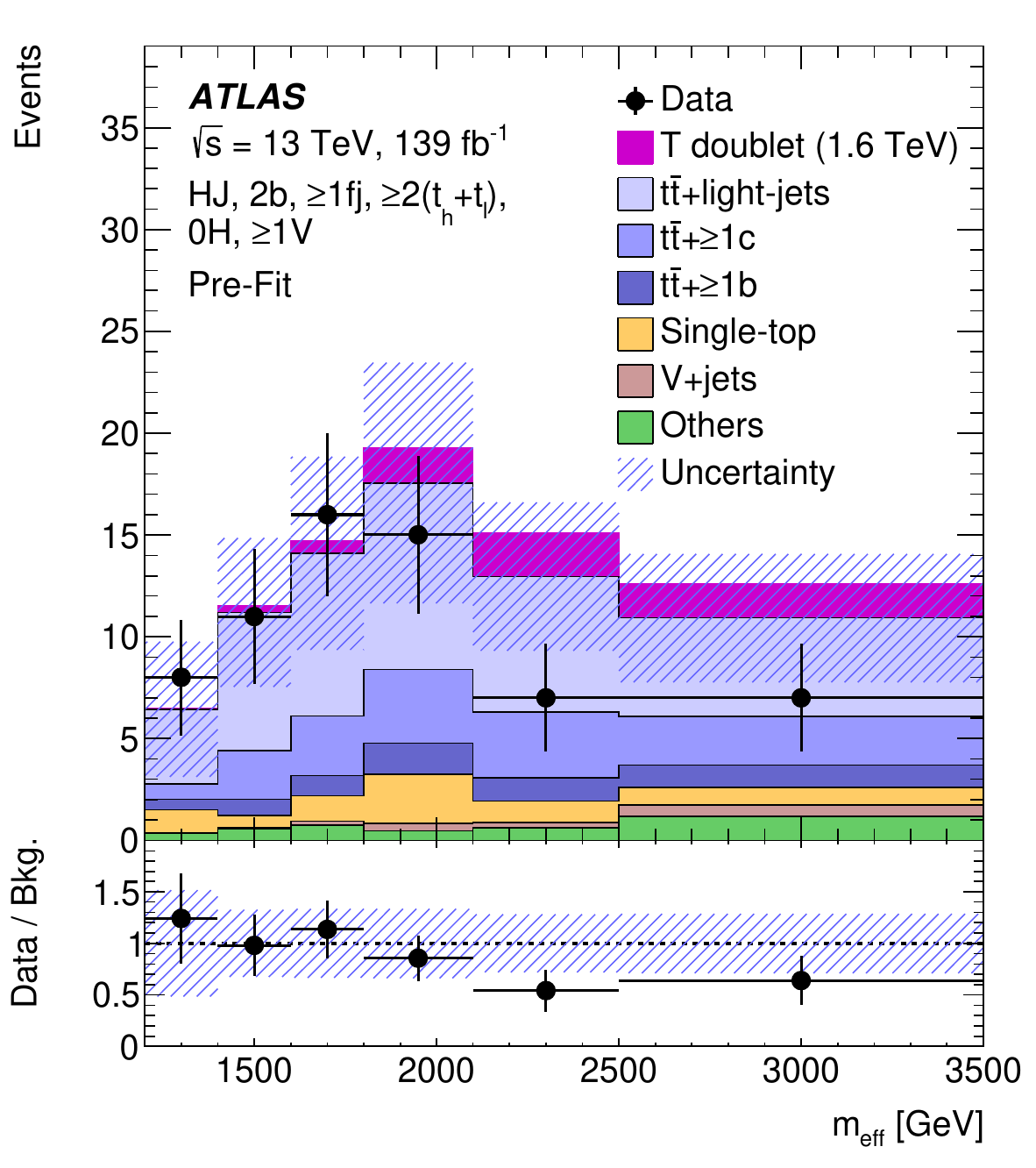}}
\subfloat[]{\includegraphics[width=0.45\textwidth]{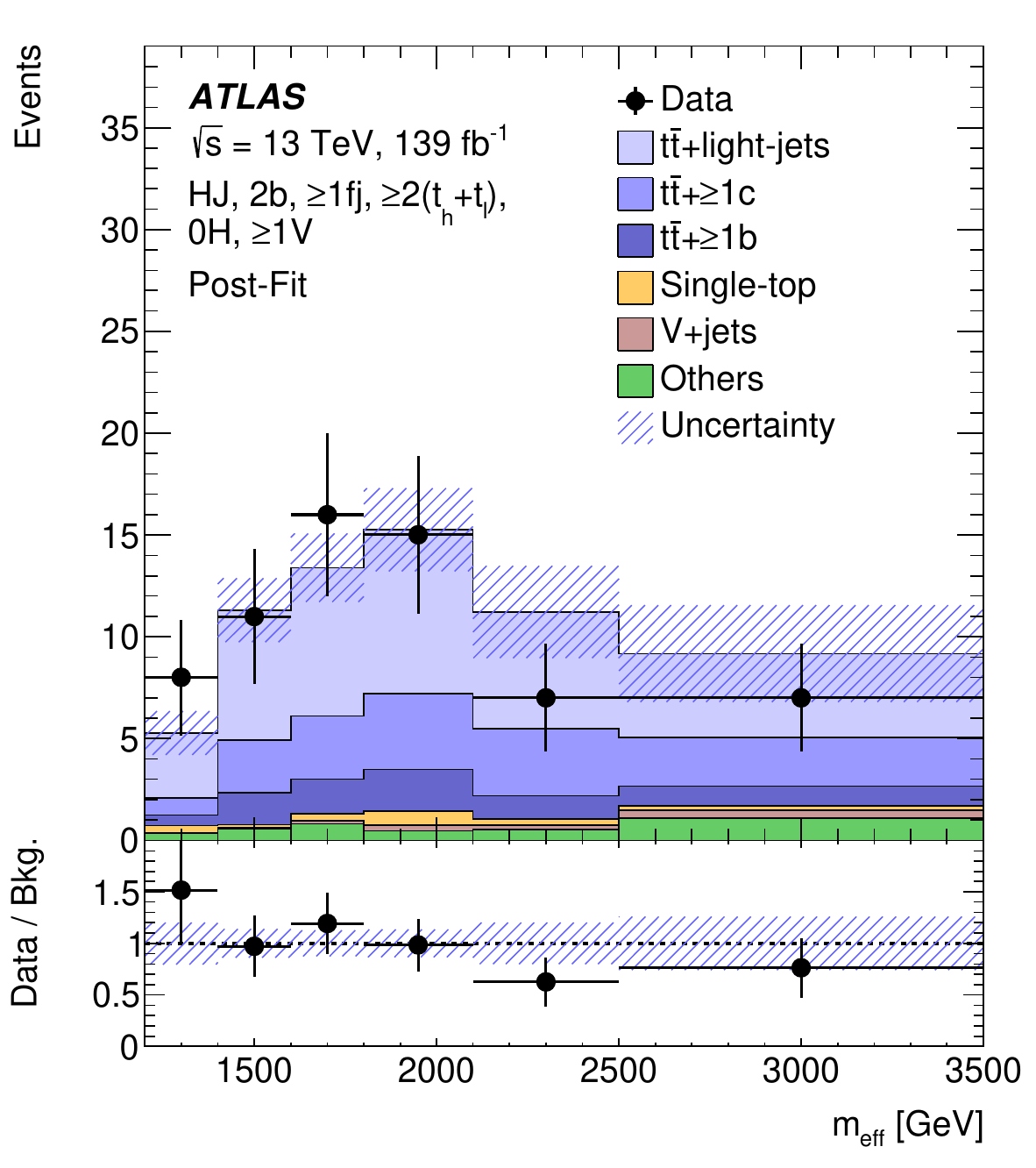}}\\
\subfloat[]{\includegraphics[width=0.45\textwidth]{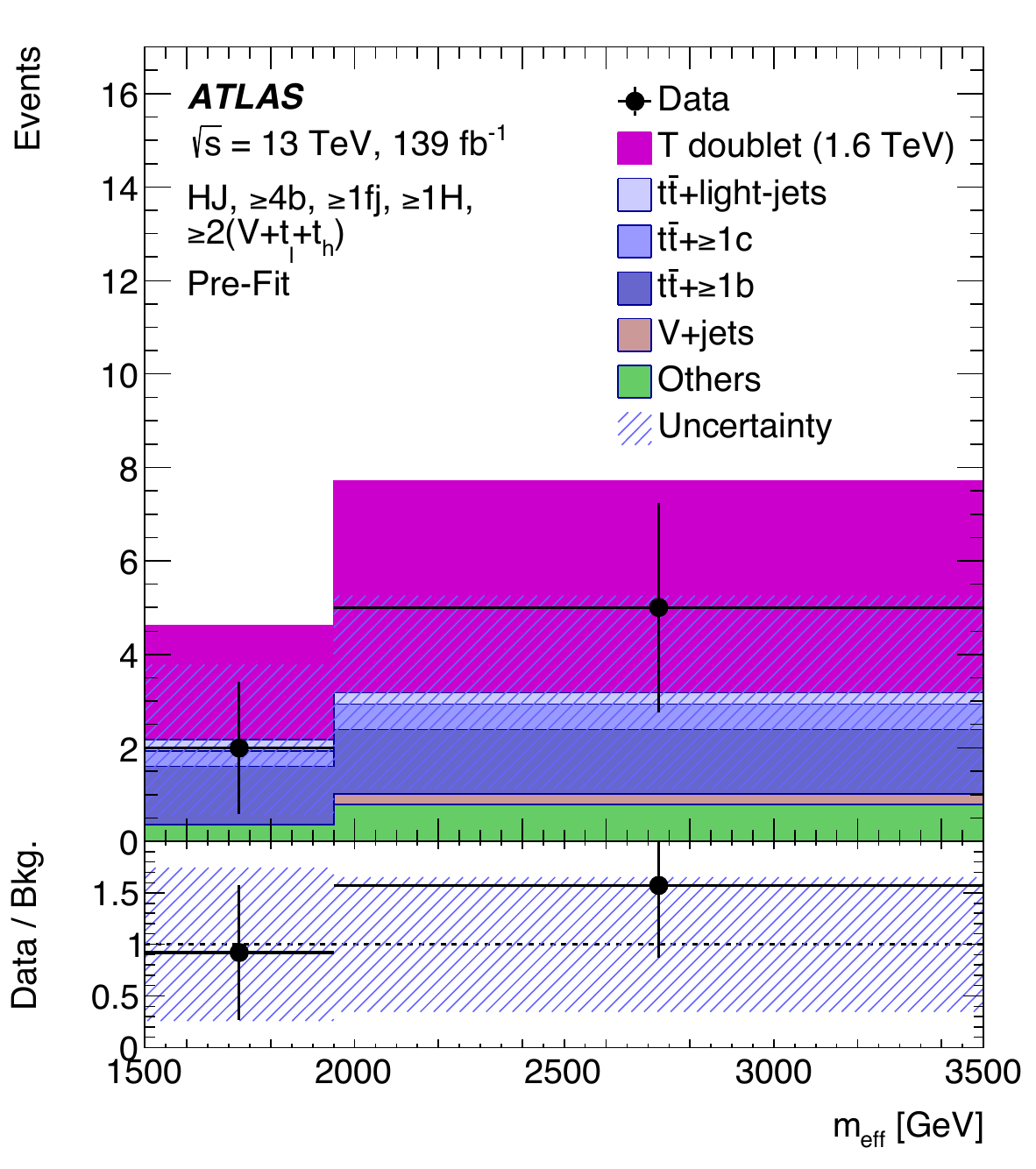}}
\subfloat[]{\includegraphics[width=0.45\textwidth]{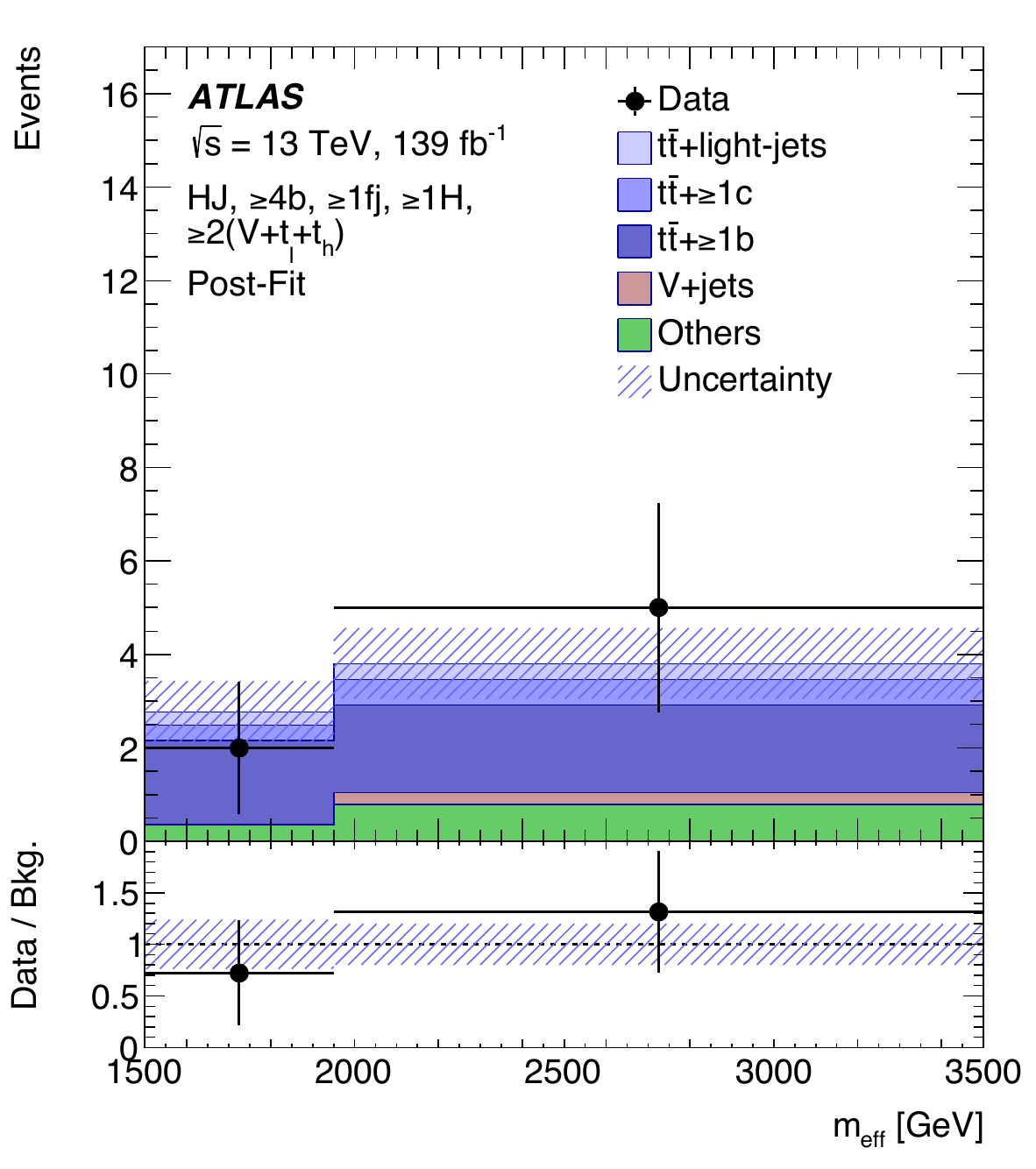}}\\
\caption{\small{Comparison between the data and prediction for the $\meff$ distribution under the background-only hypothesis, in the (HJ, 2b, $\geq$1fj, $\geq$2(t$_h$+t$_l$), 0H, $\geq$1V) region (a) pre-fit and (b) post-fit, and the (HJ, $\geq$4b, $\geq$1fj, $\geq$1H, $\geq$2(V+t$_h$+t$_l$)) region (c) pre-fit and (d) post-fit. The expected $T$ doublet signal (solid purple) for $m_{T}=1.6~\tev$ and $\kappa=0.5$ is included in the pre-fit figures. The \enquote{others} background includes the $\ttbar\ V/H$, \textit{VH}, $tZ$, \fourtop, diboson, and multijet backgrounds. The bottom panels display the ratios of data to the total background prediction. The hashed area represents the total uncertainty in the background. The last bin in each distribution contains the overflow.}}
\label{fig:prepost_HJ}
\end{figure*}
 
\begin{figure*}[h!]
\centering
\includegraphics[width=0.73\textwidth]{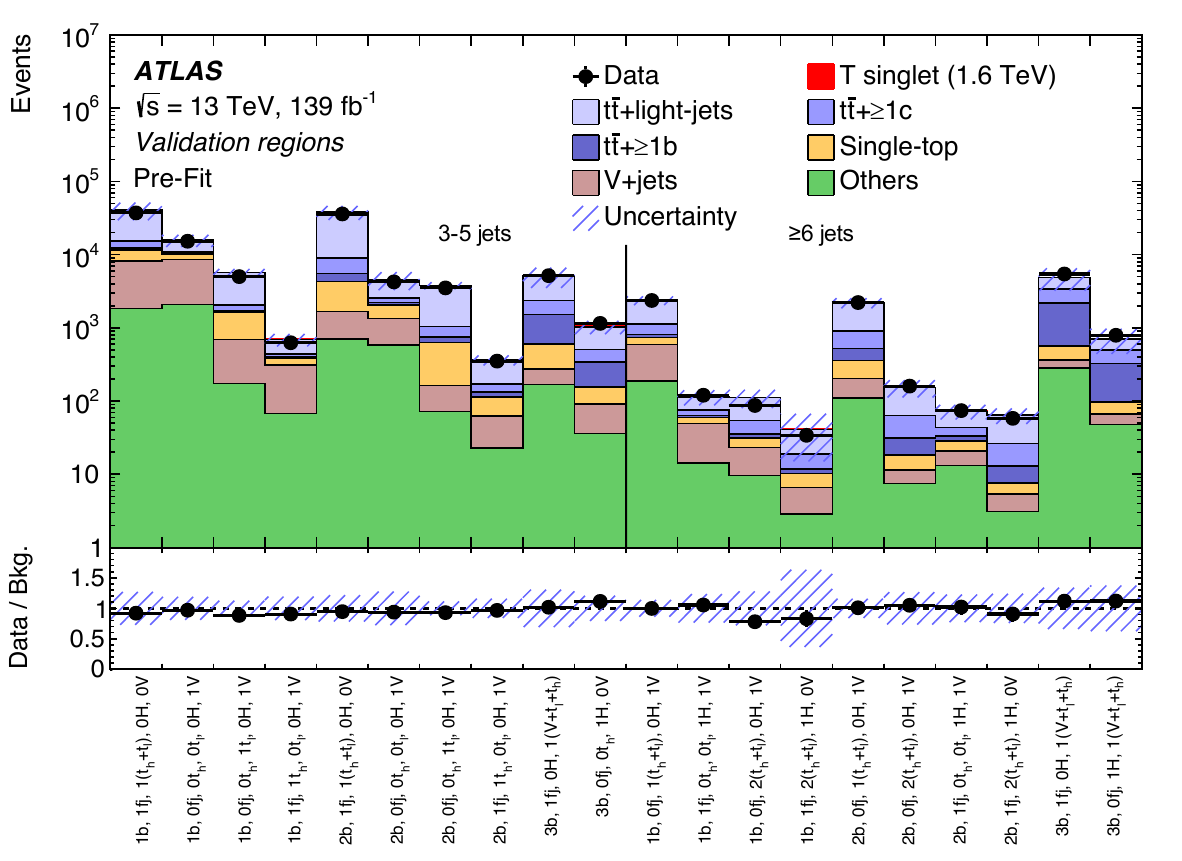}\\
\includegraphics[width=0.73\textwidth]{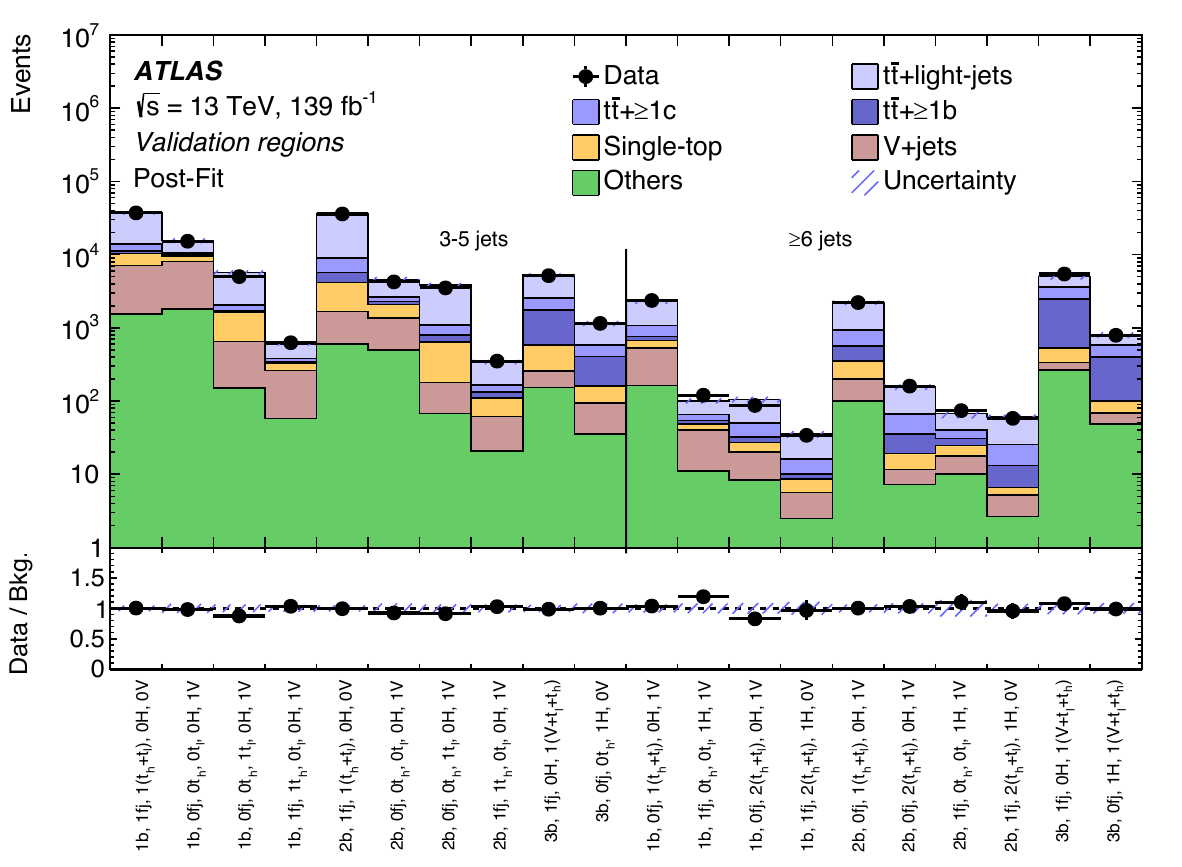}
\caption{\small{Comparison between the data and background prediction for the yields in each of the VRs considered (top) pre-fit and (bottom) post-fit, performed under the background-only hypothesis considering only the fit regions. The \enquote{others} background includes the $\ttbar\ V/H$, \textit{VH}, $tZ$, \fourtop, diboson, and multijet backgrounds. The expected $T$ singlet signal (solid red) for $m_{T}=1.6~\tev$ and $\kappa=0.5$ is included in the pre-fit figure. The bottom panels display the ratios of data to the total background prediction. The hashed area represents the total uncertainty in the background. }}
\label{fig:VR_prefit_postfit_summary}
\end{figure*}
 
\begin{figure*}[t!]
\subfloat[]{\includegraphics[width=0.45\textwidth]{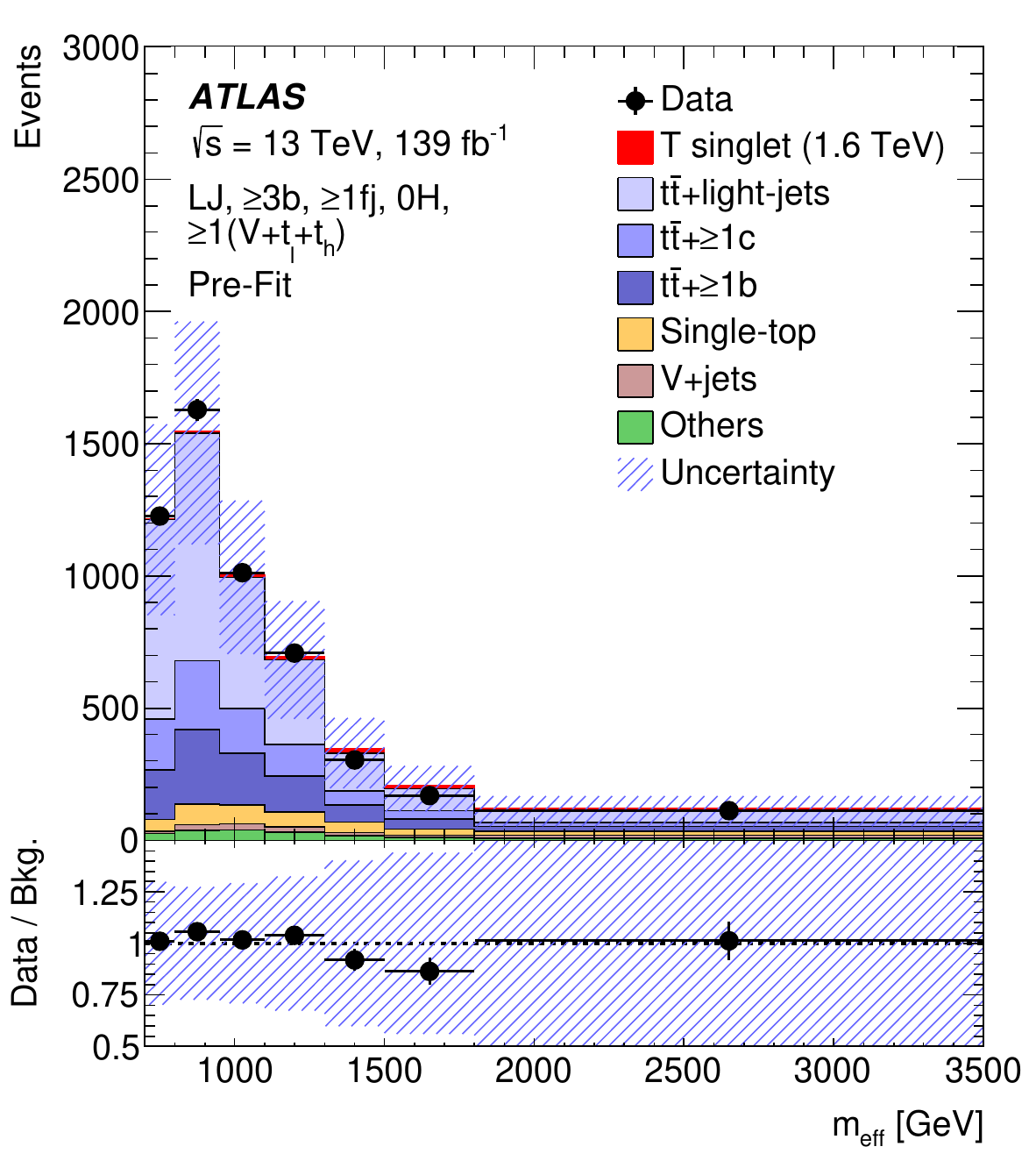}}
\subfloat[]{\includegraphics[width=0.45\textwidth]{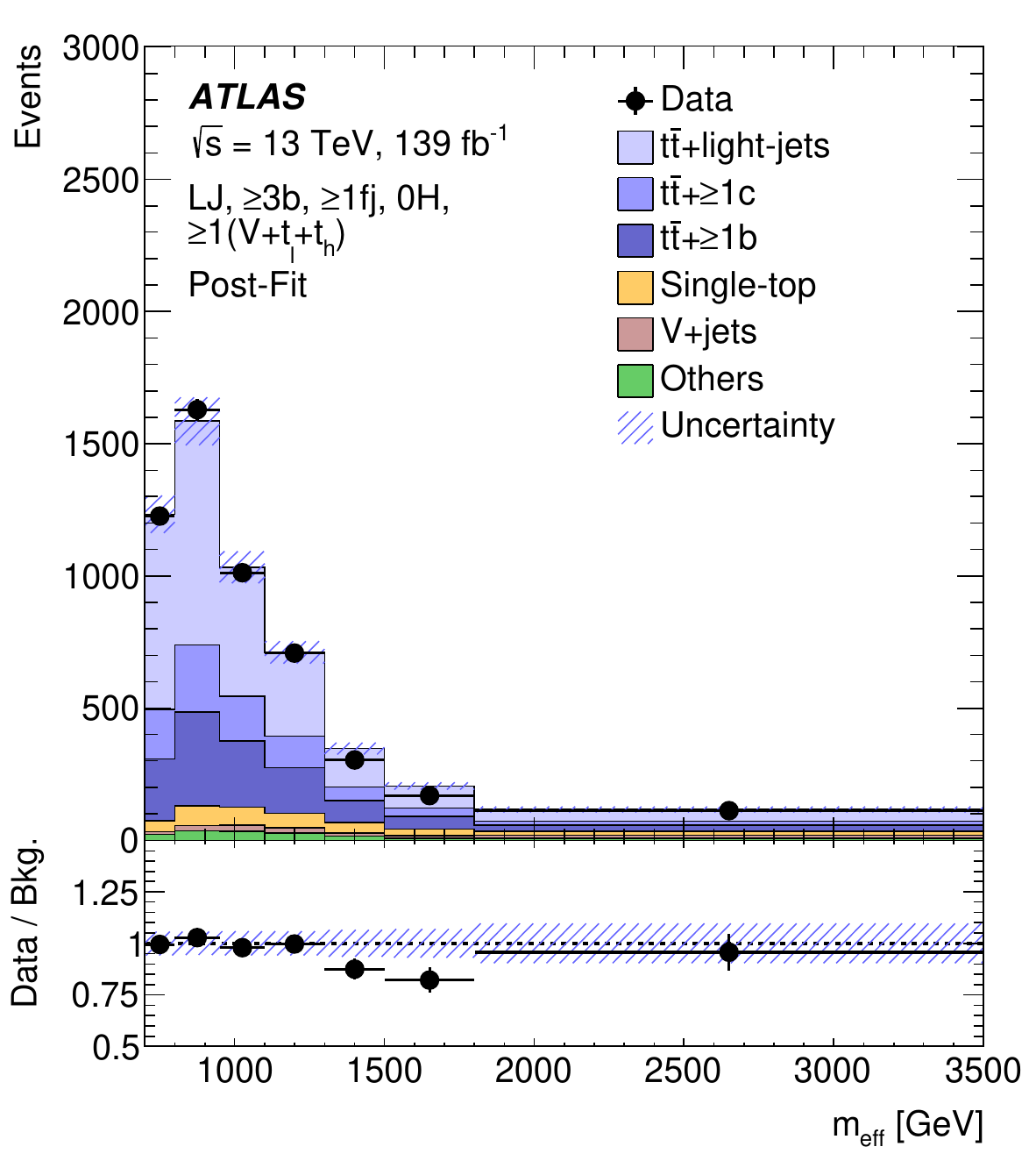}}\\
\subfloat[]{\includegraphics[width=0.45\textwidth]{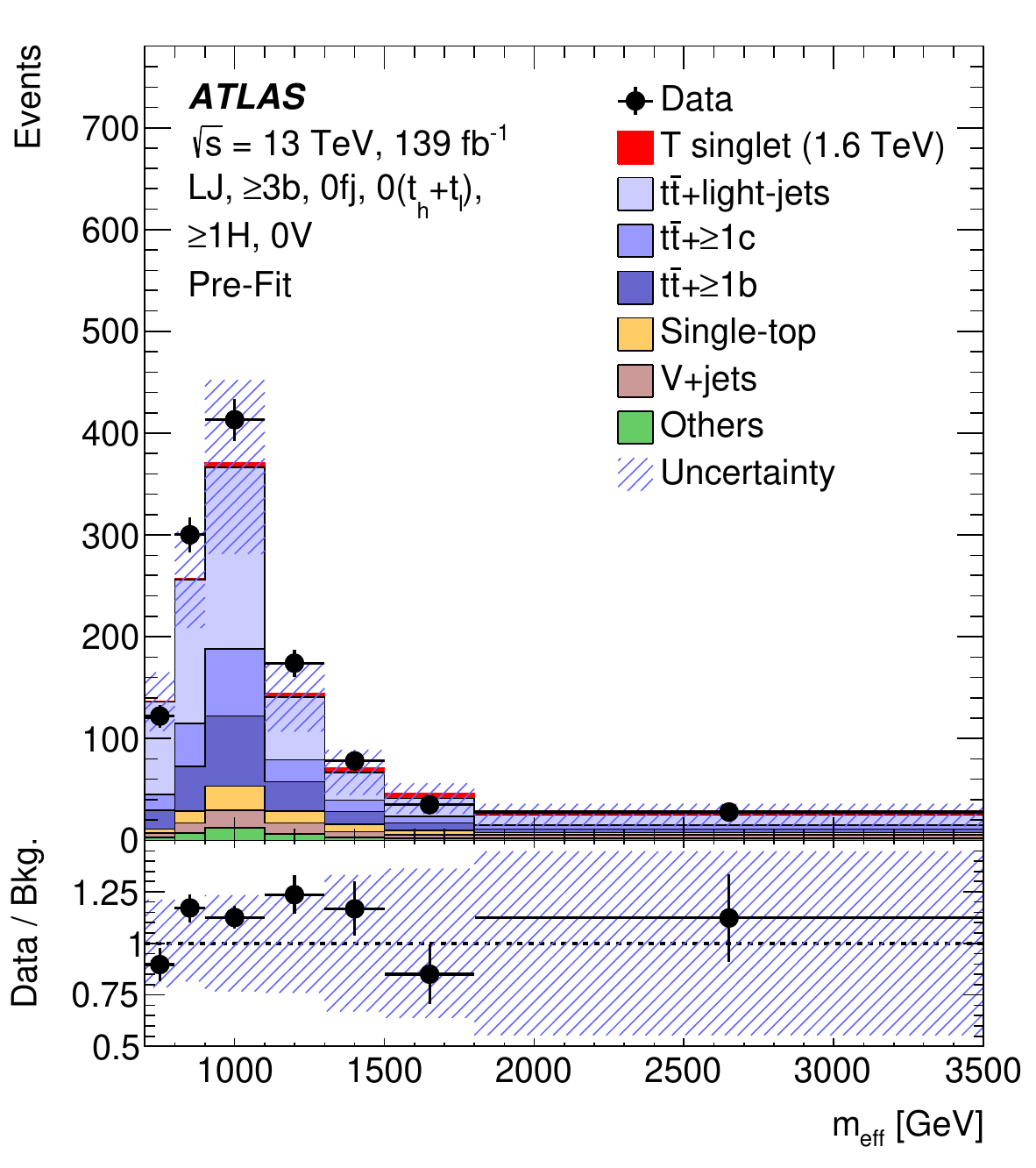}}
\subfloat[]{\includegraphics[width=0.45\textwidth]{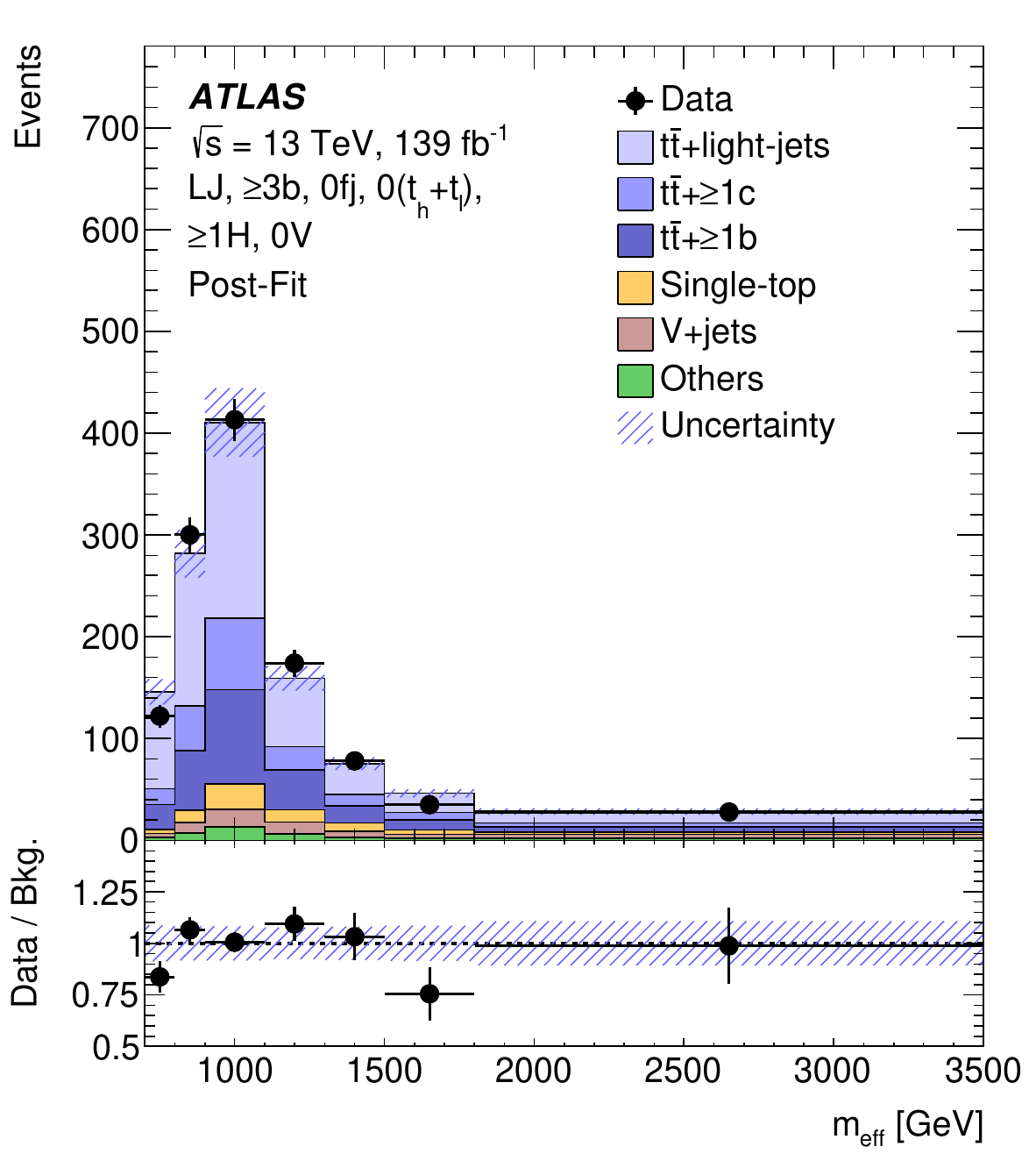}}\\
\caption{\small{Comparison between the data and prediction for the $\meff$ distribution under the background-only hypothesis, in the (LJ, $\geq$3b, $\geq$1fj, 0H, $\geq$1(V+t$_l$+t$_h$)) validation region (a) pre-fit and (b) post-fit, and the (LJ, $\geq$3b, 0fj, 0(t$_h$+t$_l$), $\geq$1H, 0V) validation region (c) pre-fit and (d) post-fit. The expected $T$ singlet signal (solid red) for $m_{T}=1.6~\tev$ and $\kappa=0.5$ is included in the pre-fit figures. The \enquote{others} background includes the $\ttbar\ V/H$, \textit{VH}, $tZ$, \fourtop, diboson, and multijet backgrounds. The bottom panels display the ratios of data to the total background prediction. The hashed area represents the total uncertainty in the background. The last bin in each distribution contains the overflow.}}
\label{fig:prepost_LJ_VR}
\end{figure*}
\begin{figure*}[t!]
\subfloat[]{\includegraphics[width=0.45\textwidth]{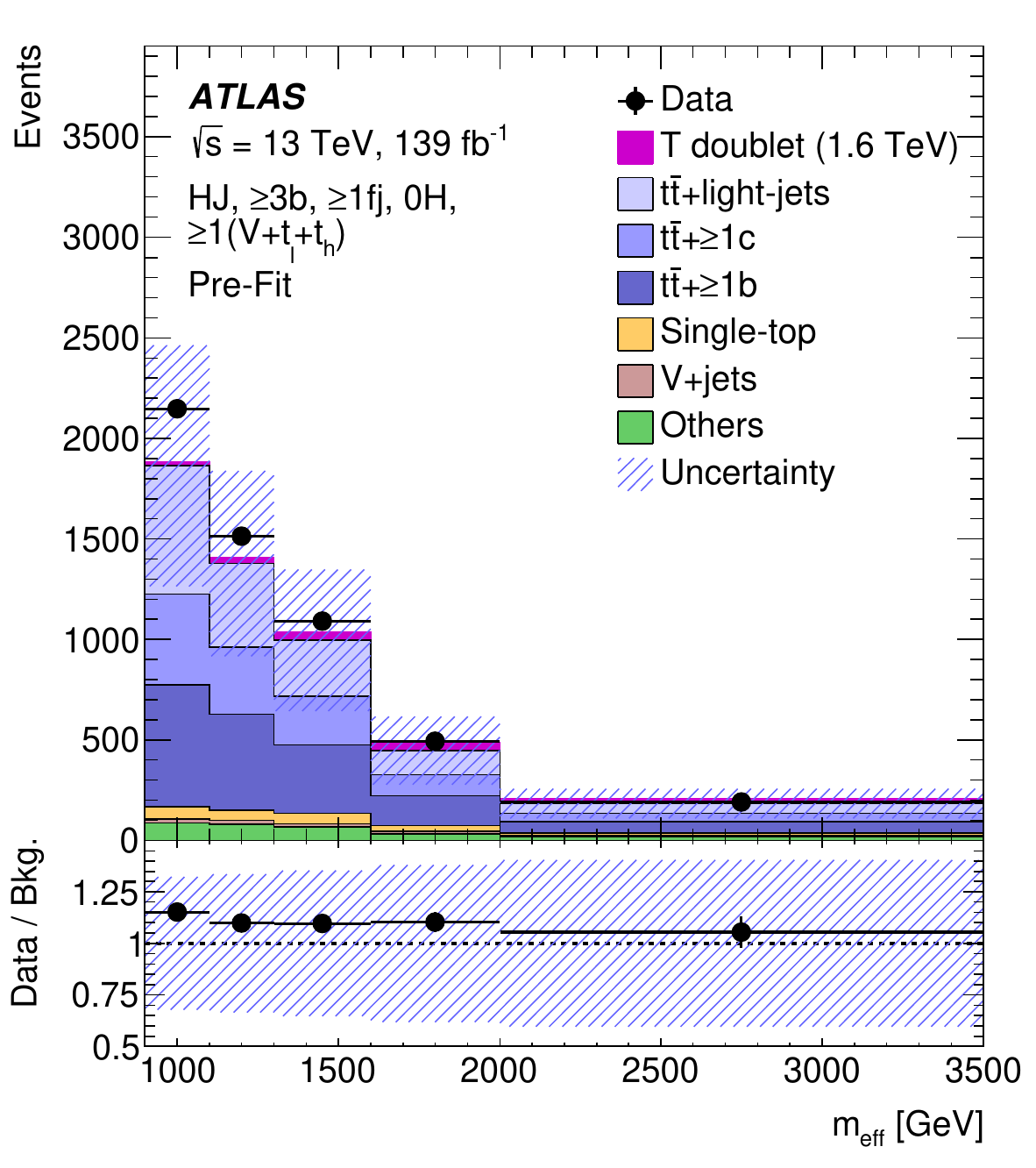}}
\subfloat[]{\includegraphics[width=0.45\textwidth]{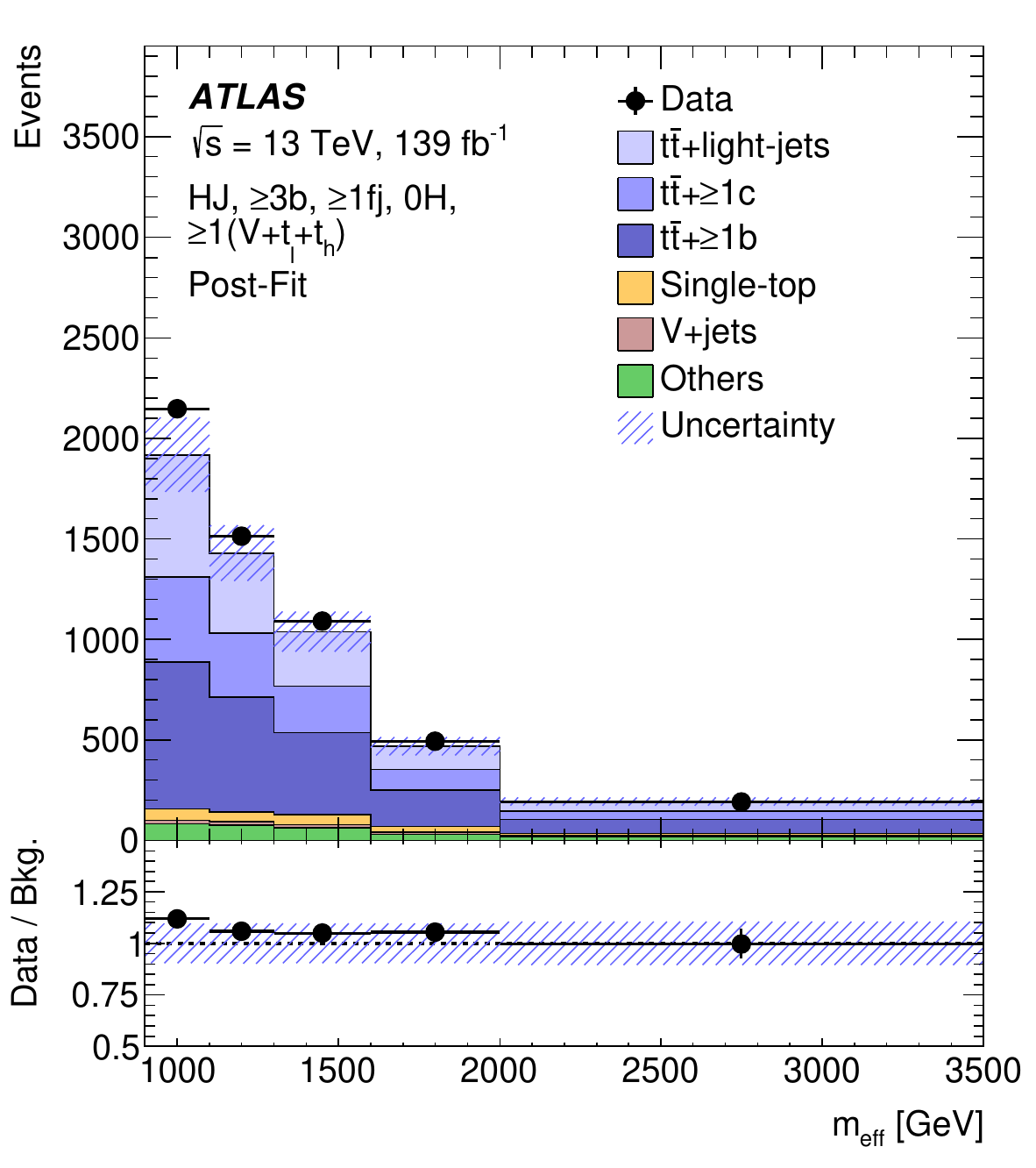}}\\
\subfloat[]{\includegraphics[width=0.45\textwidth]{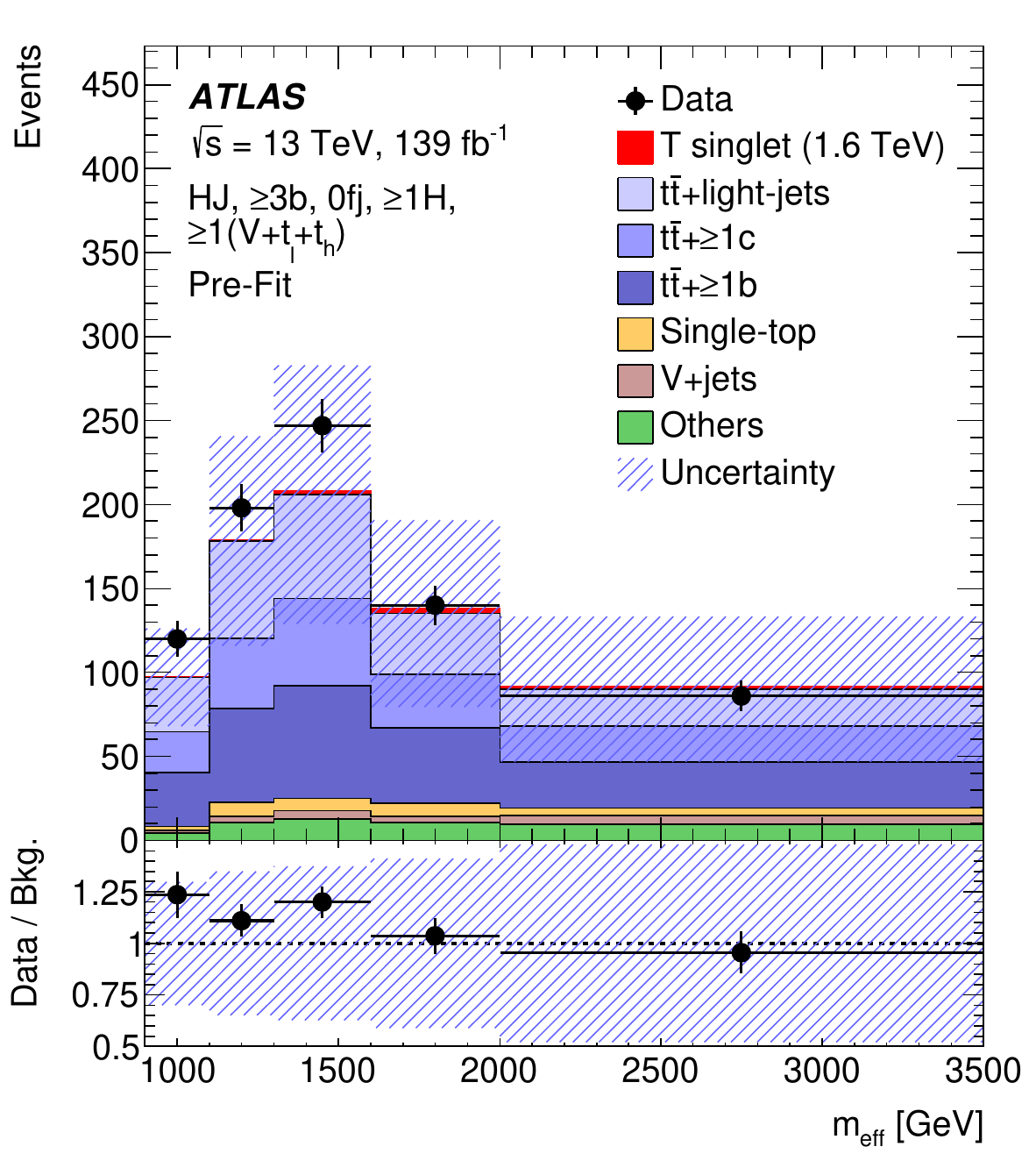}}
\subfloat[]{\includegraphics[width=0.45\textwidth]{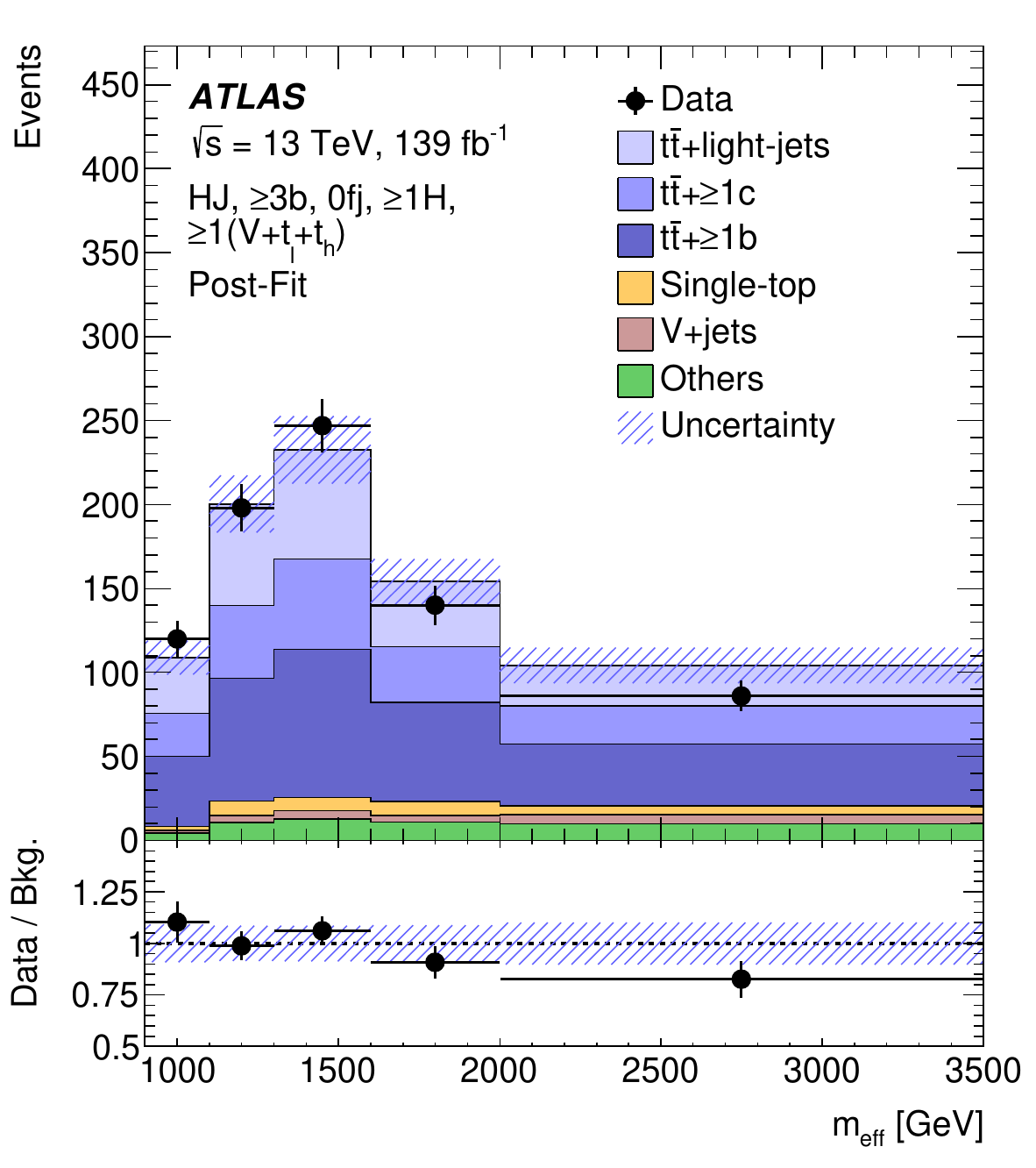}}\\
\caption{\small{Comparison between the data and prediction for the $\meff$ distribution under the background-only hypothesis, in the (HJ, $\geq$3b, $\geq$1fj, 0H, $\geq$1(V+t$_l$+t$_h$)) validation region (a) pre-fit and (b) post-fit, and the (HJ, $\geq$3b, 0fj, $\geq$1H, $\geq$1(V+t$_l$+t$_h$)) validation region (c) pre-fit and (d) post-fit. The expected $T$ doublet signal (solid purple) for $m_{T}=1.6~\tev$ and $\kappa=0.5$ is included in the pre-fit figures. The \enquote{others} background includes the $\ttbar\ V/H$, \textit{VH}, $tZ$, \fourtop, diboson, and multijet backgrounds. The bottom panels display the ratios of data to the total background prediction. The hashed area represents the total uncertainty in the background. The last bin in each distribution contains the overflow.}}
\label{fig:prepost_HJ_VR}
\end{figure*}
 
\begin{table*}[t!]
\centering
\caption{\label{tab:GroupedImpact}
Breakdown of the contributions to the uncertainties in $\mu$, shown for the singlet $T$ signal with $\kappa=0.4$ and the doublet $T$ signal with $\kappa=0.7$, both with $m_{T}=1.5~\tev$. The contributions from the different sources of uncertainty are evaluated after the fit. The $\Delta\mu$ values are obtained by repeating the fit after having fixed a certain set of nuisance parameters corresponding to a group of systematic uncertainties, and then evaluating $(\Delta\mu)^2$ by subtracting the resulting squared uncertainty of $\mu$ from its squared uncertainty found in the full fit. }
 
\begin{tabular}{lcc}
\toprule\toprule
& {$T$ singlet} & {$T$ doublet} \\
& {$m_{T} = 1.5$~\tev, $\kappa=0.4$} & {$m_{T} = 1.5$~\tev, $\kappa=0.7$} \\
\midrule
Uncertainty source & {$\Delta\mu$} & {$\Delta\mu$} \\
\midrule
Process modelling uncertainties & & \\
\hspace{0.5cm} $t\bar{t}$+light-jets                            & $\pm0.05$ & $\pm0.11$ \\
\hspace{0.5cm} $t\bar{t}$+$\geq$1$c$                            & $\pm0.03$ & $\pm0.08$ \\
\hspace{0.5cm} $t\bar{t}$+$\geq$1$b$                            & $\pm0.09$ & $\pm0.27$ \\
\hspace{0.5cm} Single-top                                       & $\pm0.12$ & $\pm0.07$ \\
\hspace{0.5cm} $W$/$Z$+jets                                     & $\pm0.03$ & $\pm0.09$ \\
Object uncertainties & & \\
\hspace{0.5cm} Jets                                             & $\pm0.07$ & $\pm0.13$ \\
\hspace{0.5cm} $b$-tagging                                      & $\pm0.03$ & $\pm0.11$ \\
Other sources                                                   & $\pm0.02$ & $\pm0.04$ \\
Background reweighting                                          & $\pm0.04$ & $\pm0.11$ \\
\midrule
Total systematic uncertainty                                    & $\pm0.23$ & $\pm0.33$ \\
\midrule
Total statistical uncertainty                                   & $\pm0.09$ & $\pm0.17$ \\
\bottomrule\bottomrule
\end{tabular}
\vspace{0.5cm}
\end{table*}
 
\clearpage
\subsection{Limits on single vector-like quark production}
\label{subsec:results_limits}
 
No significant excess above the SM prediction is found in any of the considered regions in the background-only
fit. Unconditional fits with a floating signal-strength parameter $\mu$ were also consistent with the
background-only hypothesis.
Upper limits at 95\% CL on the single-$T$ production cross section are derived in both
the singlet $(T)$ and doublet ($T~B$) scenarios.
 
The observed cross-section limits at each point in the parameter space are compared with the NLO
theoretical prediction to set exclusion limits on model parameters. Since the theory cross-section calculations
with finite-width effects and non-resonant contributions are only reliable for relative $T$-quark widths ($\Gamma_{T}/m_{T}$)
up to $\sim$50\%~\cite{Deandrea:2021vje}, results are presented only in this restricted regime.
 
The obtained limits corresponding to the singlet and doublet scenarios are shown in
Figure~\ref{fig:limits_singlet_massxs} and Figure~\ref{fig:limits_doublet_massxs}, respectively,
for a set of three values of the common coupling parameter $\kappa$, chosen to approximately span
the sensitivity range of the search in each scenario. The corresponding limits are also derived in
the mass versus coupling plane, where exclusion contours indicate the interpolated intersection
between the planes of excluded and theoretically predicted cross sections, shown in
Figure~\ref{fig:limits_2D_cont_mass_kappa} for both the singlet and doublet scenarios.
Additionally, upper limits on the production cross section of both the singlet and doublet scenarios
are derived as a function of mass and coupling, as shown in Figures~\ref{fig:limits_2D_temp_mass_kappa_TS}
and~\ref{fig:limits_2D_temp_mass_kappa_TD}.
As expected, the exclusion limits on the $T$-quark mass are generally stronger in the singlet scenario than
in the doublet scenario. All $T$-quark masses below 2.1~\TeV\ (expected 1.9~\TeV)
are excluded for singlet $T$ quarks at couplings $\kappa \geq 0.6$. At a mass of 1.6~\TeV, $\kappa$ values
above 0.3 (expected 0.41) are all excluded. By comparison, the previous ATLAS search in the all-hadronic
$T \rightarrow Ht$ channel has excluded $\kappa$ values above $\sim$0.5 (expected $\sim$0.65) at a mass of
1.6~\TeV~\cite{ATLAS:2022ozf}. In the doublet scenario, the limits on the considered mass
range extend down to coupling values of $\kappa=0.55$, corresponding to a $T$-quark mass limit of 1.0~\TeV.
Masses up to 1.68~\TeV\ are excluded at $\kappa=0.75$, at a relative $T$-quark width threshold of 50\%.
 
The expected limits on the cross section get progressively stronger at larger masses in both scenarios,
as the decay products of the $T$ quark become more boosted, and the fraction of signal in the highest
\meff\ bins increases. The limits deteriorate at larger values of $\kappa$, since this regime corresponds
to large resonance width and a larger fraction of the signal resides in the low mass regime away
from the peak of the resonance.
As can be seen, the observed limits exceed the expected limits in both benchmark scenarios in a few cases.
These deviations are larger for the singlet scenario, reaching almost 2$\sigma$ at high masses.
These findings can be ascribed to downward statistical fluctuations in a few of the most signal-sensitive bins,
most notably in the last bin of the (LJ, $\geq$4b, $\geq$1fj, 0t$_h$, $\geq$1t$_l$, $\geq$1H, 0V)
region, which has no data events, and the last few bins of the (HJ, $\geq$4b, $\geq$1fj, 0t$_h$, 1t$_l$, $\geq$1H, 0V)
region. The origin of these discrepancies was investigated, and no evidence of any systematic bias was found.
Notably, the pre- and post-fit \meff\ distributions in the corresponding validation regions, shown in
Figure~\ref{fig:prepost_LJ_VR} and Figure~\ref{fig:prepost_HJ_VR}, respectively, exhibit good agreement
between the data and expectations. Several other fit regions with kinematic features and background
compositions similar to those in the two regions mentioned above show good agreement between data and predictions. The
observed discrepancies were therefore deemed to be consistent with statistical fluctuations.
 
Finally, the results are interpreted in a more generalised representation of the parameter space, displaying
the largest excluded mass as a function of $\Gamma_{T}/m_{T}$ and the relative coupling parameter $\xi_{W}$
(Figure~\ref{fig:limits_2D_temp_xi_gamma}). As discussed in Section~\ref{sec:intro},
the $T$-quark width is determined by the $T$-quark mass $m_{T}$ and the universal coupling constant $\kappa$ and
is independent of the multiplet representation. The relative coupling parameter $\xi_{W}$ controls the
branching fraction $\BR(T \rightarrow Wb)$. For ease of representation, the limits are shown
for the assumption $\xi_{Z}~=~\xi_{H}$, which is valid for all VLQ multiplet scenarios in the phenomenological
model discussed in Section~\ref{sec:intro}. Since the relative coupling constants $\xi_{W,Z,H}$ must sum to
unity, the assumption of equal $\xi_{Z}$ and $\xi_{H}$ fully determines the values of these parameters for any
given value of $\xi_{W}$.
 
In all of the benchmarks considered in this search, only the contributions from $T$-quark
production were taken into account. In particular, limits presented in the doublet
($T~B$) scenario neglect contributions from $B$-quark production. The most relevant
signature of $B$-quark production for this analysis would occur in the $B \rightarrow Wt$
decay channel, when the top quark decays leptonically and the $W$ boson decays hadronically.
This process has a signature almost identical to the $T \rightarrow Zt$ decay signatures
considered in the search, and would therefore make very similar contributions to the fit regions.
By the same argument, contributions from $B$-quark production in the RSRs and the background
control regions are expected to be negligible.
Thus, the limits presented for the doublet scenario in this paper can be considered to be conservative.
 
\clearpage
\begin{figure*}[t!]
\centering
\subfloat[]{\includegraphics[width=0.32\textwidth]{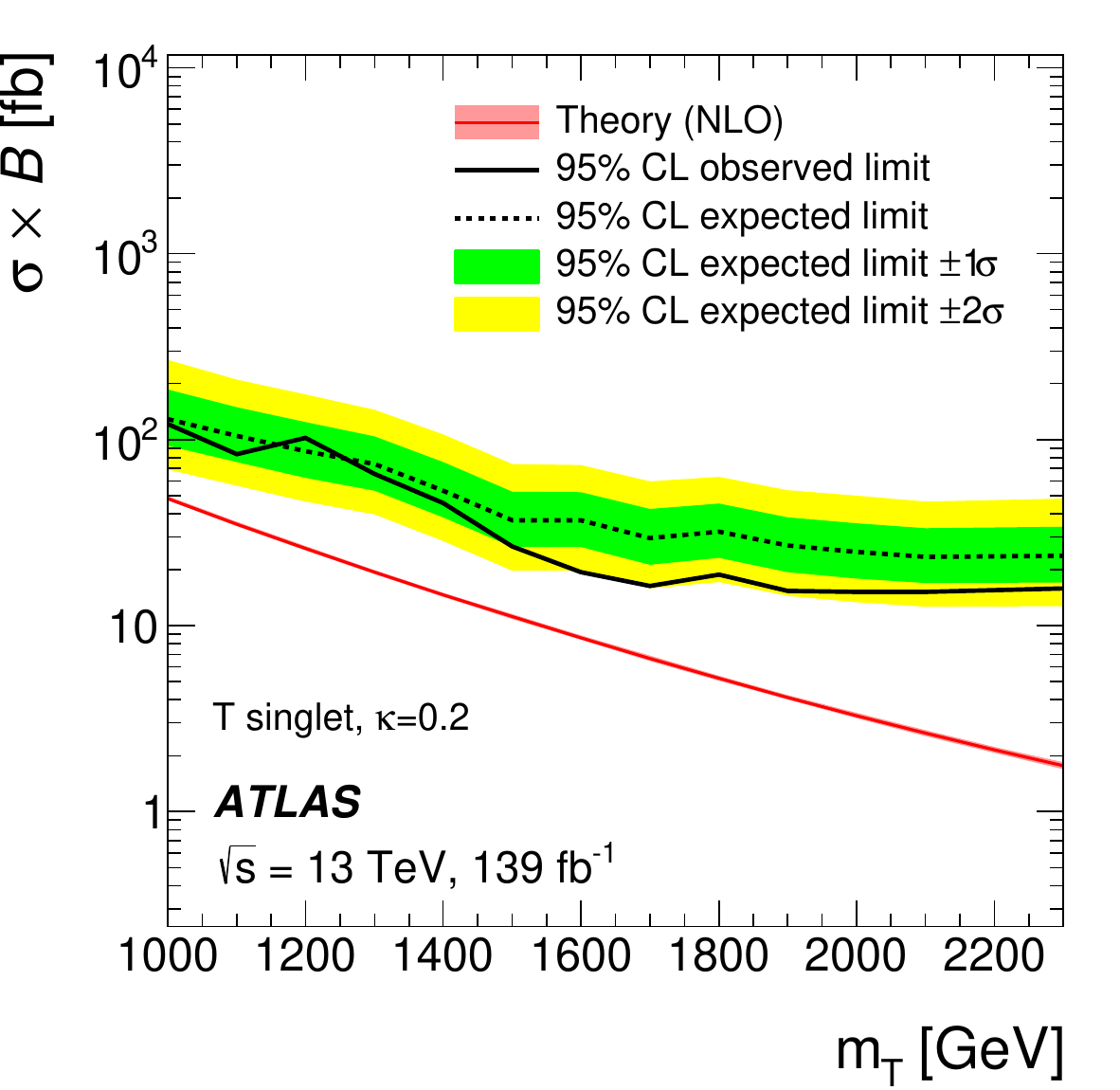}}
\subfloat[]{\includegraphics[width=0.32\textwidth]{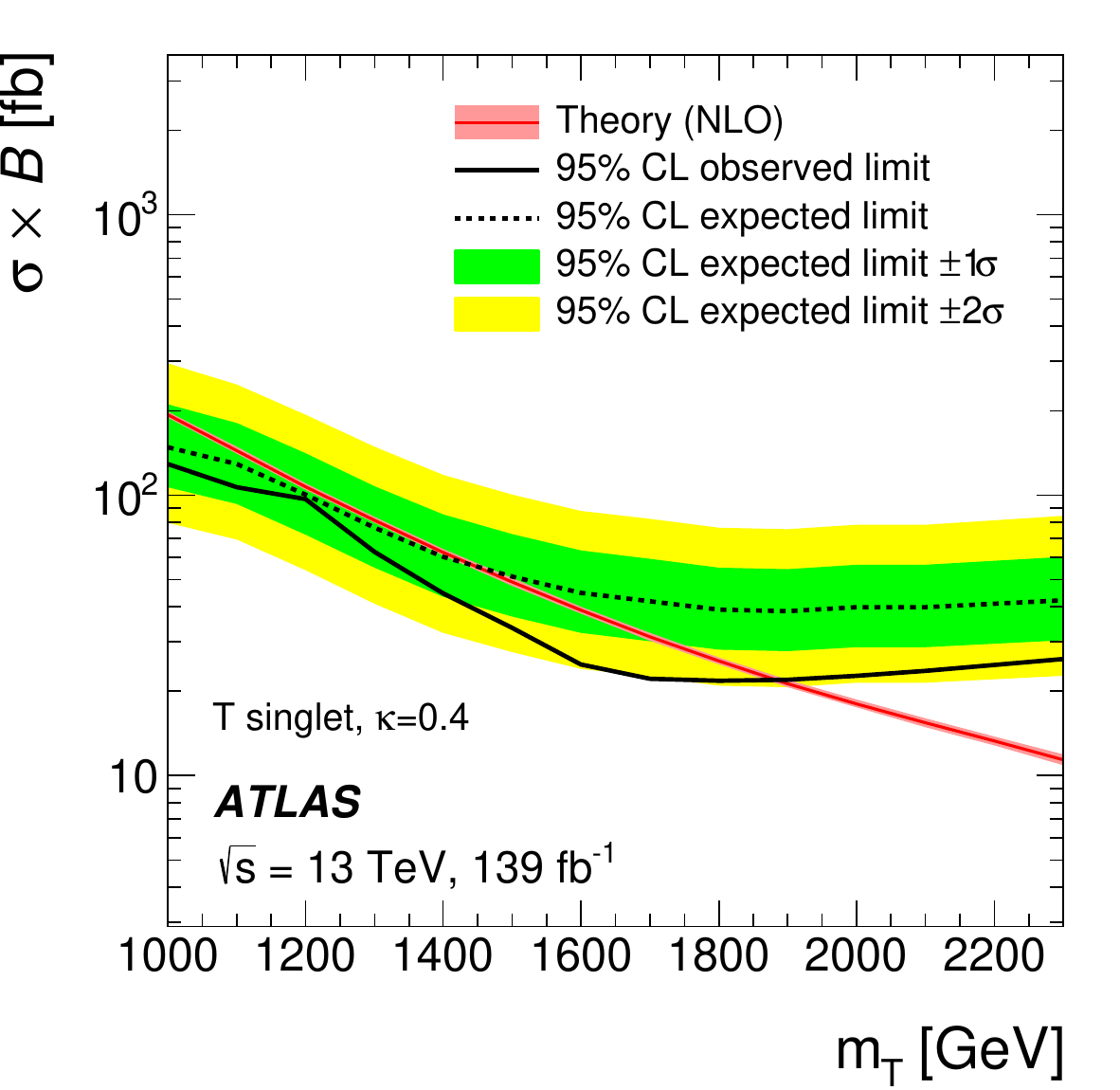}}
\subfloat[]{\includegraphics[width=0.32\textwidth]{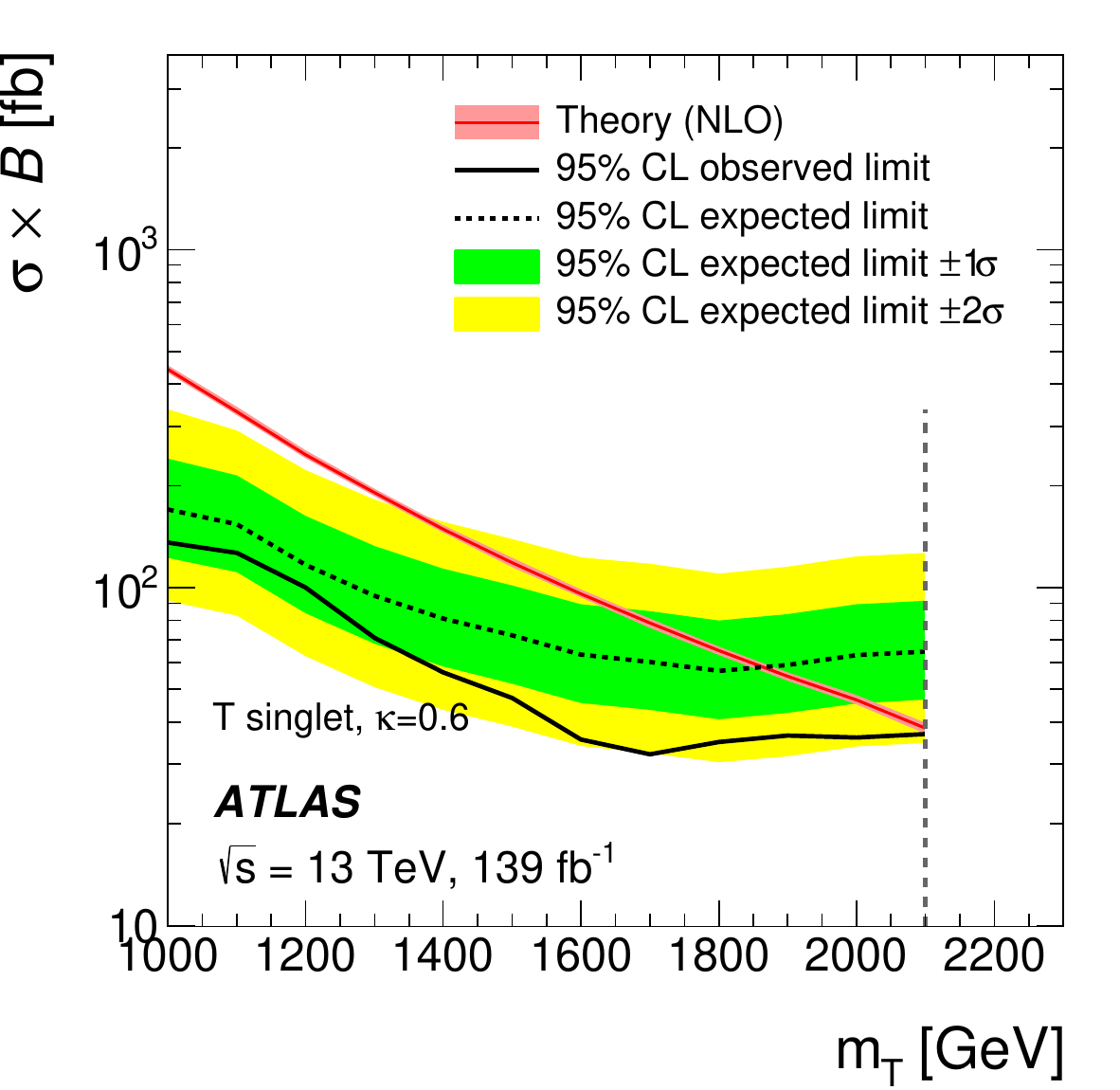}}
\caption{Observed (solid black line) and expected (dashed black line) 95\% CL upper limits on the single-$T$ production
cross section as a function of the $T$-quark mass in the singlet scenario with the common coupling parameter
(a) $\kappa=0.2$, (b) $\kappa=0.4$, and (c) $\kappa=0.6$. 
The surrounding shaded bands correspond to $\pm 1$ and $\pm 2$ standard deviations around the expected limit. The red line
shows the NLO theoretical cross-section prediction, with the surrounding shaded band representing the corresponding uncertainty.
Limits are only presented in the regime $\Gamma_{T}/m_{T} \leq 50$\%, where the theory calculations are known to be valid, as indicated by the vertical grey dashed line.}
\label{fig:limits_singlet_massxs}
\end{figure*}
\begin{figure*}[t!]
\centering
\subfloat[]{\includegraphics[width=0.32\textwidth]{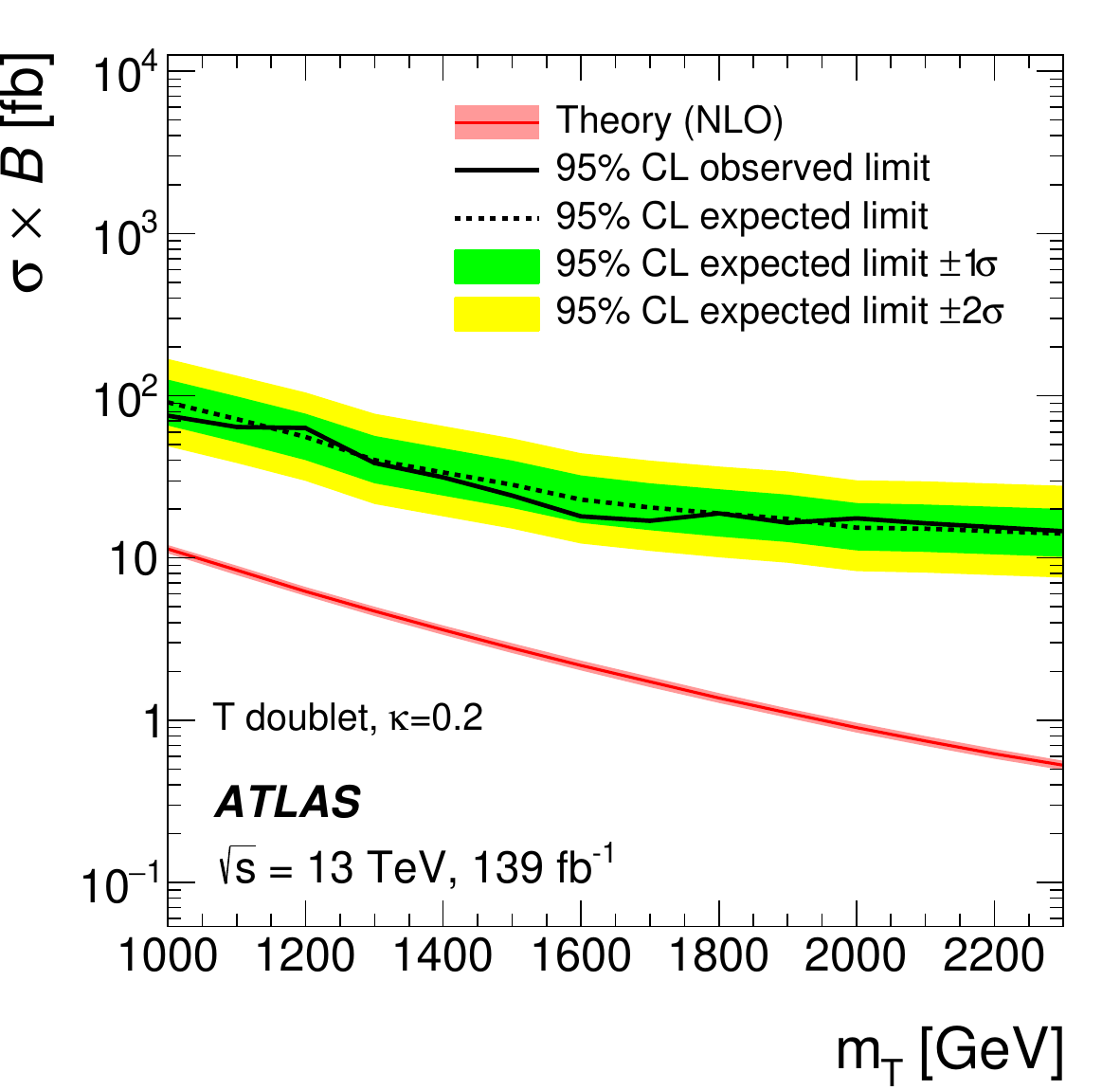}}
\subfloat[]{\includegraphics[width=0.32\textwidth]{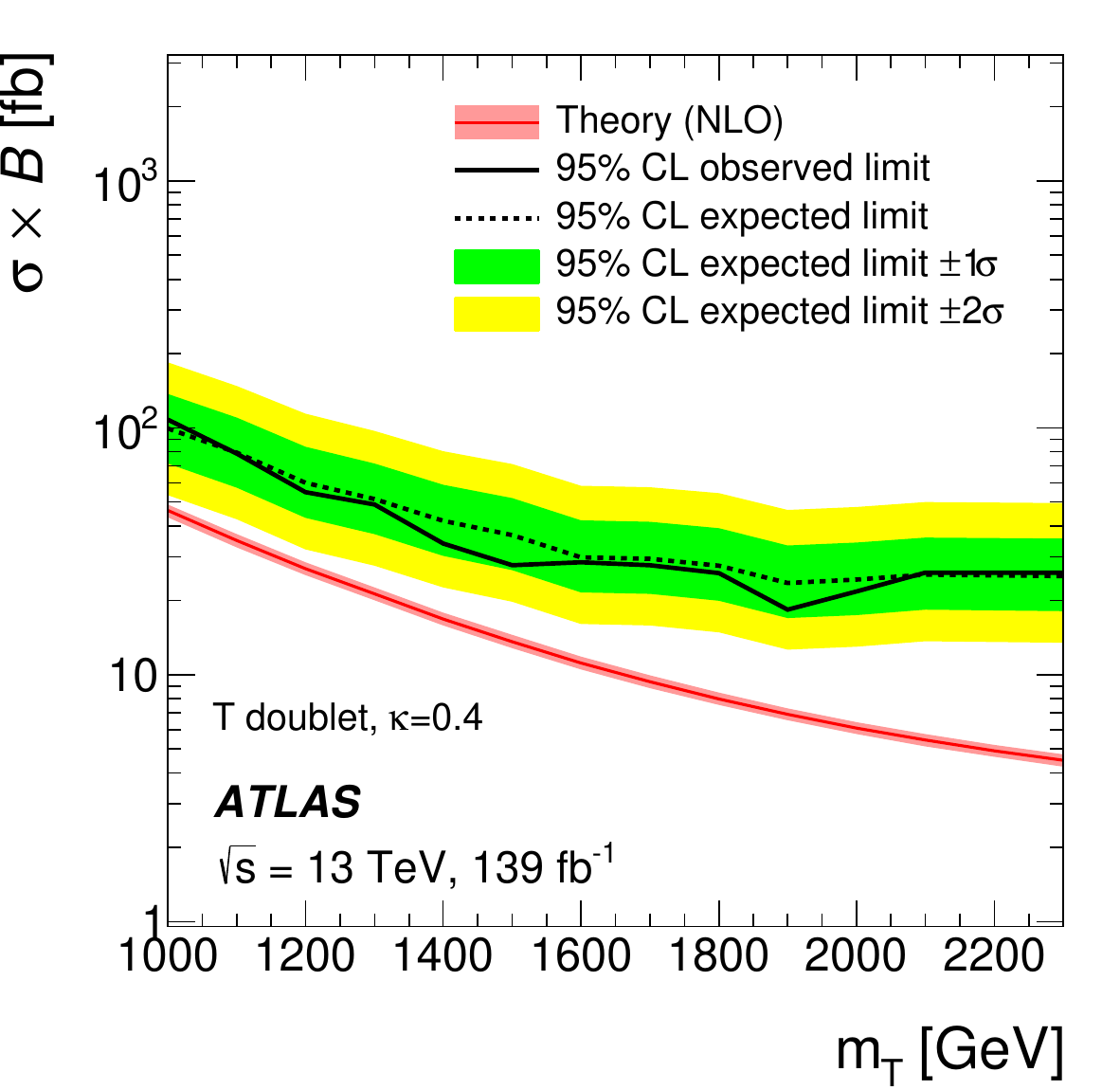}}
\subfloat[]{\includegraphics[width=0.32\textwidth]{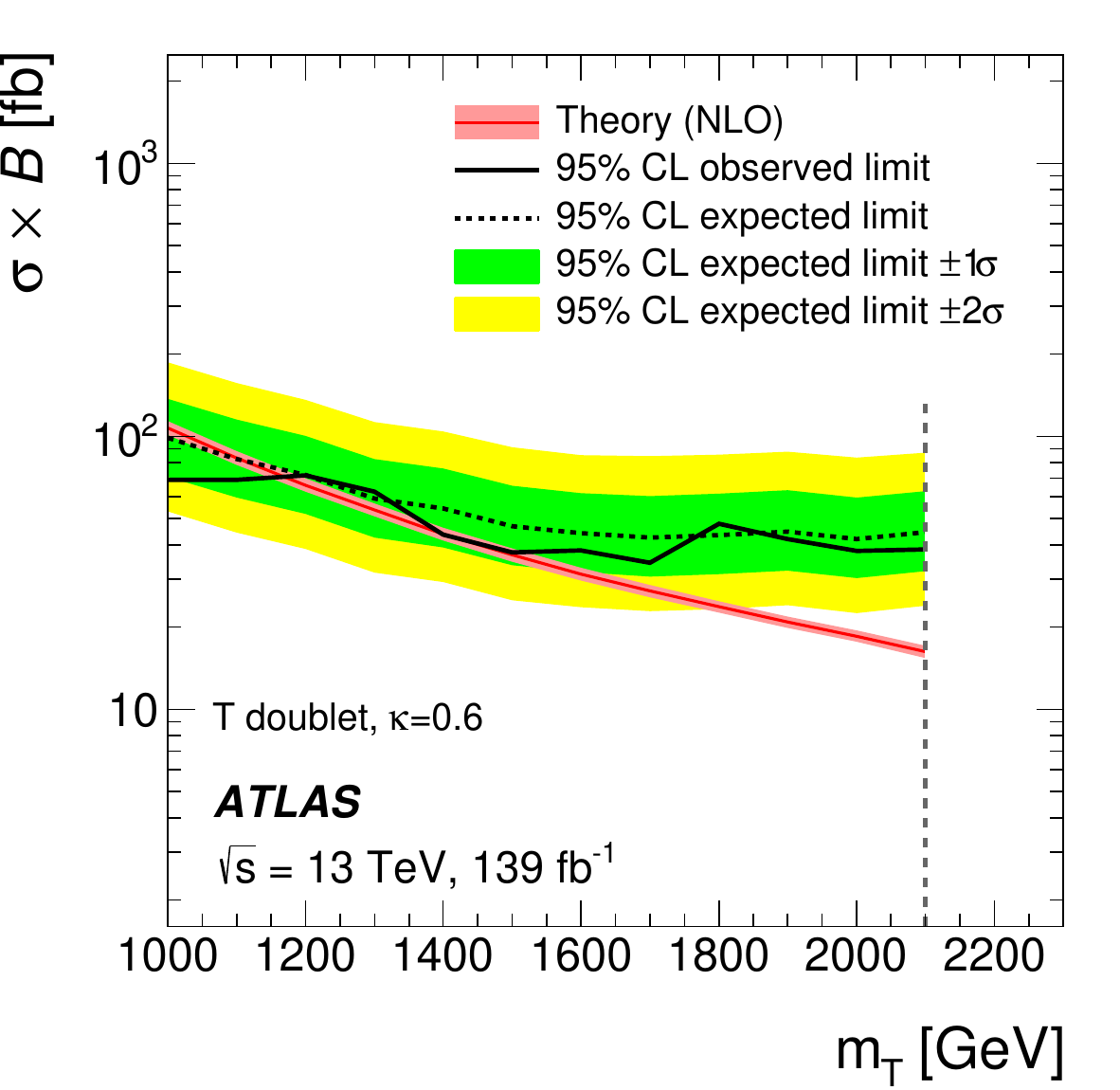}}
\caption{Observed (solid black line) and expected (dashed black line) 95\% CL upper limits on the single-$T$ production
cross section as a function of the $T$-quark mass in the doublet scenario with the common coupling parameter
(a) $\kappa=0.2$, (b) $\kappa=0.4$, and (c) $\kappa=0.6$. 
The surrounding shaded bands correspond to $\pm 1$ and $\pm 2$ standard deviations around the expected limit. The red line
shows the NLO theoretical cross-section prediction, with the surrounding shaded band representing the corresponding uncertainty.
Limits are only presented in the regime $\Gamma_{T}/m_{T} \leq 50$\%, where the theory calculations are known to be valid, as indicated by the vertical grey dashed line.}
\label{fig:limits_doublet_massxs}
\end{figure*}
\begin{figure*}[b!]
\centering
\subfloat[]{\includegraphics[width=0.48\textwidth]{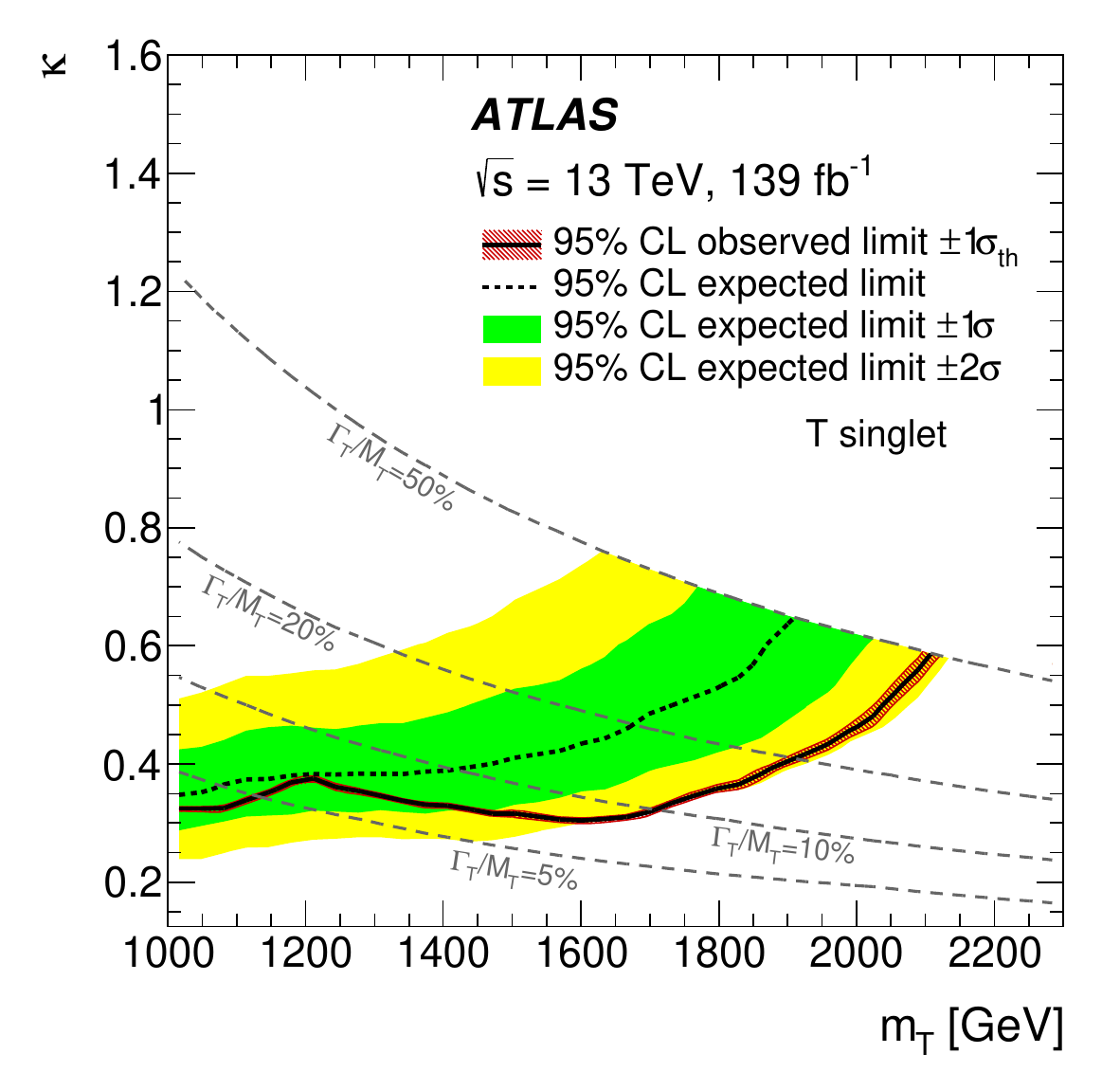}}
\hfill
\subfloat[]{\includegraphics[width=0.48\textwidth]{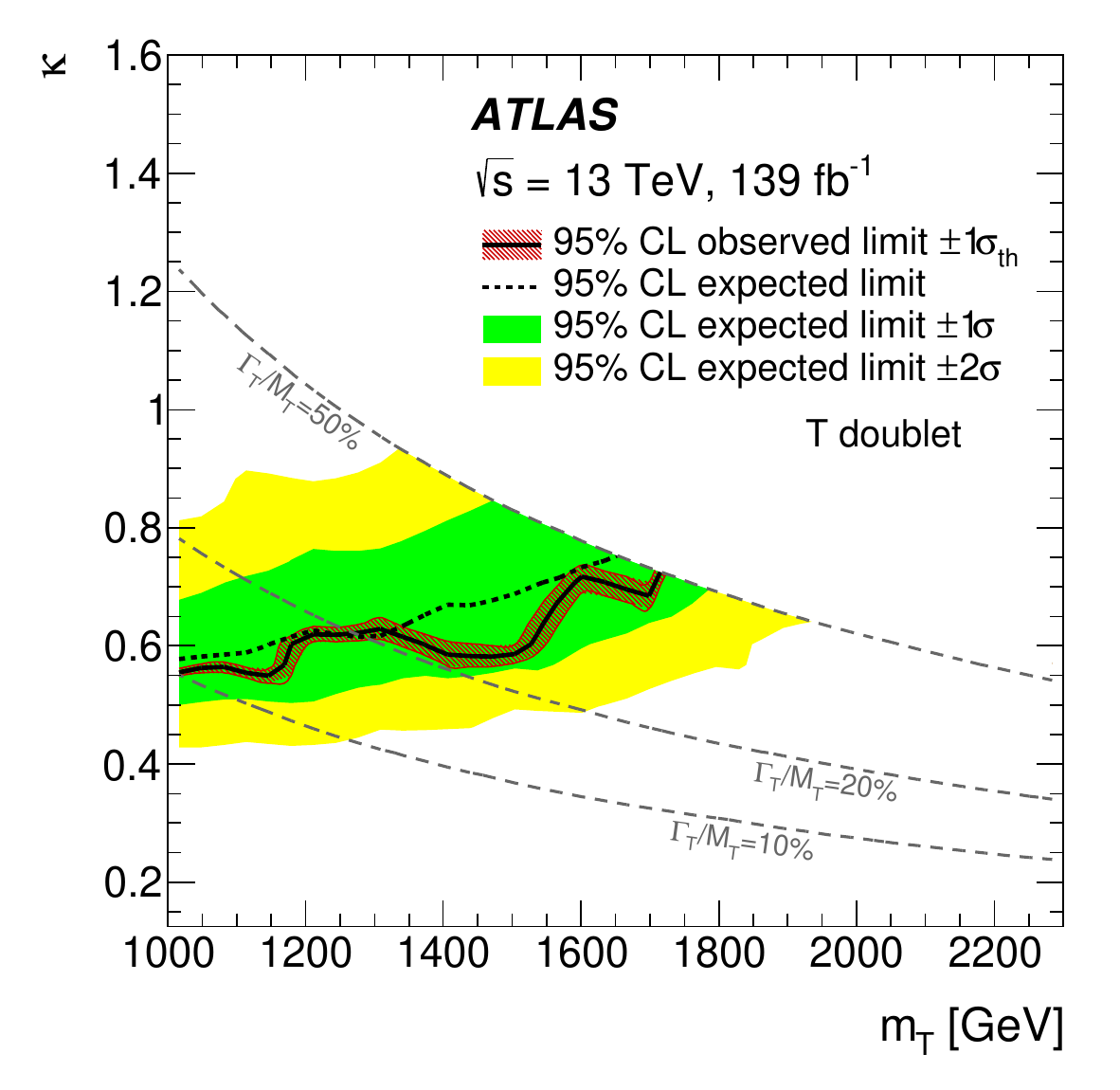}}
\caption{Observed (solid black line) and expected (dashed black line) 95\% CL exclusion limits on the universal
coupling constant $\kappa$ as a function of the $T$-quark mass in the (a) SU(2) singlet and (b) SU(2) doublet scenarios.
All values of $\kappa$ above the black contour lines are excluded at each mass point.
The shaded bands correspond to $\pm 1$ and $\pm 2$ standard deviations around the expected limit. The red hashed area around the observed limit corresponds to the theoretical uncertainty of the NLO cross-section prediction.
The grey dashed lines represent configurations of ($m_T,\,\kappa$) resulting in equal values of the relative
resonance width $\Gamma_{T}/m_{T}$. Limits are only presented in the regime $\Gamma_{T}/m_{T} \leq 50$\%, where the theory calculations are known to be valid.}
\label{fig:limits_2D_cont_mass_kappa}
\end{figure*}
\begin{figure*}[b!]
\centering
\subfloat[]{\includegraphics[width=0.48\textwidth]{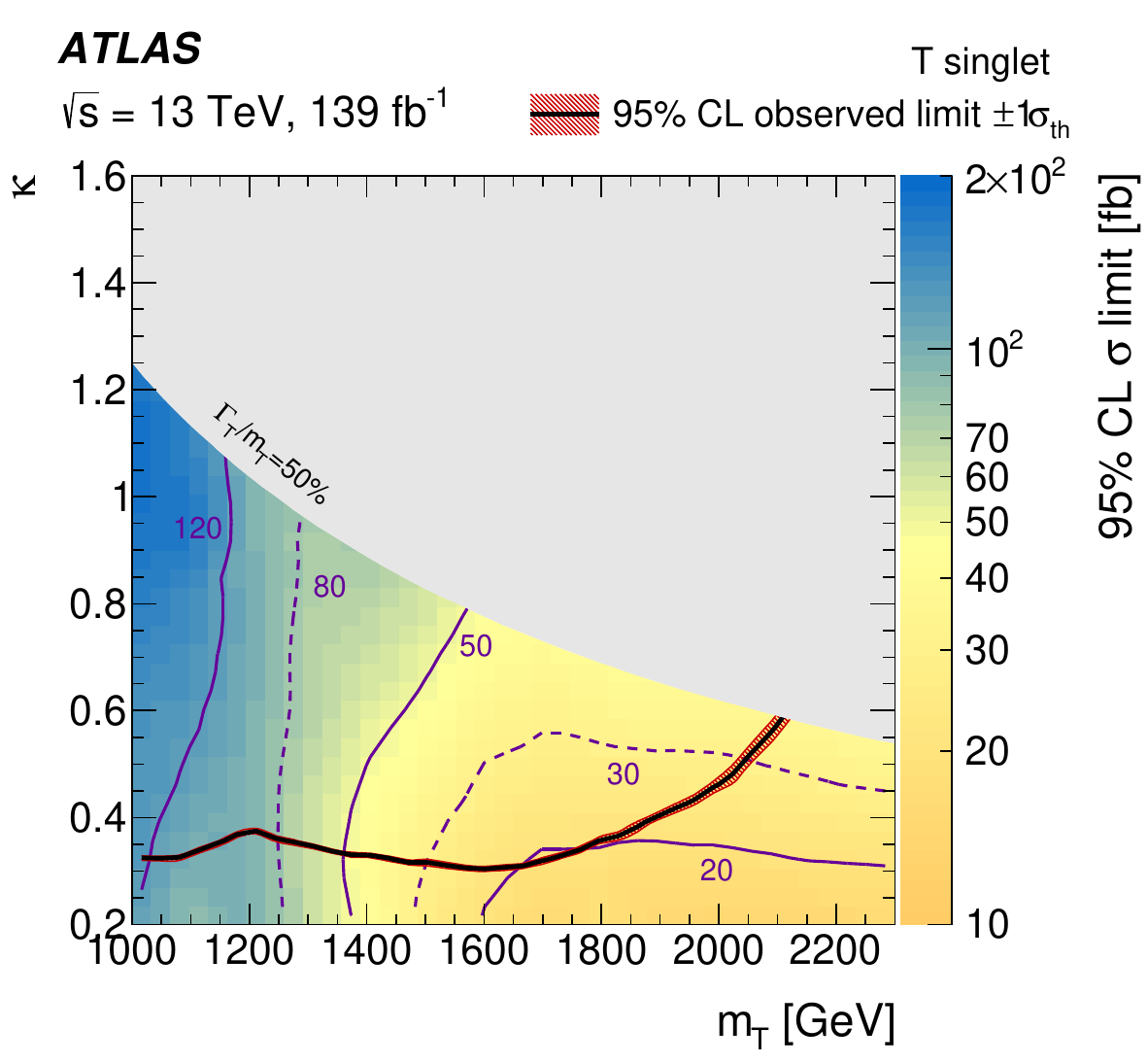}}
\hfill
\subfloat[]{\includegraphics[width=0.48\textwidth]{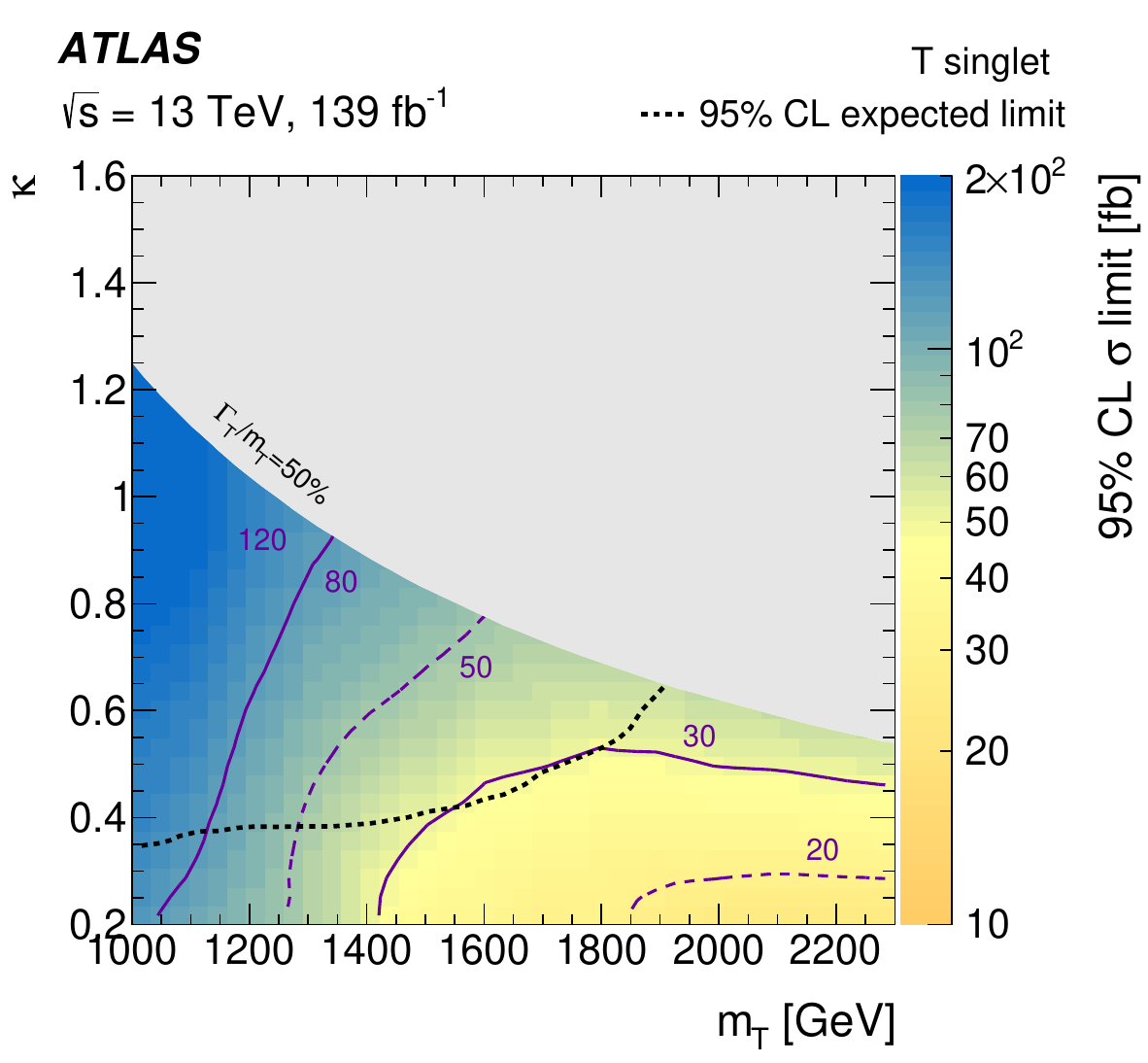}}
\caption{(a) Observed and (b) expected 95\% CL exclusion limits on the cross section times branching
ratio of single $T$-quark production as a function of the universal coupling constant $\kappa$ and the
$T$-quark mass in the SU(2) singlet scenario. The red hashed area around the observed limit corresponds to the theoretical uncertainty of the NLO cross-section prediction.
All values of $\kappa$
above the black contour line are excluded at each mass point. The purple contour lines denote exclusion
limits of equal cross section times branching ratio in units of fb. Limits are only presented in the
regime $\Gamma_{T}/m_{T} \leq 50$\%, where the theory calculations are known to be valid.}
\label{fig:limits_2D_temp_mass_kappa_TS}
\end{figure*}
\begin{figure*}[b!]
\centering
\subfloat[]{\includegraphics[width=0.48\textwidth]{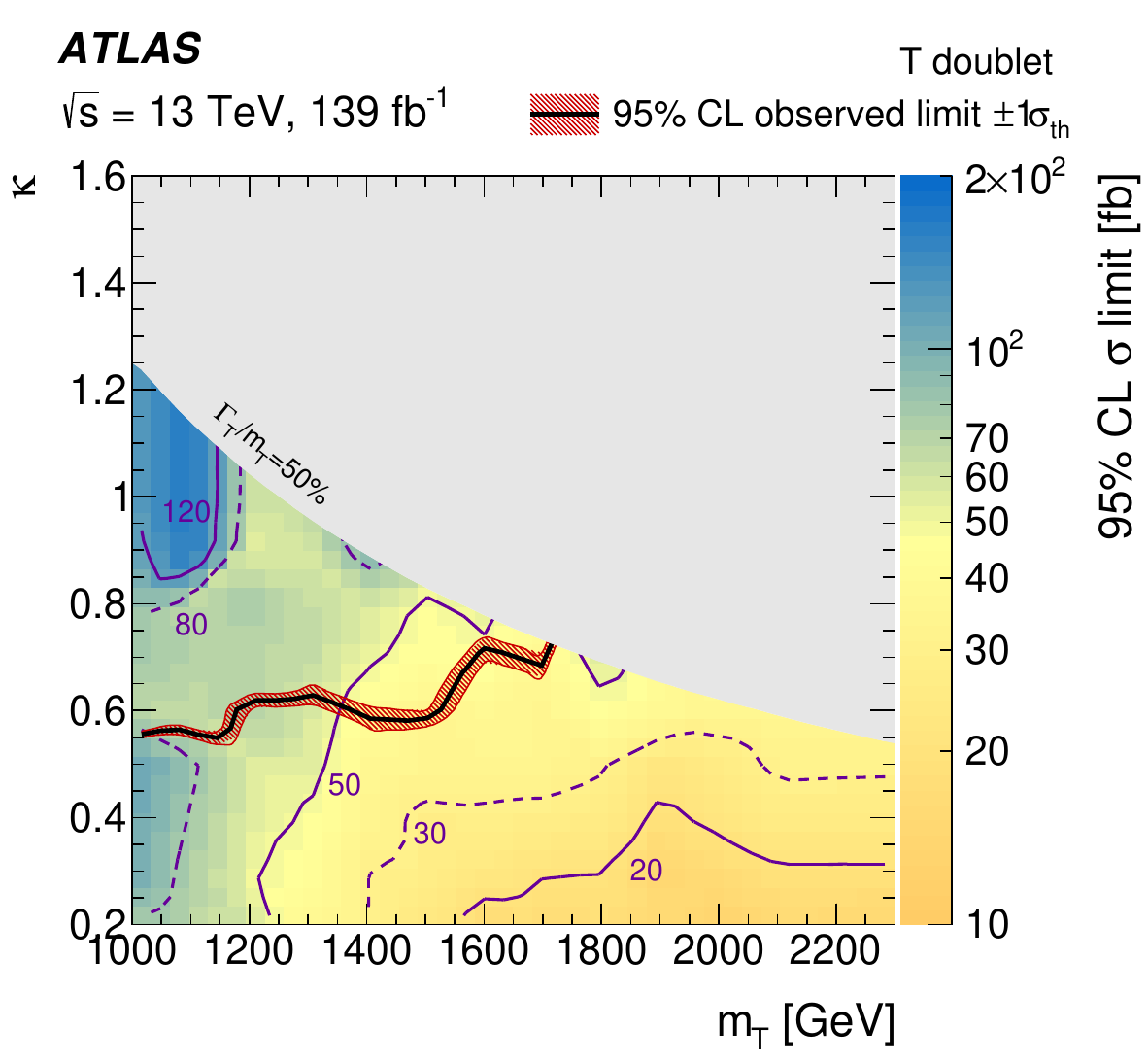}}
\hfill
\subfloat[]{\includegraphics[width=0.48\textwidth]{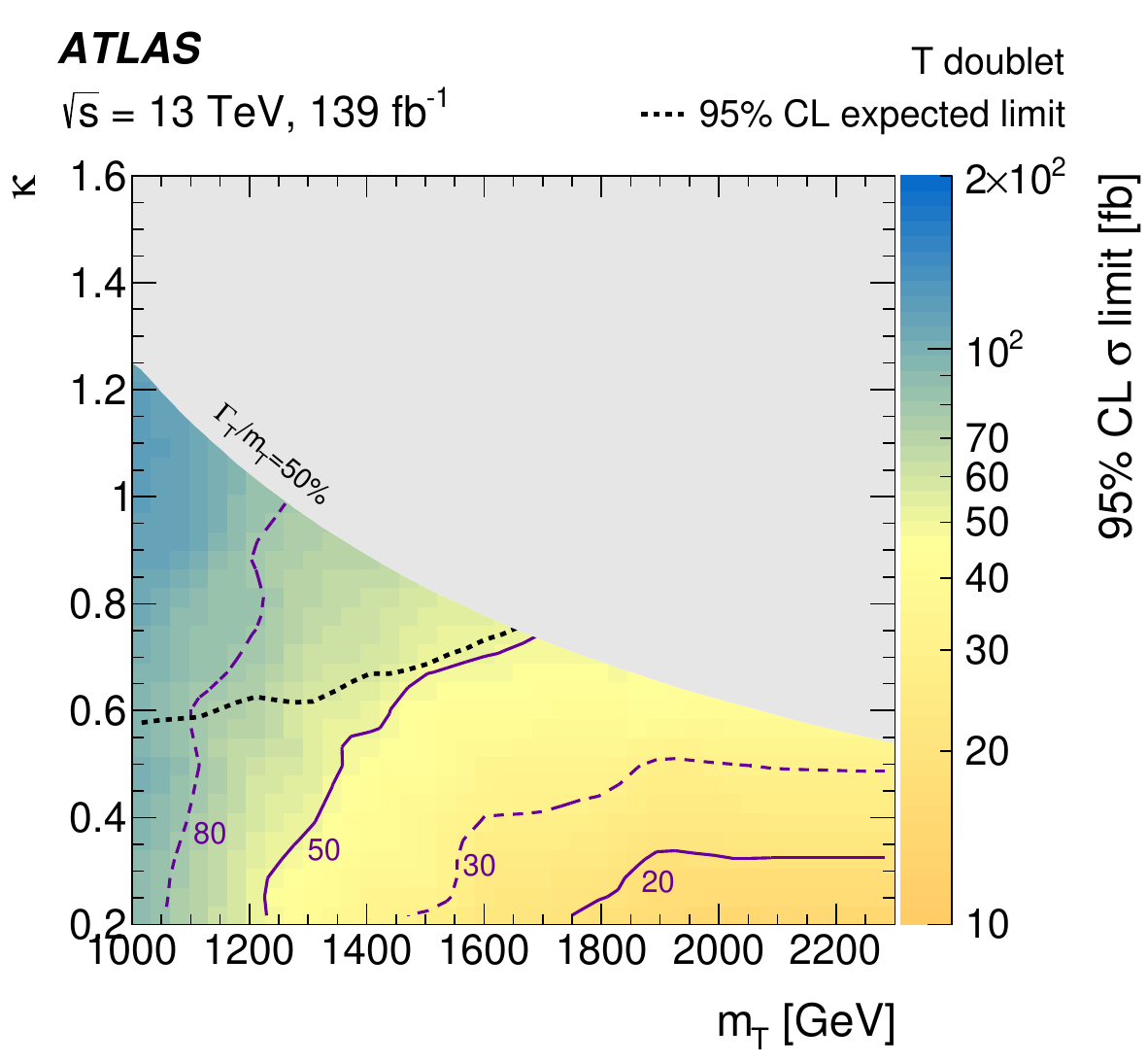}}
\caption{(a) Observed and (b) expected 95\% CL exclusion limits on the cross section times branching
ratio of single $T$-quark production as a function of the universal coupling constant $\kappa$ and the
$T$-quark mass in the SU(2) doublet scenario. The red hashed area around the observed limit corresponds to the theoretical uncertainty of the NLO cross-section prediction.
All values of $\kappa$
above the black contour line are excluded at each mass point. The purple contour lines denote exclusion
limits of equal cross section times branching ratio in units of fb. Limits are only presented in the
regime $\Gamma_{T}/m_{T} \leq 50$\%, where the theory calculations are known to be valid.}
\label{fig:limits_2D_temp_mass_kappa_TD}
\end{figure*}
\begin{figure*}[b!]
\centering
\subfloat[]{\includegraphics[width=.48\textwidth]{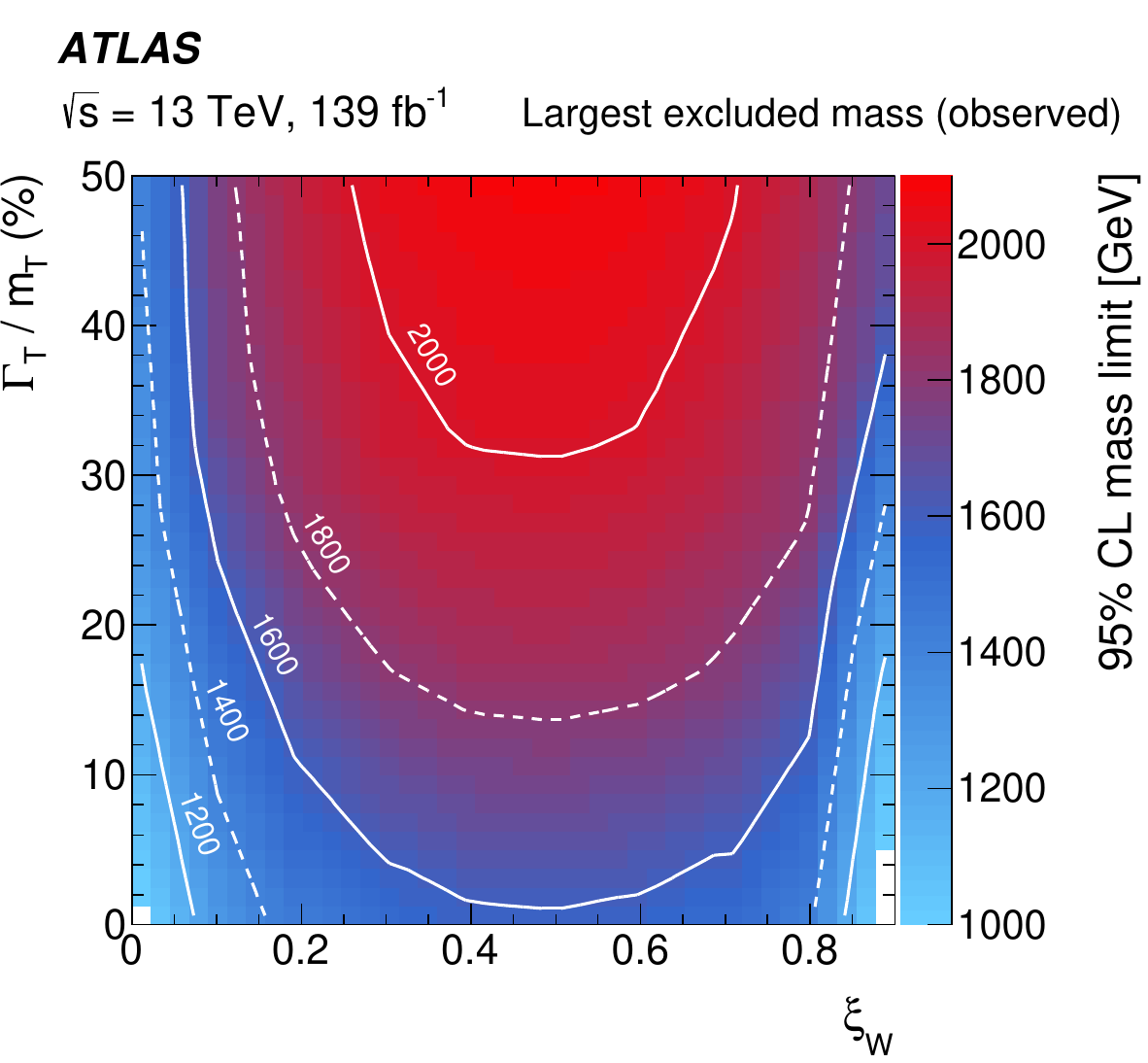}}
\hfill
\subfloat[]{\includegraphics[width=.48\textwidth]{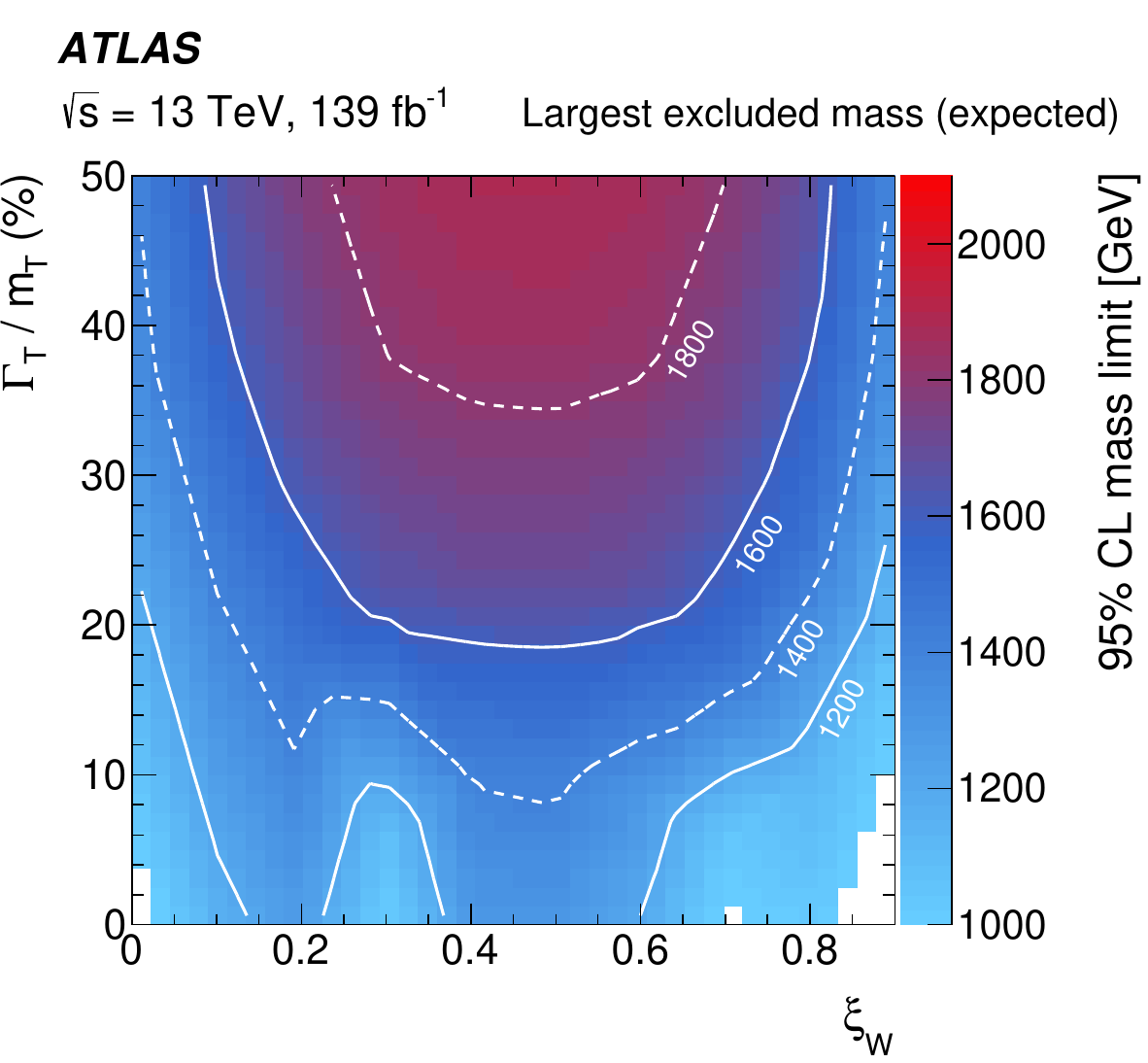}}\\
\caption{(a) Observed and (b) expected upper limits at 95\% CL on the $T$-quark
mass as a function of the relative resonance width ($\Gamma_{T}/m_{T}$) and the relative coupling parameter $\xi_{W}$,
for the assumption $\BR(T \rightarrow Ht)~=~\BR(T \rightarrow Zt)$. The white contour lines denote
exclusion limits of equal mass in units of \gev. The white regions represent points in parameter space
that are not excluded for any mass in the considered range. }
\label{fig:limits_2D_temp_xi_gamma}
\end{figure*}


\FloatBarrier

\section{Conclusion}
\label{sec:conclusion}
 
A search for the single production of up-type vector-like quarks ($T$), with subsequent decays
$T \rightarrow Ht$ with $H \rightarrow b\bar{b}$ or $T \rightarrow Zt$ with $Z \rightarrow q\bar{q}$, is
presented. The search uses $pp$ collision data at $\sqrt{s}=13$~\TeV, collected by the ATLAS detector
at the LHC during the 2015--2018 Run~2 data-taking period. The recorded dataset corresponds to an integrated
luminosity of 139~$\ifb$. Events are analysed in the lepton+jets final state, characterised by the
presence of a single lepton and multiple jets and $b$-jets in the event. The search exploits the
presence of boosted, hadronically decaying Higgs and vector bosons, and hadronically
or leptonically decaying top quarks in signal events.
 
No significant excess above Standard Model expectations is observed, and 95\% CL upper
limits are set on the production cross section of single $T$ quarks.
The results are interpreted in benchmark scenarios to set limits on the mass and universal coupling
strength ($\kappa$) of the vector-like quark. For singlet $T$ quarks, all masses below 2.1~\TeV\ are excluded at
couplings $\kappa \geq 0.6$, while the limits extend down to $\kappa=0.3$ for a $T$-quark mass of 1.6~\TeV.
For $T$ quarks in the doublet scenario, where the production cross section is much lower, coupling values
as low as $\kappa=0.55$ are excluded at a $T$-quark mass of 1.0~\TeV. The limits extend up to 1.68~\TeV\ and
$\kappa=0.75$, at a threshold of 50\% relative $T$-quark width.


\section*{Acknowledgements}


We thank CERN for the very successful operation of the LHC, as well as the
support staff from our institutions without whom ATLAS could not be
operated efficiently.
 
We acknowledge the support of
ANPCyT, Argentina;
YerPhI, Armenia;
ARC, Australia;
BMWFW and FWF, Austria;
ANAS, Azerbaijan;
CNPq and FAPESP, Brazil;
NSERC, NRC and CFI, Canada;
CERN;
ANID, Chile;
CAS, MOST and NSFC, China;
Minciencias, Colombia;
MEYS CR, Czech Republic;
DNRF and DNSRC, Denmark;
IN2P3-CNRS and CEA-DRF/IRFU, France;
SRNSFG, Georgia;
BMBF, HGF and MPG, Germany;
GSRI, Greece;
RGC and Hong Kong SAR, China;
ISF and Benoziyo Center, Israel;
INFN, Italy;
MEXT and JSPS, Japan;
CNRST, Morocco;
NWO, Netherlands;
RCN, Norway;
MEiN, Poland;
FCT, Portugal;
MNE/IFA, Romania;
MESTD, Serbia;
MSSR, Slovakia;
ARRS and MIZ\v{S}, Slovenia;
DSI/NRF, South Africa;
MICINN, Spain;
SRC and Wallenberg Foundation, Sweden;
SERI, SNSF and Cantons of Bern and Geneva, Switzerland;
MOST, Taiwan;
TENMAK, T\"urkiye;
STFC, United Kingdom;
DOE and NSF, United States of America.
In addition, individual groups and members have received support from
BCKDF, CANARIE, Compute Canada and CRC, Canada;
PRIMUS 21/SCI/017 and UNCE SCI/013, Czech Republic;
COST, ERC, ERDF, Horizon 2020 and Marie Sk{\l}odowska-Curie Actions, European Union;
Investissements d'Avenir Labex, Investissements d'Avenir Idex and ANR, France;
DFG and AvH Foundation, Germany;
Herakleitos, Thales and Aristeia programmes co-financed by EU-ESF and the Greek NSRF, Greece;
BSF-NSF and MINERVA, Israel;
Norwegian Financial Mechanism 2014-2021, Norway;
NCN and NAWA, Poland;
La Caixa Banking Foundation, CERCA Programme Generalitat de Catalunya and PROMETEO and GenT Programmes Generalitat Valenciana, Spain;
G\"{o}ran Gustafssons Stiftelse, Sweden;
The Royal Society and Leverhulme Trust, United Kingdom.
 
The crucial computing support from all WLCG partners is acknowledged gratefully, in particular from CERN, the ATLAS Tier-1 facilities at TRIUMF (Canada), NDGF (Denmark, Norway, Sweden), CC-IN2P3 (France), KIT/GridKA (Germany), INFN-CNAF (Italy), NL-T1 (Netherlands), PIC (Spain), ASGC (Taiwan), RAL (UK) and BNL (USA), the Tier-2 facilities worldwide and large non-WLCG resource providers. Major contributors of computing resources are listed in Ref.~\cite{ATL-SOFT-PUB-2021-003}.


\clearpage

\printbibliography

\clearpage
 
 
\begin{flushleft}
\hypersetup{urlcolor=black}
{\Large The ATLAS Collaboration}

\bigskip

\AtlasOrcid[0000-0002-6665-4934]{G.~Aad}$^\textrm{\scriptsize 102}$,
\AtlasOrcid[0000-0002-5888-2734]{B.~Abbott}$^\textrm{\scriptsize 120}$,
\AtlasOrcid[0000-0002-1002-1652]{K.~Abeling}$^\textrm{\scriptsize 55}$,
\AtlasOrcid[0000-0002-8496-9294]{S.H.~Abidi}$^\textrm{\scriptsize 29}$,
\AtlasOrcid[0000-0002-9987-2292]{A.~Aboulhorma}$^\textrm{\scriptsize 35e}$,
\AtlasOrcid[0000-0001-5329-6640]{H.~Abramowicz}$^\textrm{\scriptsize 151}$,
\AtlasOrcid[0000-0002-1599-2896]{H.~Abreu}$^\textrm{\scriptsize 150}$,
\AtlasOrcid[0000-0003-0403-3697]{Y.~Abulaiti}$^\textrm{\scriptsize 117}$,
\AtlasOrcid[0000-0003-0762-7204]{A.C.~Abusleme~Hoffman}$^\textrm{\scriptsize 137a}$,
\AtlasOrcid[0000-0002-8588-9157]{B.S.~Acharya}$^\textrm{\scriptsize 69a,69b,p}$
% ...
\bigskip
\\

% ...
$^{29}$Physics Department, Brookhaven National Laboratory, Upton NY; United States of America.\\
% ...
$^{35}$$^{(a)}$Facult\'e des Sciences Ain Chock, R\'eseau Universitaire de Physique des Hautes Energies - Universit\'e Hassan II, Casablanca;$^{(b)}$Facult\'{e} des Sciences, Universit\'{e} Ibn-Tofail, K\'{e}nitra;$^{(c)}$Facult\'e des Sciences Semlalia, Universit\'e Cadi Ayyad, LPHEA-Marrakech;$^{(d)}$LPMR, Facult\'e des Sciences, Universit\'e Mohamed Premier, Oujda;$^{(e)}$Facult\'e des sciences, Universit\'e Mohammed V, Rabat;$^{(f)}$Institute of Applied Physics, Mohammed VI Polytechnic University, Ben Guerir; Morocco.\\
% ...
$^{55}$II. Physikalisches Institut, Georg-August-Universit\"{a}t G\"ottingen, G\"ottingen; Germany.\\
% ...
$^{69}$$^{(a)}$INFN Gruppo Collegato di Udine, Sezione di Trieste, Udine;$^{(b)}$ICTP, Trieste;$^{(c)}$Dipartimento Politecnico di Ingegneria e Architettura, Universit\`a di Udine, Udine; Italy.\\
% ...
$^{102}$CPPM, Aix-Marseille Universit\'e, CNRS/IN2P3, Marseille; France.\\
% ...
$^{117}$Department of Physics, New York University, New York NY; United States of America.\\
% ...
$^{120}$Homer L. Dodge Department of Physics and Astronomy, University of Oklahoma, Norman OK; United States of America.\\
% ...
$^{137}$$^{(a)}$Departamento de F\'isica, Pontificia Universidad Cat\'olica de Chile, Santiago;$^{(b)}$Millennium Institute for Subatomic physics at high energy frontier (SAPHIR), Santiago;$^{(c)}$Instituto de Investigaci\'on Multidisciplinario en Ciencia y Tecnolog\'ia, y Departamento de F\'isica, Universidad de La Serena;$^{(d)}$Universidad Andres Bello, Department of Physics, Santiago;$^{(e)}$Instituto de Alta Investigaci\'on, Universidad de Tarapac\'a, Arica;$^{(f)}$Departamento de F\'isica, Universidad T\'ecnica Federico Santa Mar\'ia, Valpara\'iso; Chile.\\
% ...
$^{150}$Department of Physics, Technion, Israel Institute of Technology, Haifa; Israel.\\
$^{151}$Raymond and Beverly Sackler School of Physics and Astronomy, Tel Aviv University, Tel Aviv; Israel.\\
% ...
$^{a}$ Also Affiliated with an institute covered by a cooperation agreement with CERN.\\
$^{b}$ Also at An-Najah National University, Nablus; Palestine.\\
$^{c}$ Also at Borough of Manhattan Community College, City University of New York, New York NY; United States of America.\\
$^{d}$ Also at Bruno Kessler Foundation, Trento; Italy.\\
$^{e}$ Also at Center for High Energy Physics, Peking University; China.\\
$^{f}$ Also at Center for Interdisciplinary Research and Innovation (CIRI-AUTH), Thessaloniki ; Greece.\\
% ...
$^{p}$ Also at Department of Physics, King's College London, London; United Kingdom.\\
% ...

\end{flushleft}


\end{document}